\documentclass[useAMS,usegraphicx,usenatbib]{mn2e}

\usepackage{graphicx}
\usepackage{amssymb,amsmath}

\def\H2{{{\rm H}_2}}
\def\CO{{\rm CO}}

\def\refjnl#1{{\rm#1}}%
\newcommand\aj{\refjnl{AJ}}%
\newcommand\apj{\refjnl{ApJ}}%
\newcommand\apjl{\refjnl{ApJ}}%
\newcommand\apjs{\refjnl{ApJS}}%
\newcommand\aap{\refjnl{A\&A}}%
\newcommand\mnras{\refjnl{MNRAS}}%
\newcommand\nat{\refjnl{Nature}}%
\newcommand{\acknowledgements}{\begin{small}\section*{Acknowledgments}\end{small}}

\newcommand{\centerincludegraphics}[2][]{%
\raisebox{\dimexpr-.5\height+.5\ht\strutbox\relax}{\includegraphics[#1]{#2}}}

\begin{document}

\title[The case for a linear star formation -- $\H2$ relation]{Lessons from cosmic history: The case for a linear star formation -- $\H2$ relation}
\author[R. Feldmann]{Robert Feldmann,$^1$\thanks{Hubble fellow, feldmann@berkeley.edu}\\
$^1$ Department of Astronomy, University of California, Berkeley, CA 94720-3411, USA}

\maketitle

\begin{abstract}
Observations show that star formation in galaxies is closely correlated with the abundance of molecular hydrogen. Modeling this empirical relation from first principles proves challenging, however, and many questions regarding its properties remain open. For instance, the exact functional form of the relation is still debated and it is also unknown whether it applies at $z>4$, where $\CO$ observations are sparse. Here, we analyze how the shape of the star formation -- gas relation affects the cosmic star formation history and global galaxy properties using an analytic model that follows the average evolution of galaxies in dark matter halos across cosmic time. We show that a linear relation with an $\H2$ depletion time of $\sim{}2.5$ Gyr, as found in studies of nearby galaxies, results in good agreement with current observations of galaxies at both low and high redshift. These observations include the evolution of the cosmic star formation rate density, the $z\sim{}4-9$ UV luminosity function, the evolution of the mass -- metallicity relation, the relation between stellar and halo mass, and the gas-to-stellar mass ratios of galaxies. In contrast, the short depletion times that result from adopting a highly super-linear star formation -- gas relation lead to large star formation rates, substantial metal enrichment ($\sim{}0.1$ $Z_\odot$), and low gas-to-stellar mass ratios already at $z\gtrsim{}10$, in disagreement with observations. These results can be understood in terms of an equilibrium picture of galaxy evolution in which gas inflows, outflows, and star formation drive the metallicities and gas fractions toward equilibrium values that are determined by the ratio of the accretion time to the gas depletion time. In this picture,  the cosmic modulation of the accretion rate is the primary process that drives the evolution of stellar masses, gas masses, and metallicities of galaxies from high redshift until today.
\end{abstract}


\section{Introduction}

The relation between the star formation rate (SFR) per unit area, $\Sigma_{\rm SFR}$, and the gas surface density (of either molecular or neutral gas) has been a focal point of many galaxy evolution studies \citep{1959ApJ...129..243S, 1963ApJ...137..758S, 1989ApJ...344..685K, 1998ApJ...498..541K, 2002ApJ...569..157W, 2004ApJ...602..723H, 2004ApJ...606..271G, 2005ApJ...630..250K, 2007ApJ...669..289K, 2008AJ....136.2846B, 2008ApJ...680.1083R, 2009ApJ...697...55G, 2009ApJ...699..850K, 2010ApJ...714L.118D, 2010ApJ...714..287G, 2010ApJ...721..975O, 2010ApJ...722.1699S, 2010MNRAS.407.2091G, 2011ApJ...728...88G, 2011MNRAS.412..287N, 2011ApJ...730L..13B, 2011ApJ...730...72R, 2011ApJ...732..115F, 2012ApJ...745..183R, 2012ApJ...745..190L,2012ApJ...758..127F}. This relation holds over several orders of magnitude in $\Sigma_{\rm SFR}$ with a relatively small scatter if measured on $\sim{}$kpc sized patches of galaxies or for whole galaxies. Yet, despite these efforts, we still do not quite understand how this correlation comes about and, embarrassingly, what the precise functional form of this relation is. 

The reason for the latter is that neither SFRs nor molecular hydrogen ($\H2$) masses are directly accessible in extra-galactic observations. SFRs are inferred from a combination of ultraviolet (UV) and infrared (IR) observations, in order to capture the energy output of both obscured and unobscured star formation. Cold molecular gas is measured indirectly from emission lines of tracer molecules, such as carbon monoxide ($\CO$). While observational studies of nearby galaxies and galaxies out to $z\sim{}3$ seem to show that the relation between $\Sigma_{\rm SFR}$ and $\Sigma_\H2$ is close to linear \citep{2008AJ....136.2846B, 2010ApJ...713..686D, 2010ApJ...714L.118D, 2010MNRAS.407.2091G}, there is no clear theoretical understanding why this is so and, furthermore, there is the concern that systematics in the observables bias these results. For instance, it has been suggested that the star formation -- gas relation is super-linear with a power-law exponent of $\sim{}1.4$ on kpc scales, but that IR emission unrelated to recent star formation reduces the inferred slope, see \cite{2011ApJ...735...63L}, but cf. \cite{2011ApJ...730...72R}. Alternatively, systematic variation of the $\CO/\H2$ conversion factor with $\CO$ intensity, SFRs, or gas surface density could potentially disguise a highly super-linear relation and make it appear linear \citep{2011MNRAS.412..287N, 2012MNRAS.421.3127N}, cf. \cite{2012ApJ...758..127F}. Finally, the fitting itself has been targeted as a potential source of bias. A Bayesian re-analysis of the data presented by \cite{2008AJ....136.2846B} results in tentative evidence that the star formation -- gas relation is mildly \emph{sub}-linear and shows substantial variations from one galaxy to the next \citep{2012arXiv1210.1218S}.

Theoretical explanations of the star formation -- gas connection predict a variety of non-linear relationships, see e.g., \cite{2005ApJ...630..250K, 2009ApJ...699..850K, 2010ApJ...721..975O, 2011ApJ...731...41O, 2012ApJ...745...69K}. For instance, \cite{2009ApJ...699..850K} estimate that the power-law exponent is mildly non-linear ($\sim{}1.33$) at high gas surface densities ($\gtrsim{}100$ $M_\odot$ pc$^{-2}$),  while \cite{2011ApJ...731...41O} predict a much steeper, quadratic relation at such gas surface densities. This gas surface density regime is clearly highly relevant because it corresponds to typical gas surface densities of star forming galaxies at $z\gtrsim{}1$ and higher (e.g., \citealt{2010MNRAS.407.2091G}, see also Section \ref{sect:CSFH}). A significantly super-linear star formation -- gas relation has also been suggested by recent work based on (non-cosmological) galaxy evolution simulations that include a $\sim{}$pc-scale modeling of various stellar feedback channels \citep{2011MNRAS.417..950H, 2012arXiv1206.0011H}.

In this paper we take a very different approach to put constraints on the functional form of the star formation -- gas relation. We study how global galaxy properties vary depending on the chosen star formation -- gas relation and compare our model predictions with observations. In particular, we show that the evolution of the cosmic SFR density (SFRD; \citealt{1996ApJ...460L...1L, 1996MNRAS.283.1388M, 1999ApJ...519....1S}) at $z\sim{}4-10$ \citep{2007ApJ...670..928B, 2012arXiv1211.2230B, 2012arXiv1207.6105B} is highly sensitive to the adopted form of the star formation -- gas relation, and, hence, that it can be used to test its functional form. 

The paper is organized as follows. In Section \ref{sect:model} we describe the analytic model which underlies our predictions and present the star formation -- gas relations that enter our analysis. In Section \ref{sect:FuncForm} we first demonstrate that the model reproduces the observed stellar mass -- metallicity relation, the stellar mass -- halo mass relation, and the gas-to-stellar mass ratios of galaxies reasonably well \emph{if} a linear star formation -- gas relation is chosen. We show that predictions based on a strongly super-linear star formation -- gas relation are in conflict with observations. We then present how the cosmic star formation history is affected by the functional form of the star formation -- gas relation. Potential caveats of our approach are discussed in Section \ref{sect:Caveats}. In Section \ref{sect:Driver} we put our findings in a broader context, showing that galaxy evolution can be thought of as an evolution along a line of equilibrium points driven by the cosmic modulation of the accretion rate. We summarize our findings in Section \ref{sect:Summary} and close the paper with a comparison to previous work.

\section{The model}
\label{sect:model}

The analytic model that is used in the remainder of the paper follows closely the approach presented by \cite{2010ApJ...718.1001B} and \cite{2012ApJ...753...16K}. At its most basic level the model tracks the average evolution\footnote{Since the model does not use merger trees, it only follows the mean evolution of halos. For the quantities studied in this paper (the cosmic star formation history and the gas fraction of galaxies), such an average description is adequate.} of the dark matter, gas, and stellar content of halos using theoretically or observationally motivated accretion and transformation rates for the different mass components. For the convenience of the reader we begin by summarizing the main equations that enter the model and refer the reader to 
\cite{2012ApJ...753...16K} for more details. We then discuss the various star formation models that we explore in this work and conclude the section with a short descriptions of the actual set-up of the model. In the following, we adopt cosmological parameters consistent with WMAP 7-year data \citep{2011ApJS..192...18K}: $\Omega_{\rm m}=0.27$, $\Omega_{\rm b}=0.046$, $\Omega_\Lambda=0.73$, $h=0.7$, and $\sigma_8=0.81$.

\subsection{Basic equations}
\label{sect:BasicEquations}

{\it Accretion rates:}
The total (baryon plus dark matter) accretion rate at time $t$ onto a halo of a given mass $M_{\rm halo}$ is \citep{2006MNRAS.372..933N, 2008MNRAS.383..615N, 2012ApJ...753...16K}
\begin{equation}
\dot{M}_{\rm halo} = 1.06\times{}10^{12} M_\odot \left(\frac{M_{\rm halo}}{10^{12} M_\odot}\right)^{1.14} \frac{\dot{D}(t)} {D(t)^2}.
\label{eq:dotMhalo}
\end{equation}
$D(t)$ is the linear growth factor normalized to unity at the present epoch. The (average) evolution of the mass of a halo can be computed from its initial mass and the time- and mass-dependent accretion rate (\ref{eq:dotMhalo}).
The corresponding accretion rates of the gas and stellar components are
\begin{flalign}
\dot{M}_{\rm g, in} &= \epsilon_{\rm in} f_{\rm g, in}  f_{\rm b} \, \dot{M}_{\rm halo}, \label{eq:dotMgin} \\
\dot{M}_{\rm *, in} &= (1 - f_{\rm g, in})  f_{\rm b} \, \dot{M}_{\rm halo}.  \label{eq:dotMstarin}
\end{flalign}
Here, $\epsilon_{\rm in}$ specifies which fraction of the accreted gas reaches the galaxy at the center of the halo, $f_{\rm b}= \Omega_{\rm b}/\Omega_{\rm m}$ is the cosmic baryon fraction, and  $f_{\rm g, in}$ is the gas fraction (in units of  $f_{\rm b}$) of the accreted matter. 
We assume that accretion of gas onto the disk is completely quenched ($\epsilon_{\rm in}=0$) when the halo mass is above the virial shock mass $M_{\rm shock}(z)/10^{12} M_\odot=\max(2, 10^{12} M_\odot/(3M_{*,{\rm PS}}))$ \citep{2009Natur.457..451D, 2012ApJ...753...16K}. Here, $M_{*,{\rm PS}}(z)$ is the Press-Schechter mass at the given redshift, see \cite{2002MNRAS.336..112M}. Accretion is also quenched when the halo mass is so low that photo-ionization limits the baryon fraction in the halos to 50\% of $f_{\rm b}$ or less \citep{2006MNRAS.371..401H, 2008MNRAS.390..920O}. Specifically we assume $\epsilon_{\rm in}=0$ if the halo mass is below the mass $M_{\rm c}(z)$ given in Fig. 3 of \cite{2008MNRAS.390..920O}. We adopt $M_{\rm c}=0$ for $z\geq{}9$. We assume that accretion is not quenched ($\epsilon_{\rm in}=1$) if the halo mass lies between $M_{\rm c}$ and $M_{\rm shock}$. We compute $f_{\rm g, in}$ as described by \cite{2012ApJ...753...16K}. In brief, we first compute which fraction of the accretion rate is due to accreted halos of a given mass. The model provides us directly with the gas fraction of a halo of a given mass and we then calculate  $f_{\rm g, in}$ as the accretion-weighted mean gas fraction (in units of $f_{\rm b}$) of all accreted halos plus an additional non-halo contribution of 20\%.

{\it Evolution of the gas and stellar mass:}
Besides accretion, the baryonic components are subject to additional physical processes, such as star formation, stellar mass loss, and galactic outflows which we consider next. Specifically, we assume that the gas and stellar masses change according to
\begin{flalign}
\dot{M}_{\rm g} &= \dot{M}_{\rm g, in} - (1-R+\epsilon_{\rm out})\dot{M}_{\rm *, form}, \label{eq:dotMg} \\
\dot{M}_{\rm *} &= \dot{M}_{\rm *, in} + (1-R)\dot{M}_{\rm *, form}. \label{eq:dotMstar}
\end{flalign}
$\dot{M}_{\rm *, form}$ is the SFR of the galaxy (see below). $R$ is the fraction of the stellar mass of a single stellar population that is recycled to the ISM via, e.g., stellar winds. For the IMF assumed in this work \citep{2005ASSL..327...41C}, $R=0.46$ is appropriate. The model makes the simplifying assumption of instant recycling which is, however, adequate for our purposes \citep{2012ApJ...753...16K}. The parameter $\epsilon_{\rm out}$ is included to model large scale galactic outflows. We follow \cite{2012ApJ...753...16K} and assume $\epsilon_{\rm out}=1$, i.e., a gas outflow rate equal to the SFR, based on the estimates by \citep{2008ApJ...674..151E}. We discuss the impact of varying this and other model parameters in Section \ref{sect:Caveats}.

{\it The gas disk:}
For the modeling of the star formation (see below) we need to know how the gas is distributed in the galaxy. We make the ansatz that the gas forms an exponential disk, see e.g., \cite{2012ApJ...756..183B}, 
\begin{equation}
\Sigma_{\rm g}(r) = \Sigma_{\rm g}(0) \exp^{-r/R_d}.
\end{equation}
We assume that the scale radius $R_{\rm d}$ is set by the initial, pre-collapse angular momentum of the gas \citep{1998MNRAS.295..319M}. Specifically, 
\begin{equation}
R_{\rm d} = 0.5 \lambda R_{\rm vir},
\label{eq:Rd}
\end{equation}
where $R_{\rm vir}(M_{\rm halo}, z)$ and $\lambda=0.07$ \citep{2012ApJ...753...16K} are the virial radius and (median) spin parameter of the halo, respectively. The pre-factor takes into account that the half-mass radius of an exponential disk is $\sim{}1.7R_{\rm d}$. The predictions of equation (\ref{eq:Rd}) are in reasonable agreement with observations. \cite{2012ApJ...756..183B} find a mean scale radius of $6.1\pm{}0.7$ kpc (with significant galaxy-to-galaxy variations) for a sample of roughly MW-like galaxies in the local Universe, while equation (\ref{eq:Rd}) predicts $R_{\rm d}\sim{}9$ kpc for a $10^{12}$ $M_\odot$ halo at $z=0$. Reducing $\lambda$ to a more traditional value $\lambda\sim{}0.04-0.05$ \citep{1987ApJ...319..575B, 2001ApJ...555..240B, 2011MNRAS.410.1660D}, would lead to even better agreement, see also Section \ref{sect:Caveats}.

{\it Molecular hydrogen:}
The abundance of molecular gas plays an important role in many of the star formation models that we consider in the paper (see below).
We compute the $\H2$ surface density $\Sigma_\H2=f_\H2\Sigma_{\rm g}$ following \cite{2008ApJ...689..865K, 2009ApJ...693..216K}:
\begin{flalign}
f_\H2(\Sigma_{\rm g},Z) &= \begin{cases}1 -  \frac{0.75 s}{1+0.25 s}, & \mbox{if }s<2 \\ 0, & \mbox{otherwise} \end{cases} \label{eq:fH2} \\
s&=\frac{\ln( 1 + 0.6\chi + 0.01\chi^2)}{0.6 \tau_c}\\
\chi &= \frac{3.1}{4.1}\left(1+3.1 (Z/Z_\odot)^{0.365} \right)\\
\tau_c&=320\, c\, \frac{\Sigma_{\rm g}}{{\rm g} {\rm cm}^{-2}} Z/Z_\odot,\,c=5. \label{eq:tauc}
\end{flalign}
Here, $Z=M_{\rm Z}/M_{\rm g}$ is the average metallicity of the gas and we effectively assume that there are no strong metallicity gradients. We note that gas disks in nearby galaxies often show a moderate radial metallicity gradient of the order of $\sim{}0.05-0.1$ dex per kpc  \citep{1994ApJ...420...87Z}, corresponding to a factor $\sim{}1.5-4$ decrease in oxygen abundance from the center to the optical edge of the disk \citep{2010ApJS..190..233M}. We discuss how metallicity gradients affect our results in Section \ref{sect:Caveats}.

{\it Metallicity evolution:}
 The dependence of the $\H2$ abundance on the metallicity\footnote{More precisely, it is the dust-to-gas ratio that regulates the abundance of $\H2$. We assume that the dust-to-gas ratio scales proportional with metallicity as observed for galaxies enriched to $Z\gtrsim{}0.1\,Z_\odot$ \cite{2011ApJ...737...12L}.} of the ISM requires that we keep track of the build-up of metals in each galaxy. 
In the instantaneous recycling approximation and ignoring inflows and outflows the total metal production rate is proportional to both the yield $y$ of the stellar population and the star formation rate, i.e., $\dot{M}_{\rm Z, tot} = y(1-R)\dot{M}_{\rm *,form}$. Metals that are locked in stellar remnants do not contribute to the metallicity of the ISM, which therefore evolves according to $\dot{M}_{\rm Z} = y(1-R)\dot{M}_{\rm *,form}-Z(1-R)\dot{M}_{\rm *,form}$. As suggested by \cite{2012ApJ...753...16K} we add three processes to the basic picture. First, a fraction $\zeta(M_{\rm halo})=0.9\,\exp(-M_{\rm halo}/3\times{}10^{11} M_\odot)$ of the newly produced metals (predominantly in a hot, shocked phase) is removed from small halos by supernovae driven outflows. Second, cold and warm gas of the ISM and the associated metals are lost by galactic winds at a rate $\epsilon_{\rm out}(M_{\rm Z}/M_{\rm g})\dot{M}_{\rm *,form}$. Third, inflows from the IGM add metals at the rate  $Z_{\rm IGM}\dot{M}_{\rm g,in}$. Hence, the total metal mass of a galaxy changes according to
\begin{flalign}
\dot{M}_{\rm Z} = \, & \left[y(1-R)(1-\zeta)-(1-R+\epsilon_{\rm out})M_{\rm Z}/M_{\rm g}\right]\dot{M}_{\rm *,form} \nonumber \\
&+ Z_{\rm IGM}\dot{M}_{\rm g,in}.
\label{eq:dotMZ}
\end{flalign}
A single pair instability supernova can enrich halos at high redshifts to metallicities $\sim{}2\times{}10^{-5}$ \citep{2012ApJ...745...50W}. We therefore adopt the value $Z_{\rm IGM}=2\times{}10^{-5}$ as the metallicity of both the initial gas and the gas accreted onto halos.

{\it Differences to \cite{2012ApJ...753...16K}:}
Our implementation of the model follows closely the approach presented by \cite{2012ApJ...753...16K}. The only noteworthy differences are the following.
\begin{itemize}
\item We added a mass threshold below which gas accretion is quenched due to photo-ionization.
\item The metallicity evolution of the ISM (Equation \ref{eq:dotMZ}) accounts for the locking-up of metals in stellar remnants. Mathematically this amounts to replacing $\epsilon_{\rm out}$ in Equation 23 of \cite{2012ApJ...753...16K} with $\epsilon_{\rm out}+(1-R)$ and, hence, is equivalent to a change of the parameter $\epsilon_{\rm out}$.
We show in Section \ref{sect:Caveats} that varying $\epsilon_{\rm out}$ by a factor of a few does not strongly affect the SFRDs at $z>2$.
\item We use a variety of different star formation -- gas relations (see next section).
\end{itemize}
 
\subsection{Star formation}
We calculate the SFR of a galaxy by converting a gas surface density (either the total gas or the $\H2$ surface density) into a SFR surface density with the help of an empirically motivated or theoretically derived star formation -- gas relation. The SFR of a galaxy is then simply the integral of the SFR surface density over the area of the gas disk of the galaxy.

A variety of star formation -- gas relations have been suggested based on theoretical modeling and/or observations. The relationship often takes the form of a power-law with an exponent that ranges from approximately 1 (linear) to 2 (quadratic). In this paper we primarily focus on these two limiting cases for the following reasons. First, it allows us to extract more easily the systematic variations of galaxy properties resulting from the change of the star formation -- gas relation. Second, relations that fall between the two limiting cases do not add anything qualitatively new, but rather result in intermediate predictions for galaxy properties. Hence, much can be learned already from a study of the extreme cases. Finally, both a linear and a quadratic star formation -- gas relation are supported by various lines of evidence. It is thus clearly of high interest to test their specific implications for the evolution of galaxy properties. 

An approximately linear relation is empirically motivated by observational studies of the $\Sigma_{\rm SFR}-\Sigma_\H2$ relation in the local universe \citep{2008AJ....136.2846B, 2011ApJ...730L..13B} and out to, at least, $z\sim{}2$ \citep{2010MNRAS.407.2091G}. The depletion time in nearby galaxies is $\sim{}2-3$ Gyr \citep{2011ApJ...730L..13B}, but whether this is also true at high redshift is not so clear, see, e.g., \cite{2012arXiv1211.5743T}. High redshift samples typically include only galaxies with large star formation rates and gas masses, and, hence, in case of a slightly super-linear star formation -- gas relation it is difficult to distinguish a change in the depletion time with redshift from a change in depletion time due to the increase in gas surface density. Uncertainties related to the $\CO/\H2$ conversion factor and the limited spatial resolution of $\CO$ observations further complicate an accurate measurement of the depletion time at high redshift. 

Our default star formation -- gas relation is thus a \emph{linear relation} with a constant molecular depletion time of 2.5 Gyr \citep{2011ApJ...730L..13B}:
\begin{equation}
\Sigma_{\rm SFR} = \frac{\Sigma_\H2}{2.5\,{\rm Gyr}}.
\label{eq:SGlin}
\end{equation}
The star formation -- gas relation is assumed to hold at all redshifts and for all galaxies. Some galaxies may not follow this relation, e.g., massively starbursting galaxies that lie above the so-called main sequence of star formation, but their contribution to the cosmic star formation budget is estimated to be limited ($\sim{}10\%$ or less, see e.g., \citealt{2011ApJ...739L..40R}). 

A quadratic star formation -- gas has been proposed based on analytic and numerical models of feedback-regulated star formation, e.g., \cite{2011ApJ...731...41O}. Typically the quadratic relationship is expected to hold only at sufficiently high gas surface densities ($\gtrsim{}50$ $M_\odot$ pc$^{-2}$), because at lower surface densities the star formation -- gas relation reduces to effectively counting the number of molecular clouds in a beam and, hence, should still follow a linear relation. Some theoretical studies that aim at constraining the conversion factor between  $\CO$ emission and $\H2$ surface density also support the idea of an intrinsically quadratic star formation -- gas relation (\citealt{2012MNRAS.421.3127N}, but cf. \citealt{2012ApJ...758..127F}). 

We therefore also explore the implications of the following star formation -- gas relation that is quadratic at gas surface densities $\geq{}50$ $M_\odot$ pc$^{-2}$, but linear at smaller surface densities \citep{2011ApJ...731...41O}:
 \begin{equation}
 \Sigma_{\rm SFR} = \begin{cases}  0.1\,M_\odot\,{\rm yr}^{-1}\,{\rm kpc}^{-2} \left[\frac{\Sigma_\H2}{M_\odot{\rm pc}^{-2}}\right]^2 & \mbox{if }\Sigma_{\rm g}\geq{}50\,M_\odot\,{\rm pc}^{-2}  \\
                                                                \frac{\Sigma_\H2}{2\,{\rm Gyr}} & \mbox{if }\Sigma_{\rm g}<50\,M_\odot\,{\rm pc}^{-2}.\end{cases}
 \label{eq:SGquad}
 \end{equation}
The $z\gtrsim{}3$ predictions of this star formation -- gas relation are effectively indistinguishable from the predictions of a relation that extends the quadratic relationship to $\Sigma_{\rm g}<50\,M_\odot\,{\rm pc}^{-2}$, because surface densities of star forming galaxies at high redshift typically exceed $50\,M_\odot\,{\rm pc}^{-2}$, see Section \ref{sect:CSFH}. In the remainder of this paper we will refer to the relation given by (\ref{eq:SGquad}) as the \emph{strongly super-linear} star formation -- gas relation.

In Section \ref{sect:CSFH} we also test a few additional star formation -- gas relations suggested in the literature and explore how they affect the cosmic star formation history. These additional star formation -- gas relations are:

\begin{itemize}

\item A relation with a moderately sub-linear exponent $0.84$ as suggested by \cite{2012arXiv1210.1218S}:
\begin{equation}
\Sigma_{\rm SFR} = \left(\frac{\Sigma_\H2}{10\,M_\odot\,{\rm pc}^{-2}}\right)^{0.84}\frac{10\,M_\odot\,{\rm pc}^{-2}}{\rm 2.5\,Gyr}
\label{eq:SGS12}
\end{equation}

\item A relation suggested by \cite{2009ApJ...699..850K}:
\begin{equation}
\Sigma_{\rm SFR} = \frac{\Sigma_\H2}{\rm 2.6 Gyr} \begin{cases} \left(\frac{\Sigma_{\rm g}}{\rm 85 M_\odot {\rm pc}^{-2}}\right)^{-0.33}, & \mbox{if } \Sigma_{\rm g}<\rm 85\,M_\odot\,{\rm pc}^{-2} \\
\left(\frac{\Sigma_{\rm g}}{\rm 85 M_\odot {\rm pc}^{-2}}\right)^{0.33}, & \mbox{if } \Sigma_{\rm g}\geq{}\rm 85\,M_\odot\,{\rm pc}^{-2}. \end{cases}
\label{eq:SGK09}
\end{equation}

\item The ``universal'' star formation relation suggested by \cite{2012ApJ...745...69K}:
\begin{equation}
\Sigma_{\rm SFR} = 0.01 \frac{\Sigma_\H2}{t_{\rm ff}}, \textrm{ where } t_{\rm ff}=\min(t_{\rm ff, GMC}, t_{\rm ff, T}),
\label{eq:SGK12}
\end{equation}
and
\begin{flalign}
t_{\rm ff, GMC} &= \frac{\pi^{1/4}}{\sqrt{8}}\frac{\sigma}{G(\Sigma_{\rm GMC}^3\Sigma_{\rm g})^{1/4}},\,\Sigma_{\rm GMC}=85 M_\odot {\rm pc}^{-2}, \label{eq:tffGMC}\\
t_{\rm ff, T} &= \frac{\pi}{8}\frac{1}{\Omega} \label{eq:tffT}
\end{flalign}
In Equation (\ref{eq:tffGMC}), $\sigma$ is the gas velocity dispersion on scales comparable to the disk scale height (few 100 pc). We set it to 8 km/s, adequate for disk galaxies in the local universe \citep{2008AJ....136.2563W}. High redshift galaxies may have larger velocity dispersions \citep{2010Natur.463..781T}, but since typically $t_{\rm ff, GMC} > t_{\rm ff, T}$ for these galaxies \citep{2012ApJ...745...69K}, the choice of $\sigma$ does not matter in this case. Equation (\ref{eq:tffT}) includes the angular velocity of galactic rotation $\Omega$. We chose $\Omega=V_{\rm rot}/(1.7\,R_{\rm d})$ as our fiducial value. Here, 1.7 $R_{\rm d}$ is the half-mass radius of the gas disk and $V_{\rm rot}$ is the rotation velocity of the disk at 1.7 $R_{\rm d}$. The latter we equate with $V_{\rm max,h}$, the maximum circular velocity of a dark matter halo with NFW profile \citep{1997ApJ...490..493N} that hosts the given galaxy. We determine $V_{\rm max,h}$ for a halo of given mass and at a given redshift using the fits (converted to our definition of halo mass) by \cite{2012MNRAS.423.3018P} based on high resolution N-body simulations \citep{2011ApJ...740..102K}. The ansatz $V_{\rm rot}\approx{}V_{\rm max,h}$ has a firm observational justification, at least for disk galaxies in the local universe \citep{2012MNRAS.425.2610R}. In compact star formation galaxies at high redshift, however, it is conceivable that $V_{\rm rot}$ exceeds $V_{\rm max,h}$. If the latter is true, we potentially overestimate $t_{\rm ff, T}$ and, hence, potentially underestimate the SFR at high redshift. As we will show in Section \ref{sect:CSFH}, this only increases the conflict between this particular star formation -- gas relation and observations.

\end{itemize}

\begin{figure*}
\begin{tabular}{cc}
\includegraphics[width=80mm]{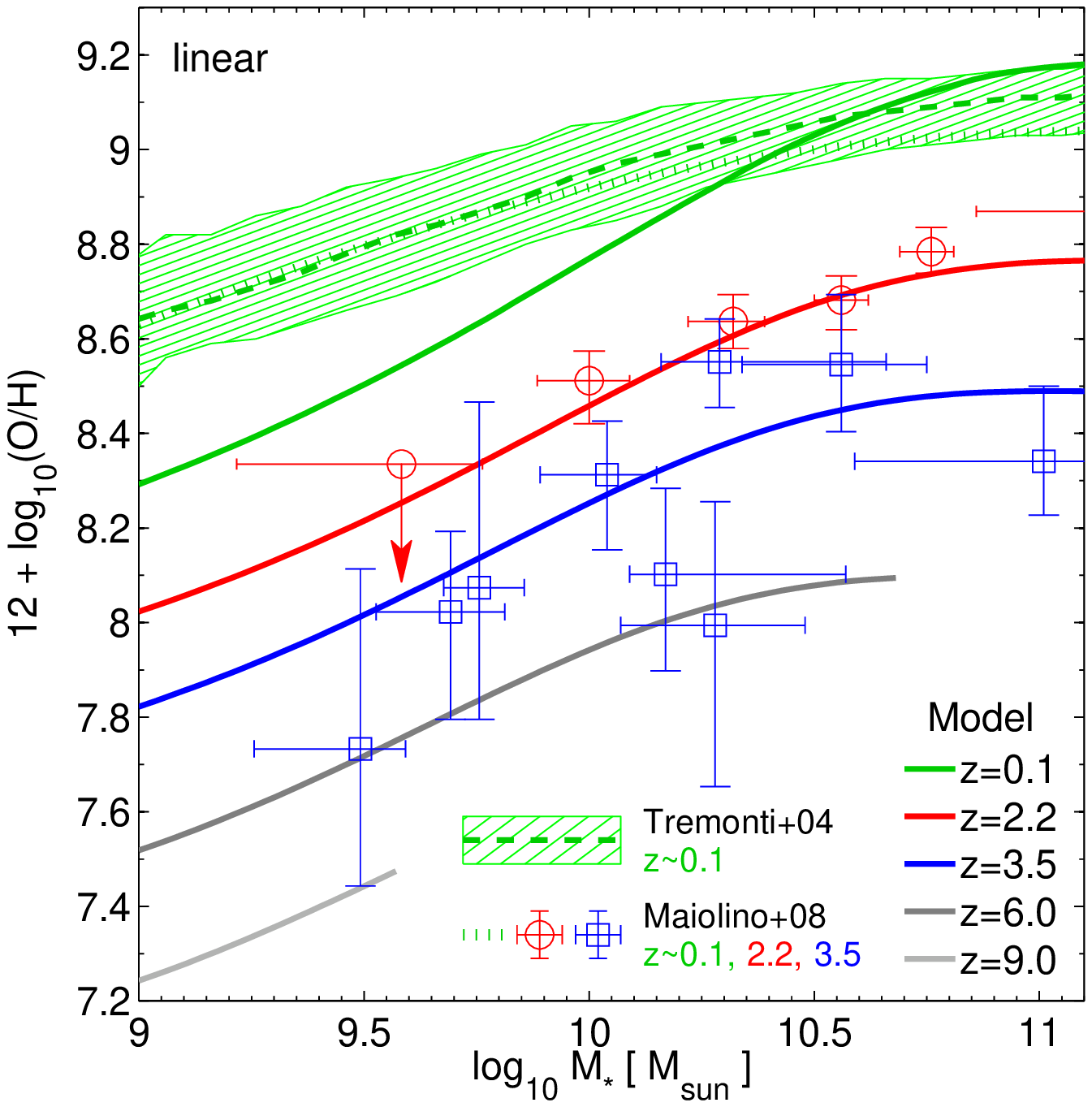} &
\includegraphics[width=80mm]{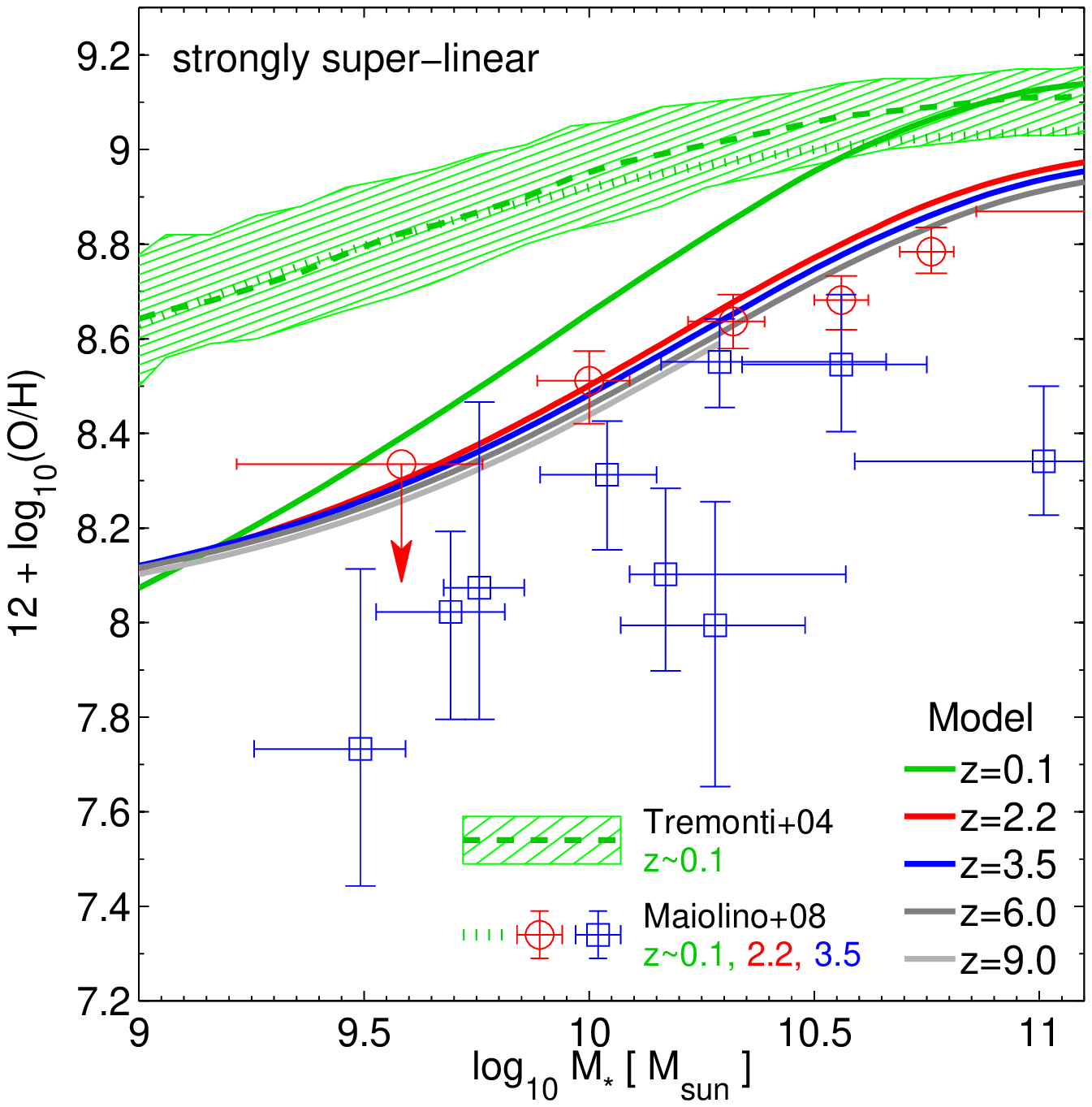} 
\end{tabular}
\caption{Metallicity vs stellar mass. (Left) Model predictions based on a linear $\Sigma_{\rm SFR}-\Sigma_\H2$ relation, Equation (\ref{eq:SGlin}). (Right) Model predictions based on the strongly super-linear relation given by Equation (\ref{eq:SGquad}). Solid lines are the model predictions for $z=0.1$, $z=2.2$, $z=3.5$, $z=6.0$, and $z=9.0$ (see legend). The dot-dashed line (median) and the shaded region (1-$\sigma$ scatter) at the top show the observed low redshift ($z\sim{}0.1$) mass-metallicity relation from SDSS \protect\citep{2004ApJ...613..898T}. The dotted line shows the results of a re-analysis of the \protect\cite{2004ApJ...613..898T} study by \protect\cite{2008ApJ...681.1183K}. Circles and squares are observational estimates for $z\sim{}2.2$ (based on stacked spectra; \protect\citealt{2006ApJ...644..813E}) and for $z\sim{}3.5$ (individual galaxies), respectively, by \protect\cite{2008A&A...488..463M}, who use the same IMF and metallicity calibrations as \protect\cite{2008ApJ...681.1183K}. Predictions based on a linear $\Sigma_{\rm SFR}-\Sigma_\H2$ relation with a constant $\sim{}2.5$ Gyr depletion time are in good agreement with observations of the mass-metallicity relation at $z\gtrsim{}2$.}
\label{fig:MZrelation}
\end{figure*}

\subsection{Running the model}

We have implemented the model described in the previous sections as a GNU Octave\footnote{http://www.gnu.org/software/octave} program. We adopt the fiducial choice of \cite{2012ApJ...753...16K} (see their Table 1) for the model parameters, and note that most values are justified based on either theoretical grounds (e.g., the recycling fraction $R$ is a consequence of the adopted IMF) or on observations (e.g., $f_{\rm b}$, $y$). We discuss the robustness of our predictions to changes in the parameter values in Section \ref{sect:Caveats}.

We assume that each halo initially contains a fraction $f_{\rm b}\sim{}0.17$ of its total mass in baryons. The large majority of the baryons is assumed to be in a gaseous component (90\%) with an initial metallicity of $Z_{\rm IGM}=2\times{}10^{-5}$. The remaining 10\%  are locked up in a stellar component formed during mergers at high redshift \citep{2011ApJ...739L..40R}. We checked that decreasing the initial stellar contribution from 10\% to 1\% (and increasing the initial gas contributions correspondingly) does not change any of our results in a significant way. We start the model at $z=30$ and run it forward until $z=0$ in 300 steps spaced equally in values of the expansion factor $a=1/(1+z)$. The model follows the concurrent evolution of 300 halos of different mass ranging from $3.2\times{}10^{5}$ $M_\odot$ to $1.3\times{}10^{8}$ $M_\odot$ at $z=30$ and from $1.4\times{}10^{7}$ $M_\odot$ to $8\times{}10^{15}$ $M_\odot$ at $z=0$ in a logarithmic spacing. 

\begin{figure*}
\begin{tabular}{cc}
\includegraphics[width=80mm]{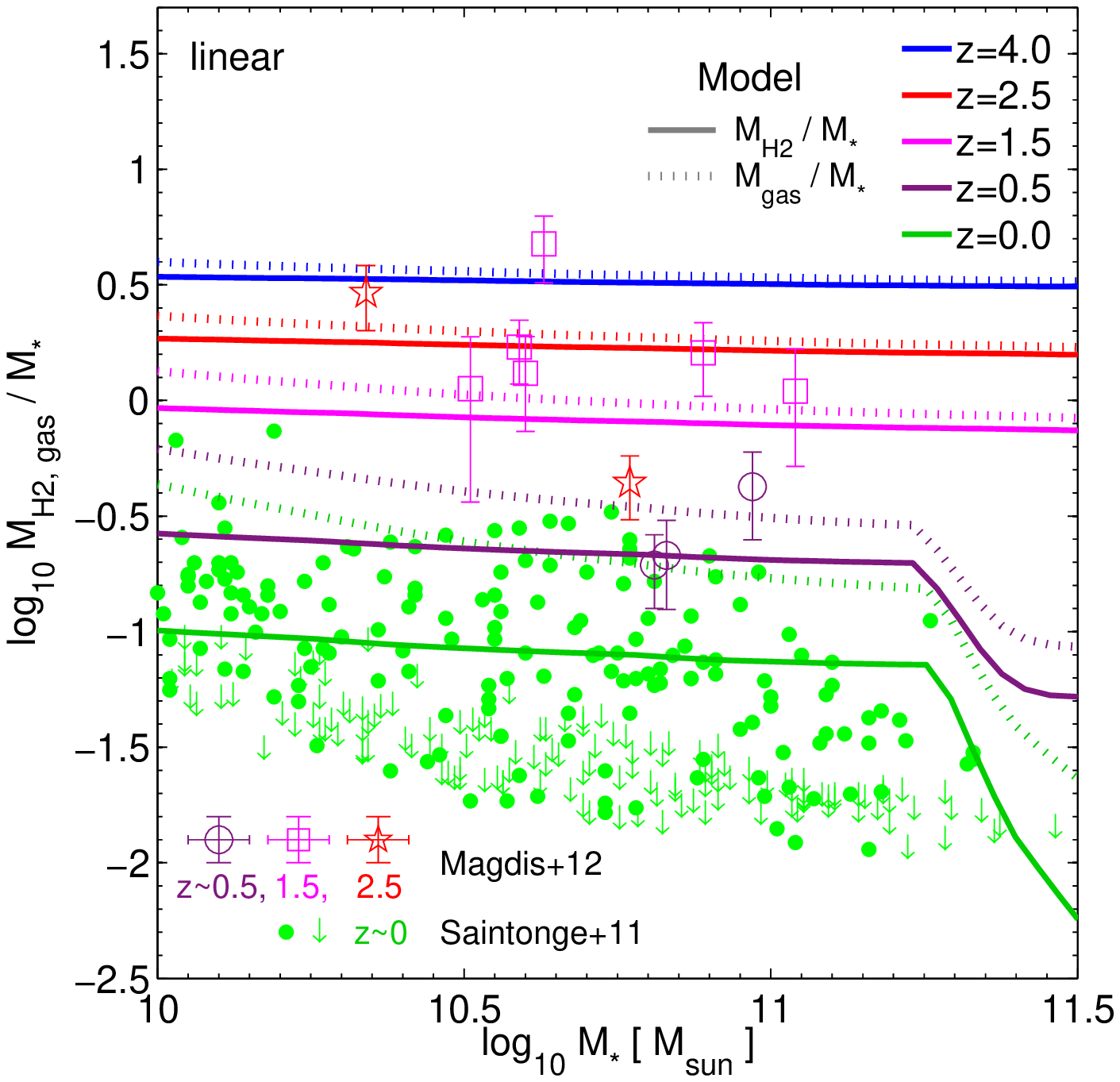} &
\includegraphics[width=80mm]{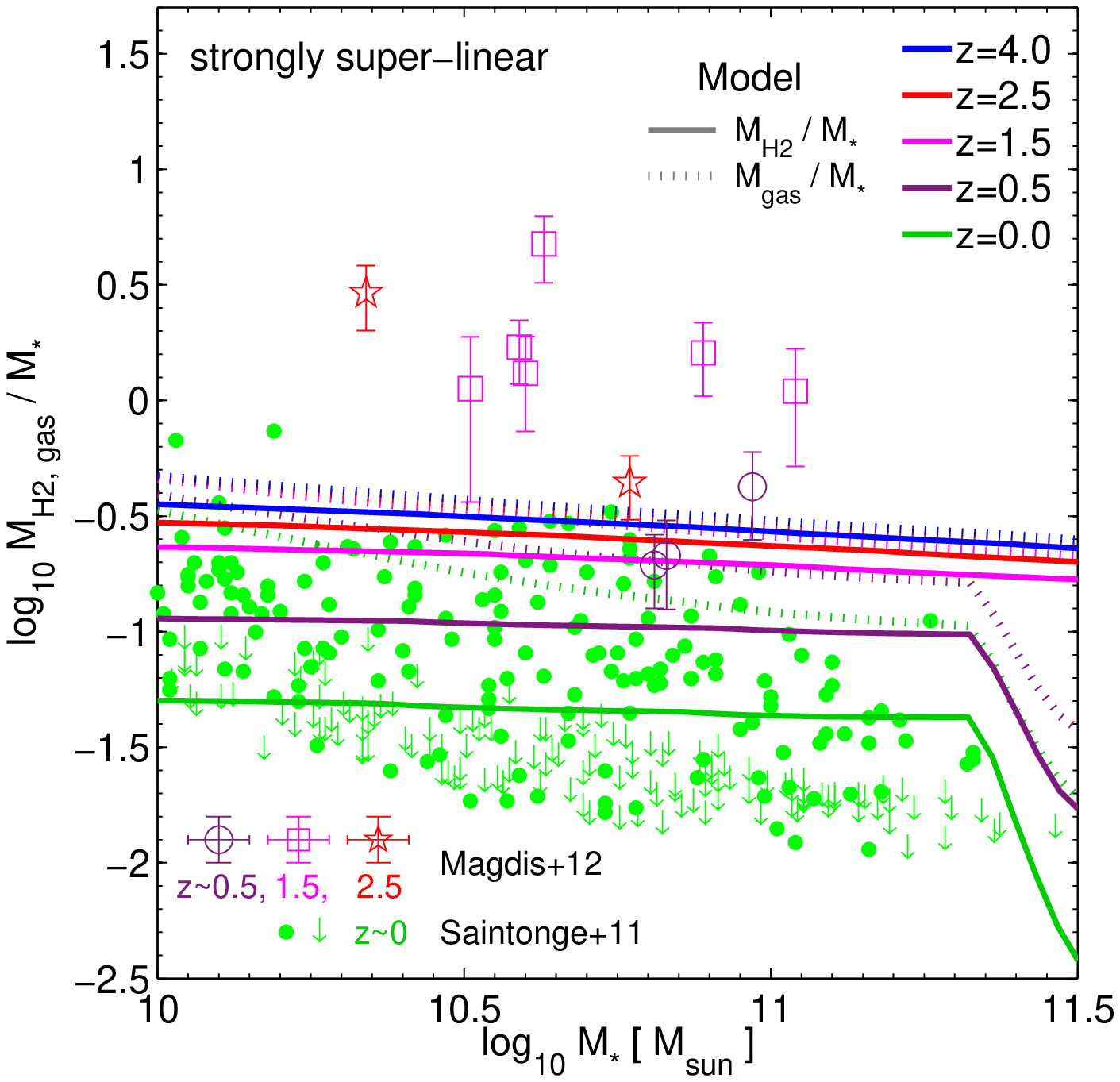}
\end{tabular}
\caption{Gas-to-stellar mass ratio in galaxies as function of their stellar mass. (Left) Model predictions based on a linear $\Sigma_{\rm SFR}-\Sigma_\H2$ relation, Equation (\ref{eq:SGlin}). (Right) Model predictions based on the strongly super-linear relation given by Equation (\ref{eq:SGquad}). Solid lines show the predicted molecular gas-to-stellar mass ratio, while dotted lines show the total gas-to-stellar mass ratio. Different colors correspond to different redshifts as indicated in the legend. The circles (arrows) show $>3-\sigma$ detections (upper limits) of the $\H2$ fraction of galaxies in the local universe \citep{2011MNRAS.415...32S}. Circles ($z\sim{}0.5$), squares ($z\sim{}1.5$), and stars ($z\sim{}2.5$) are measurements of cold gas masses for a sample of galaxies at higher redshift \citep{2012ApJ...760....6M}. These latter observations are based on dust infrared emission and are thus not affected by uncertainties related to the $\CO$-$\H2$ conversion. A linear $\Sigma_{\rm SFR}-\Sigma_\H2$ relation predicts a large increase in the gas fractions up to $\sim{}2$ in good agreement with the observed evolution, followed by only a modest change toward higher redshift, cf. \protect\cite{2012ApJ...758L...9M}.}
\label{fig:MH2}
\end{figure*}

\section{The functional form of the star formation -- gas relation}
\label{sect:FuncForm}

In this section, we analyze the importance of the functional form of the star formation -- gas relation for various galaxy properties and for the cosmic star formation history in the context of the model of Section \ref{sect:BasicEquations}. We will show that a strongly non-linear star formation -- gas relation results in predictions that are in conflict with observations, while predictions based on a linear star formation -- gas relation match the various observational data better and in most cases remarkably well.

\subsection{Metallicities, stellar and gas masses}
\label{sect:ZSG}

In Fig.~\ref{fig:MZrelation} we show the evolution of the stellar mass -- metallicity relation as predicted by the model for a linear and a strongly super-linear $\Sigma_{\rm SFR}-\Sigma_\H2$ relation, see Equations (\ref{eq:SGlin}) and (\ref{eq:SGquad}). We compare these predictions with observations of galaxies at $z\sim{}0$ \citep{2004ApJ...613..898T,2008ApJ...681.1183K}, at $z\sim{}2$ \citep{2006ApJ...644..813E, 2008A&A...488..463M}, and at $z\sim{}3.5$  \citep{2008A&A...488..463M}. Predictions based on a linear relation are in excellent agreement with the observed mass -- metallicity relation at $z\gtrsim{}2$. At low redshifts and low stellar masses the model somewhat underpredicts the metallicities. We suspect that this is a result of the particular modeling of the metal enriched blow-outs (parametrized by the $\zeta$ parameter), which we plan to address in future work. A strongly super-linear relation, however, severely underpredicts the evolution of the mass -- metallicity relation at high redshift. The evolution at $z<2$ is similar to the one for the linear star formation -- gas relation, because as those times typical gas surface densities  in galaxies falls below $\sim{}50$ $M_\odot$ pc$^{-2}$ and the star formation -- gas relation is effectively linear, see Equation (\ref{eq:SGquad}). In fact, if we had used a purely quadratic relation down to low gas surface densities, the mass -- metallicity relation would not evolve between $z=9$ and $z=0$ - in obvious disagreement with observations.

In Fig.~\ref{fig:MH2} we compare the predicted gas-to-stellar mass ratios with observations spanning the redshift range $z\sim{}0-3$. A strongly super-linear star formation -- gas relation underpredicts both the total gas-to-stellar mass ratio (see dotted lines) and the molecular gas-to-stellar mass ratio (see solid lines). At $z\sim{}0.5$ the model predictions are off by a factor $\sim{}3$. The disagreement becomes even more severe at higher redshift. At $z\sim{}2$ the observed gas-to-stellar mass ratios exceed the predicted ones by an order of magnitude. In contrast, we find good agreement between model predictions and observations for both the normalization and the redshift evolution of the gas-to-stellar mass ratio if the star formation -- gas relation is linear. For instance, the predicted fraction $M_\H2/(M_\H2+M_*)$ is 39\% at $z=1.5$ and 54\% at $z=2.5$, compared with $34\%$ at $z\sim{}1.2$ and $44\%$ at $z\sim{}2.3$ as found by \cite{2010Natur.463..781T} based on $\CO$ observations of a sample of massive star forming galaxies. It has been argued that the observed gas masses are potentially overestimated, because of systematic variations of the $\CO/\H2$ conversion factor with $\CO$ surface brightness that is unaccounted for \citep{2012MNRAS.426.1178N}. We therefore also compare our predictions with observations that estimate gas masses based on the thermal emission of dust \citep{2012ApJ...760....6M} to circumvent this potential issue. While this observational method also has its challenges, it is encouraging that we find again good agreement if the star formation -- gas relation is linear, but stark disagreement if the relation is strongly super-linear.

\begin{figure*}
\begin{tabular}{cc}
\includegraphics[width=80mm]{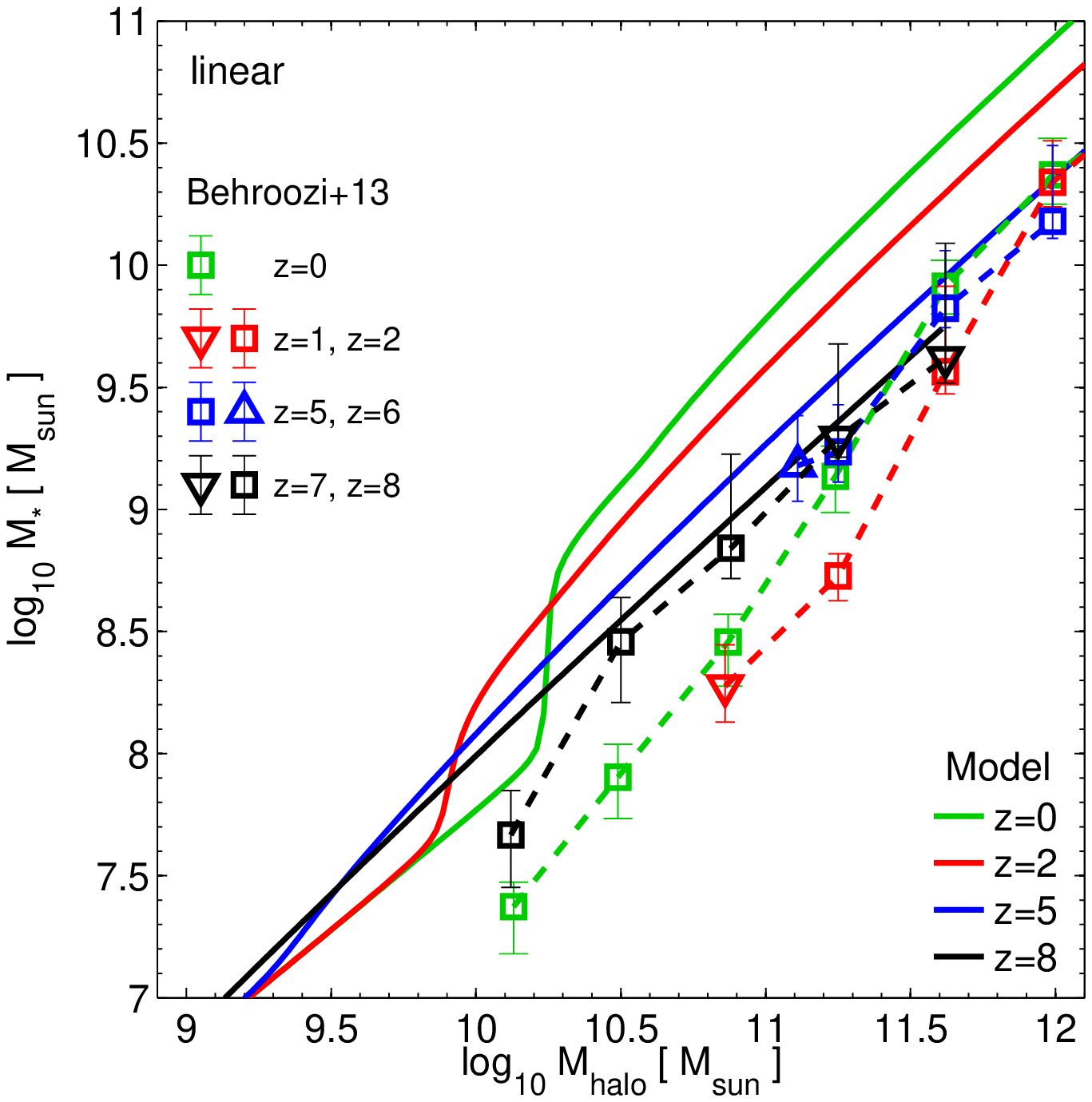} &
\includegraphics[width=80mm]{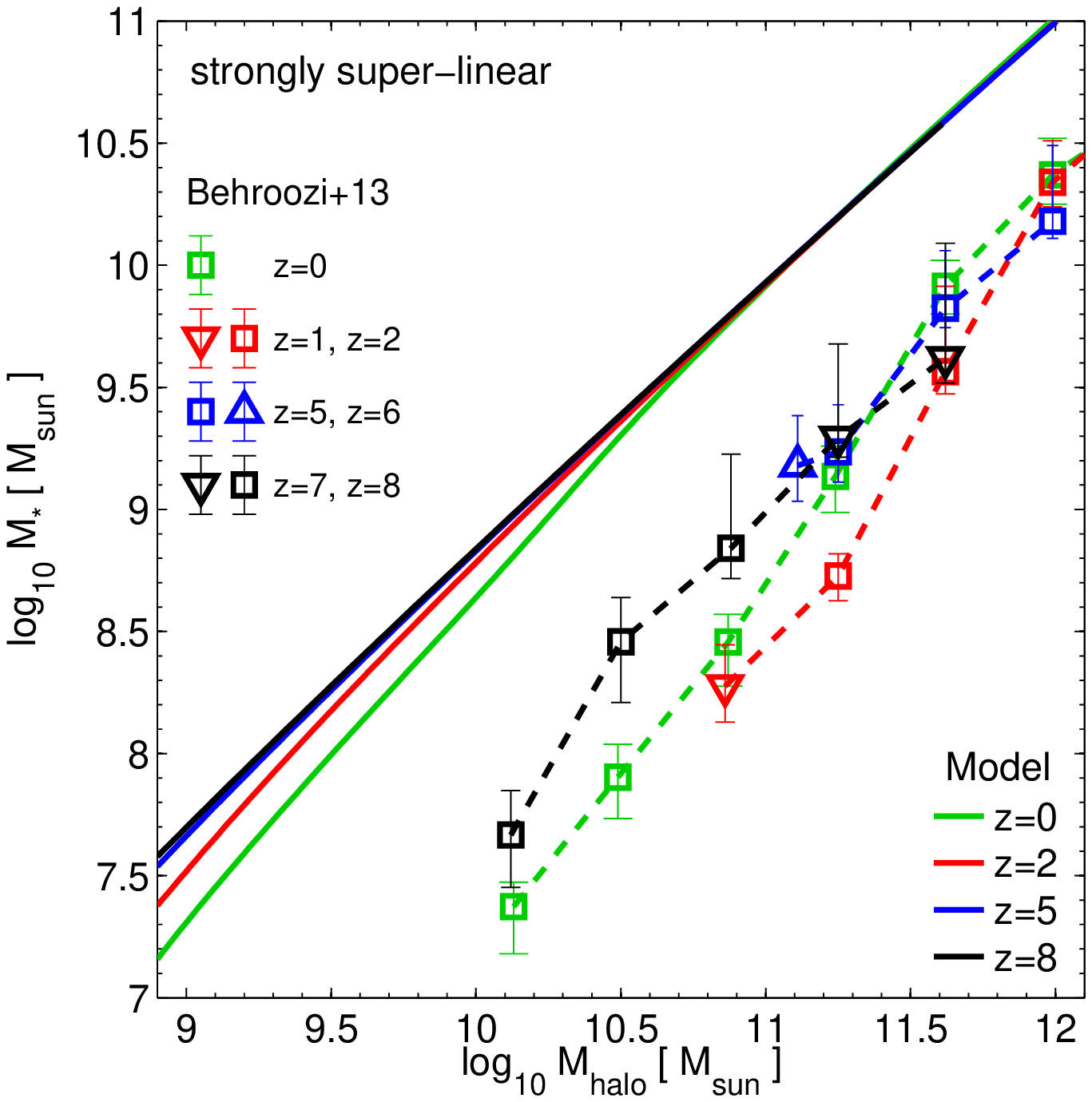} 
\end{tabular}
\caption{Relation between halo mass and stellar mass of galaxies. (Left) Model predictions based on a linear $\Sigma_{\rm SFR}-\Sigma_\H2$ relation, Equation (\ref{eq:SGlin}). (Right) Model predictions based on the strongly super-linear relation given by Equation (\ref{eq:SGquad}). Solid lines are the model predictions for $z=0$, $z=2$, $z=5$, and $z=8$ (see legend). Symbols (color coded according to redshift, see legend) are empirical estimates based on the abundance matching technique \citep{2012arXiv1209.3013B}. Because the evolution of the $M_{\rm halo}-M_*$ relation is weak, we combine, for a given redshift $z$, the abundance matching results from the  $z-1$ (downward pointing triangles), $z$ (squares), and $z+1$ (upward pointing triangles) redshift bins.}
\label{fig:MhaloMstar}
\end{figure*}

\begin{figure}
\begin{tabular}{c}
\includegraphics[width=80mm]{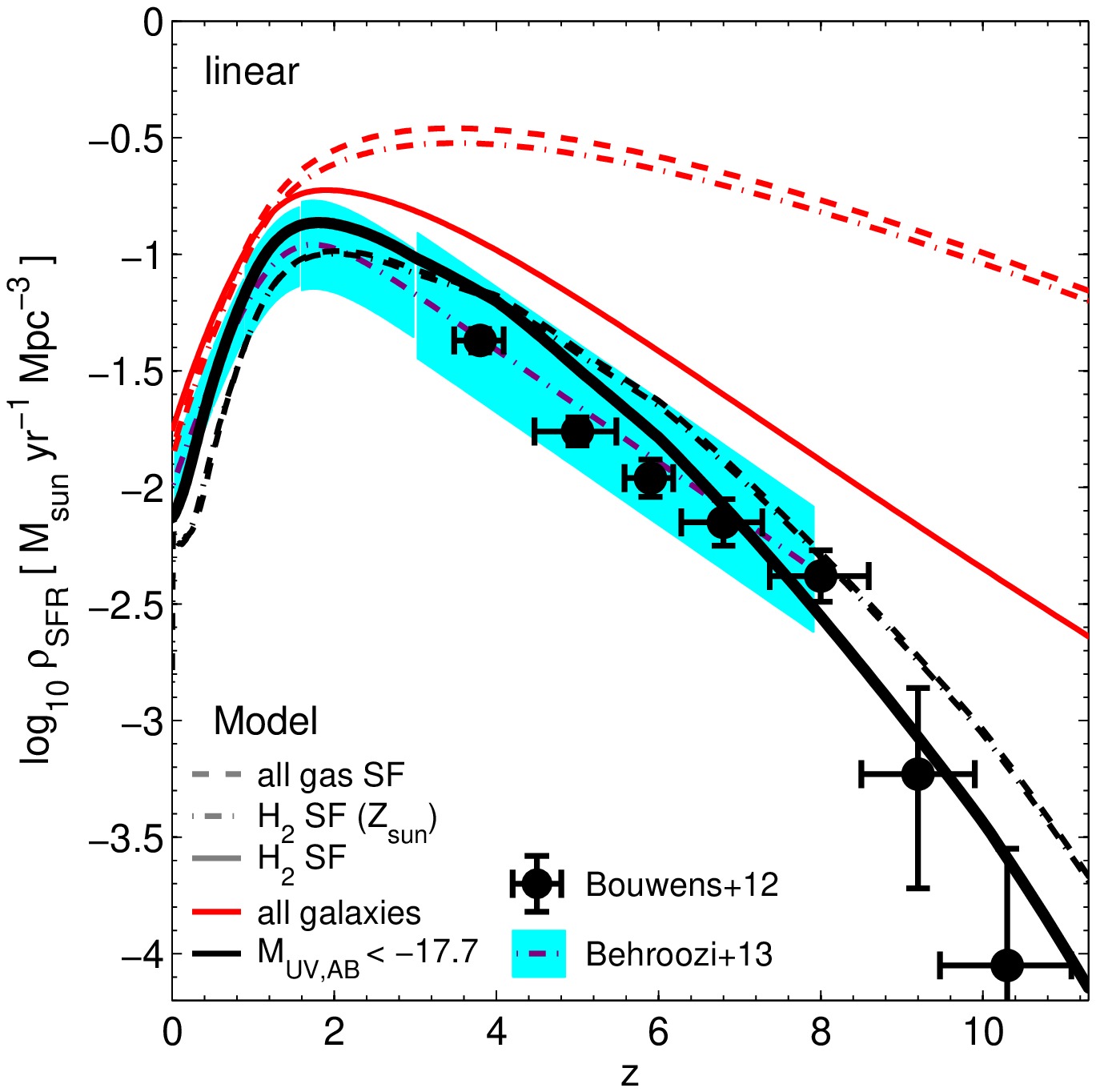}
\end{tabular}
\caption{
The cosmic star formation history as predicted by the model for a linear $\Sigma_{\rm SFR}-\Sigma_\H2$ relation (Equation \ref{eq:SGlin}; solid and dot-dashed lines) and a linear $\Sigma_{\rm SFR}-\Sigma_{\rm g}$ relation (Equation \ref{eq:SGlin} but replacing $\Sigma_\H2$ with $\Sigma_{\rm g}$; dashed lines), respectively, with a depletion time of 2.5 Gyr \protect\citep{2011ApJ...730L..13B}. The $\H2$ abundance is computed either based on the current gas metallicity (solid lines) or by assuming that the gas has solar metallicity (dot-dashed lines). Black lines (lower three curves) show the contribution of galaxies bright enough to be accessible to current observations, while red lines (upper three curves) include all galaxies down to the photo-ionization limit \protect\citep{2008MNRAS.390..920O}. The contribution of galaxies below the current detection limit depends sensitively on whether star formation correlates preferentially with $\H2$ or with the total gas mass in galaxies. Points with error bars are observational constraints from \protect\cite{2012ApJ...754...83B} and should be compared with the black lines. The dot-dashed line and the shaded region in the background show the mean and the systematic uncertainties of the cosmic star formation history as estimated by \protect\cite{2012arXiv1207.6105B} using abundance matching and a compilation of observational data from the literature. Predictions based on a linear $\Sigma_{\rm SFR}-\Sigma_\H2$ relation with a $\sim{}2.5$ Gyr $\H2$ depletion time are in good agreement with the observed cosmic star formation history.}
\label{fig:MadauPlot}
\end{figure}

The relation between stellar mass and halo mass of a galaxy is also sensitive to the functional form of the star formation -- gas relation, see Fig.~\ref{fig:MhaloMstar}. Again, a linear star formation -- gas relation leads to much better agreement with abundance matching results than a strongly super-linear star formation -- gas relation. The agreement is particularly good at high redshifts $z>4$, where, as we show in the next section, the cosmic star formation history is sensitive to changes in the star formation -- gas relation. 

\subsection{The Cosmic Star Formation History}
\label{sect:CSFH}

In Fig.~\ref{fig:MadauPlot} we present the cosmic star formation history as predicted by the model for a linear star formation -- $\H2$ relation (Equation \ref{eq:SGlin}). Here, the black solid curve shows the contribution of all galaxies sufficiently bright (M$_{\rm UV, AB}<-17.7$) to be accessible to current observations at $z\ge{}4$. We estimate the luminosity of a galaxy based on its SFR and dust attenuation. For the latter we use Table 6 of \cite{2012ApJ...754...83B} and we extrapolate the multiplicative dust extinction to 1 (no dust extinction) at $z=10$. For $z<3$ we assume that the dust extinction remains constant at its value at $z=3$. The red solid curve shows the contribution of all galaxies down to the photo-ionization limit \citep{2008MNRAS.390..920O} to the SFRD. The corresponding dashed curves show what happens if we assume that \emph{all} gas in the disk is molecular. The figure also includes compilations of observational data by \cite{2012ApJ...745..110O}, \cite{2012ApJ...754...83B}, and \cite{2012arXiv1207.6105B}.

An important result of Fig.~\ref{fig:MadauPlot} is that a linear star formation -- gas relation predicts a cosmic star formation history that is in good agreement with observational data. Interestingly, given the current observational limits the cosmic star formation history at $z>2$ changes only by a moderate amount depending on whether star formation is based on molecular hydrogen or not. We can rephrase this result, namely that quenching of star formation in low metallicity and hence $\H2$-poor galaxies at high redshift is not very important for galaxies accessible to observations (M$_{\rm UV, AB}<-17.7$). It does, however, play a crucial role for fainter galaxies. In fact, considering all galaxies the cosmic star formation rate density at $z\sim{}10$ is suppressed by 1.5 orders of magnitude if star formation is allowed to occur only in the molecular gas instead of all the gas in the disk. Obviously, this has implications for the re-ionization contribution by faint galaxies at such redshifts, see e.g., \cite{2012MNRAS.423..862K}.

\begin{figure}
\begin{tabular}{c}
\includegraphics[width=80mm]{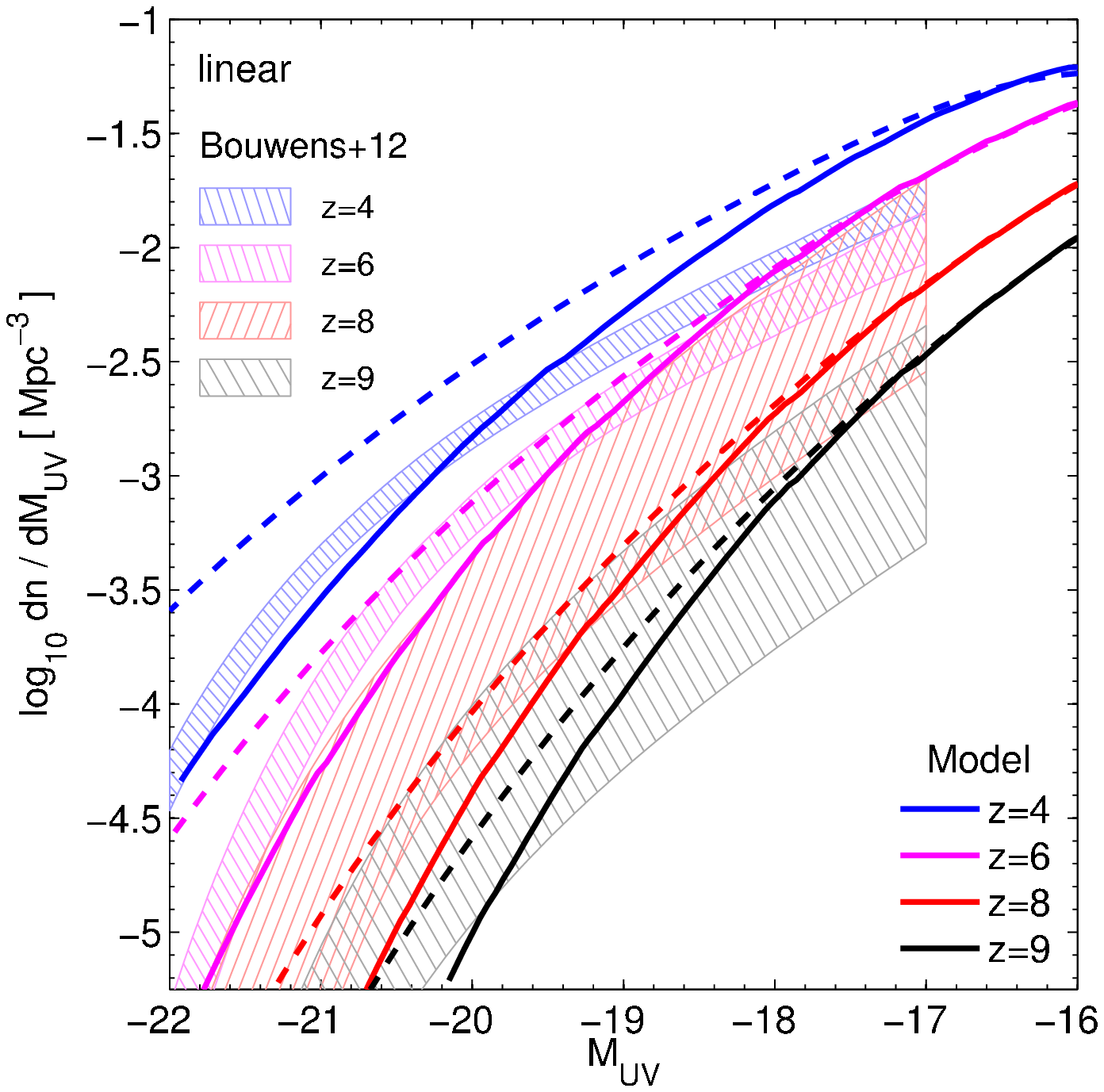}
\end{tabular}
\caption{The UV luminosity function as predicted by the model for a linear $\Sigma_{\rm SFR}-\Sigma_\H2$ relation with a 2.5 Gyr $\H2$ depletion time. SFRs are converted into UV magnitudes according to $M_{\rm UV} =-2.5\log_{10}({\rm SFR}/f_{\rm dust}\,[M_\odot\,{\rm yr}^{-1}]) -18.16$. A 0.1-125 $M_\odot$ Salpeter initial mass function (IMF) and a constant SFR over the last $\gtrsim{}100$ Myr is assumed for the conversion. Solid lines correspond to UV magnitudes that include extinction by dust ($f_{\rm dust}>1$), while dashed lines show the extinction free predictions ($f_{\rm dust}=1$). We use the multiplicative dust attenuation factor $f_{\rm dust}(z)$ as given in Fig. 14 of \protect\cite{2012ApJ...754...83B} for $z=4$ and $z=6$ (the latter is assumed to hold for $z\sim{}6-9$), and extend it to faint magnitudes such that $f_{\rm dust}\rightarrow{}1$. 
The hatched regions are measurements of the UV luminosity functions by \protect\cite{2012arXiv1211.2230B}. We use the Schechter function fits provided in their Table 3 and vary the fit parameters within their quoted error bars to create the hatched regions. The agreement between model predictions and observations is reasonable given the simplicity of the model, the uncertainties in the dust extinction corrections and/or changes in normalization due to cosmic variance.
}
\label{fig:UVlum}
\end{figure}

In Fig.~\ref{fig:UVlum}, we show the predicted UV luminosity function and its evolution with redshift for a linear $\Sigma_{\rm SFR}-\Sigma_\H2$ relation and compare it with observations \citep{2012arXiv1211.2230B}. The agreement is excellent considering the simplicity of the model and the various observational uncertainties. At $z\sim{}8$ the observed luminosity function appears to have a somewhat larger normalization compared with our model predictions, while no such offset is seen at the other redshifts, which could indicate that the $z\sim{}8$ observations are affected by cosmic variance. At low luminosities (not shown) the UV luminosity function turns over as a result of the suppression of star formation in low mass halos. Unfortunately, the magnitude at which the turn-over takes place depends sensitively on model parameters, e.g., on the disk sizes (parametrized by the spin parameter $\lambda$), and is therefore not a robust prediction of the model. For instance, at $z\sim{}4$ the UV luminosity function is predicted to break at $M_{\rm UV}\sim{}-15.5$ if we use a linear $\Sigma_{\rm SFR}-\Sigma_\H2$ relation with our default choice of model parameters. If, however, we decrease $\lambda$ from 0.07 to 0.05 the bright end of the luminosity function remains unaffected, while the turn-over shifts to $M_{\rm UV}\sim{}-12.5$.

\begin{figure*}
\begin{tabular}{cc}
\includegraphics[width=80mm]{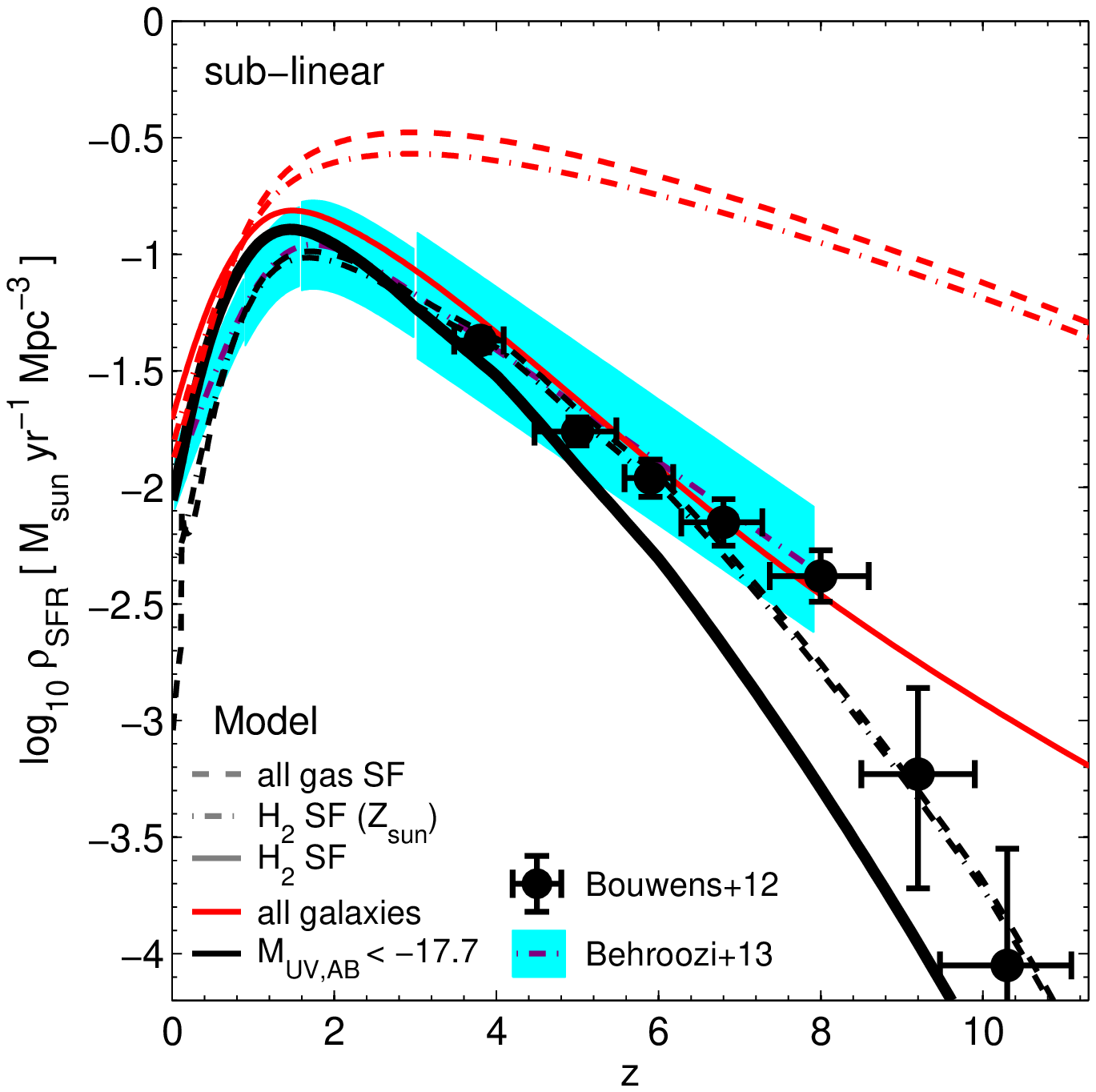} & 
\includegraphics[width=80mm]{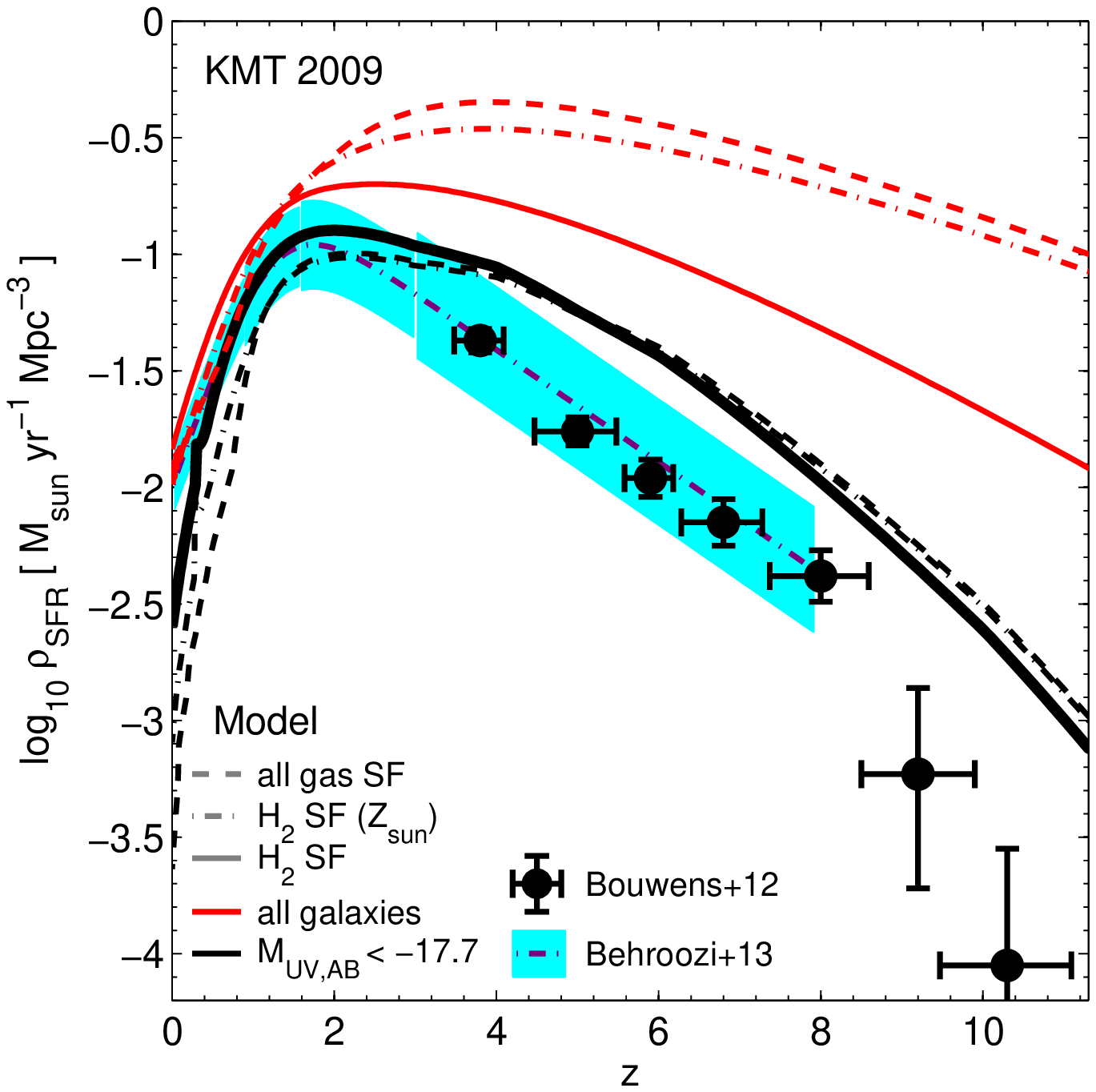} \\
\includegraphics[width=80mm]{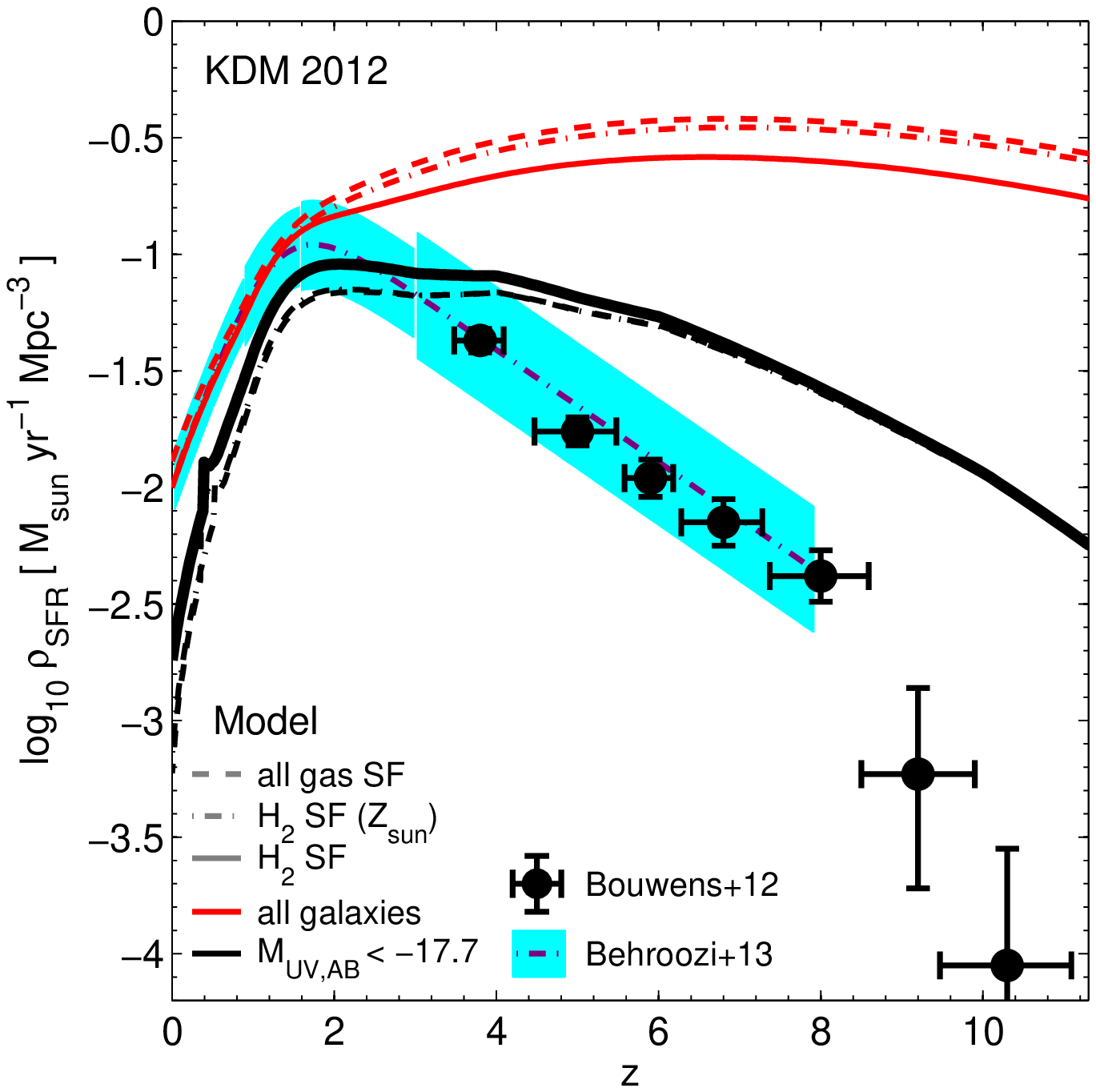} &
\includegraphics[width=80mm]{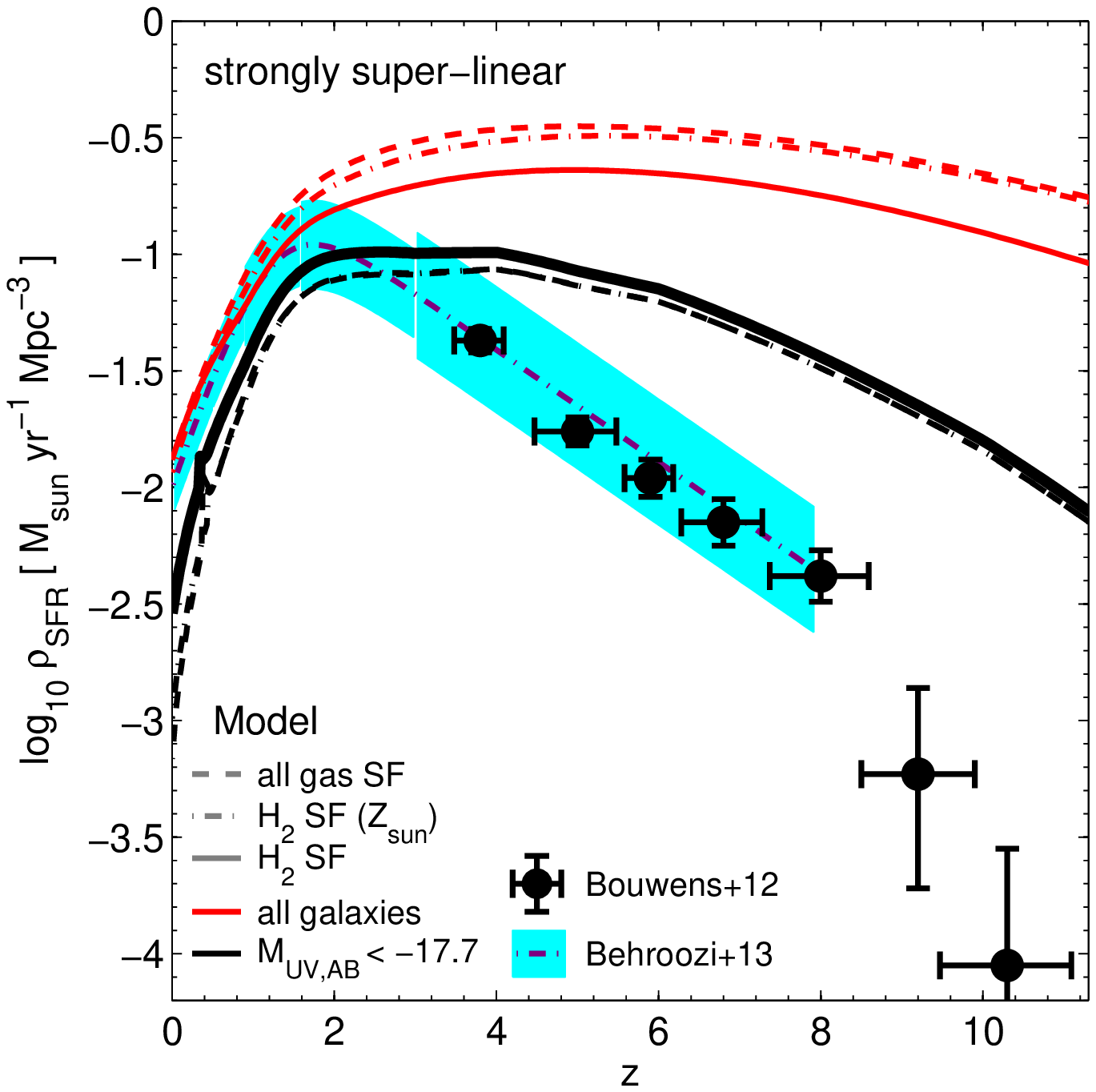}
\end{tabular}
\caption{Same as Fig.~\ref{fig:MadauPlot}, but for non-linear $\Sigma_{\rm SFR}-\Sigma_{\rm g}$ and $\Sigma_{\rm SFR}-\Sigma_\H2$ relations. (Top left) Predictions for a moderately sub-linear relation, see Equation (\ref{eq:SGS12}). (Top right) Predictions for the moderately super-linear relation suggested by \protect\cite{2009ApJ...699..850K}, see Equation (\ref{eq:SGK09}). (Bottom left) Predictions for the ``universal'' star formation -- gas relation suggested by \protect\cite{2012ApJ...745...69K}, see Equation (\ref{eq:SGK12}). (Bottom right) Predictions for a strongly super-linear relation \protect\citep{2011ApJ...731...41O}, see Equation (\ref{eq:SGquad}). The observable cosmic star formation history (black lines) is highly sensitive to the functional form of the star formation -- gas relation.}
\label{fig:MadauPlot2}
\end{figure*}

In Fig.~\ref{fig:MadauPlot2} we show plots analogous to Fig.~\ref{fig:MadauPlot} for four non-linear star formation -- gas relation: a moderately sub-linear relation (Equation \ref{eq:SGS12}), the moderately super-linear relation postulated by \cite{2009ApJ...699..850K} (Equation \ref{eq:SGK09}), the ``universal relation'' proposed by \cite{2012ApJ...745...69K} (Equation \ref{eq:SGK12}), and the strongly super-linear relation (Equation \ref{eq:SGquad}). The figure clearly demonstrates that a super-linear star formation -- gas relation results in SFRDs higher than observed at $z\sim{}4-10$. The discrepancy is particularly large for strongly non-linear star formation--gas relations, such as those given by Equations (\ref{eq:SGquad}) and (\ref{eq:SGK12}). These relations result in SFRDs that are inconsistent with observations. For instance, at $z=8$ ($z=10$) they are one (two) orders of magnitude above observations. Moderately non-linear relations such as those given by Equation (\ref{eq:SGK09}) or (\ref{eq:SGK12}) are in tension with observations, but are probably not excluded given the uncertainties of the model.

Similar to Fig.~\ref{fig:MadauPlot}, the \emph{observable} SFRD changes little whether star formation is tied to the abundance of molecular gas or to all the gas in the disk. Specifically, a sub-linear star formation -- gas relation leads to a difference no larger than $\sim{}0.5$ dex, while there is almost no discernible change if the relation is super-linear. Hence, we find that quenching of star formation in low metallicity and, thus, $\H2$-poor galaxies does not have a significant impact on the observable cosmic star formation history. In contrast, the predicted \emph{total} star formation history that includes all galaxies below the observational limits can change noticeably depending on whether the star formation -- gas relation is based on $\H2$ or all the gas. For instance, the difference is $\sim{}1$ dex for the moderately super-linear star formation -- gas relation suggested by \cite{2009ApJ...699..850K}. The difference is even larger for a linear or sub-linear star formation -- gas relation.

\subsection{Caveats}
\label{sect:Caveats}

\begin{figure*}
\begin{tabular}{ccc}
\includegraphics[width=55mm]{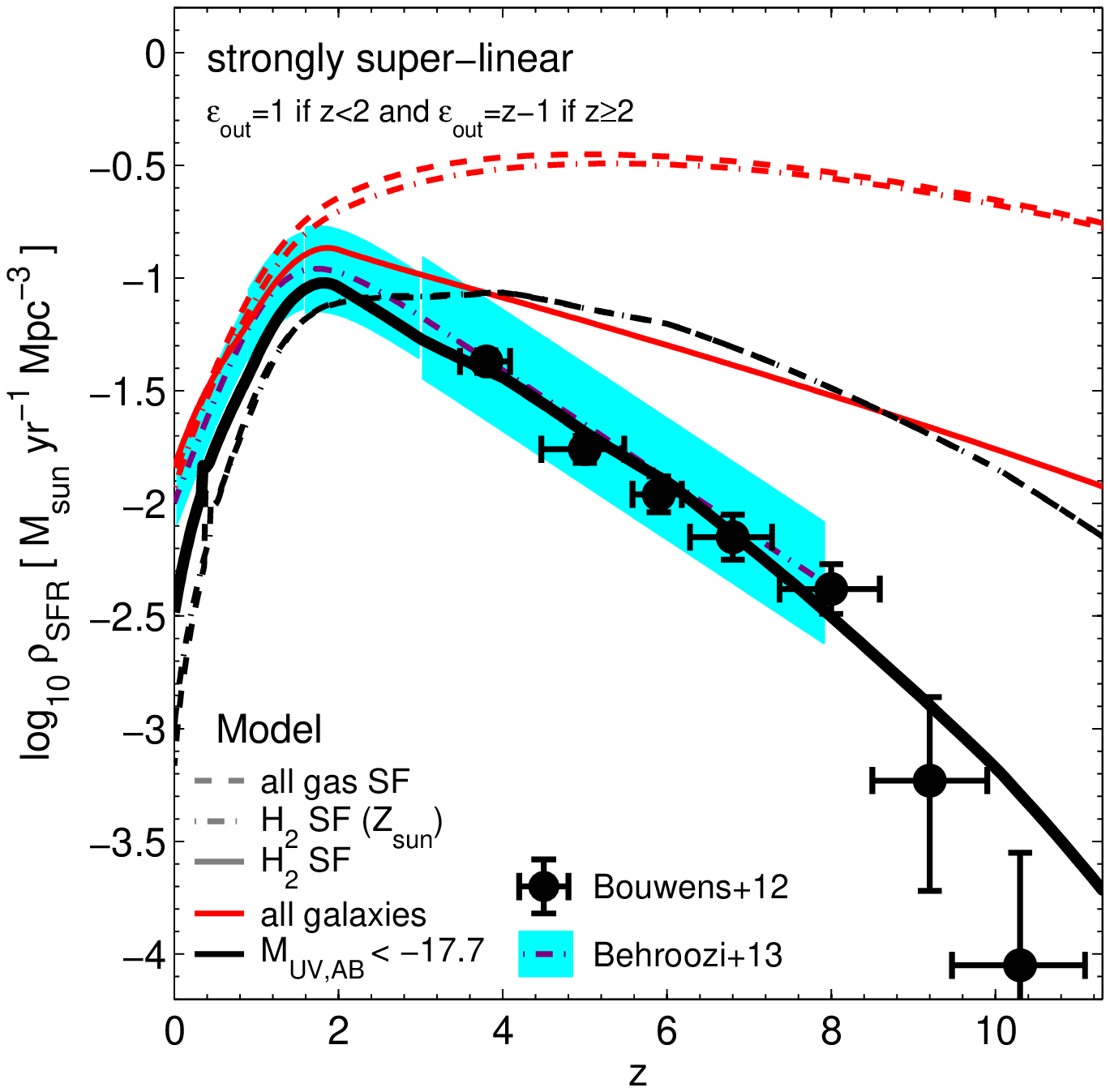} &
\includegraphics[width=56.2mm]{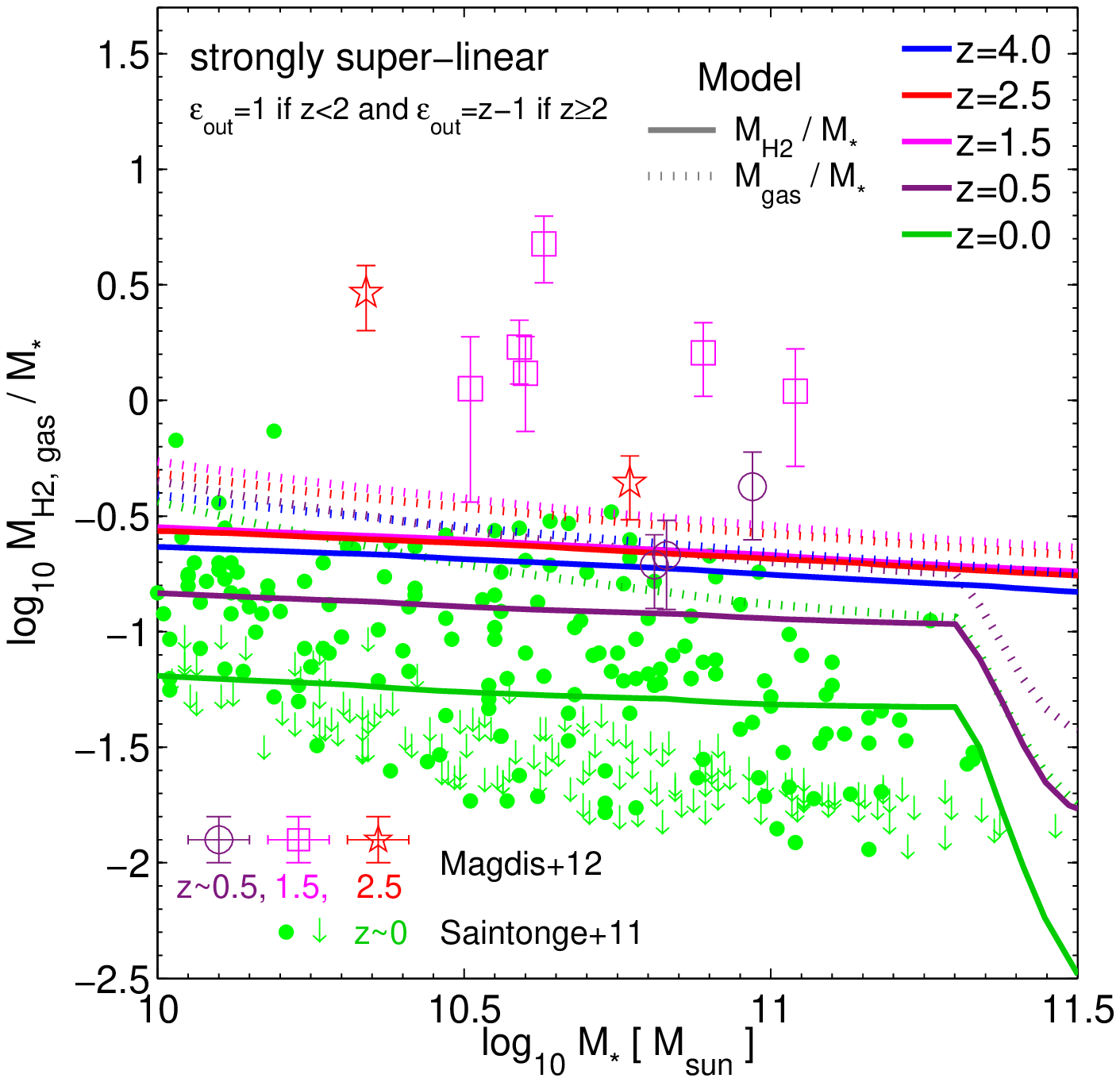} &
\includegraphics[width=55mm]{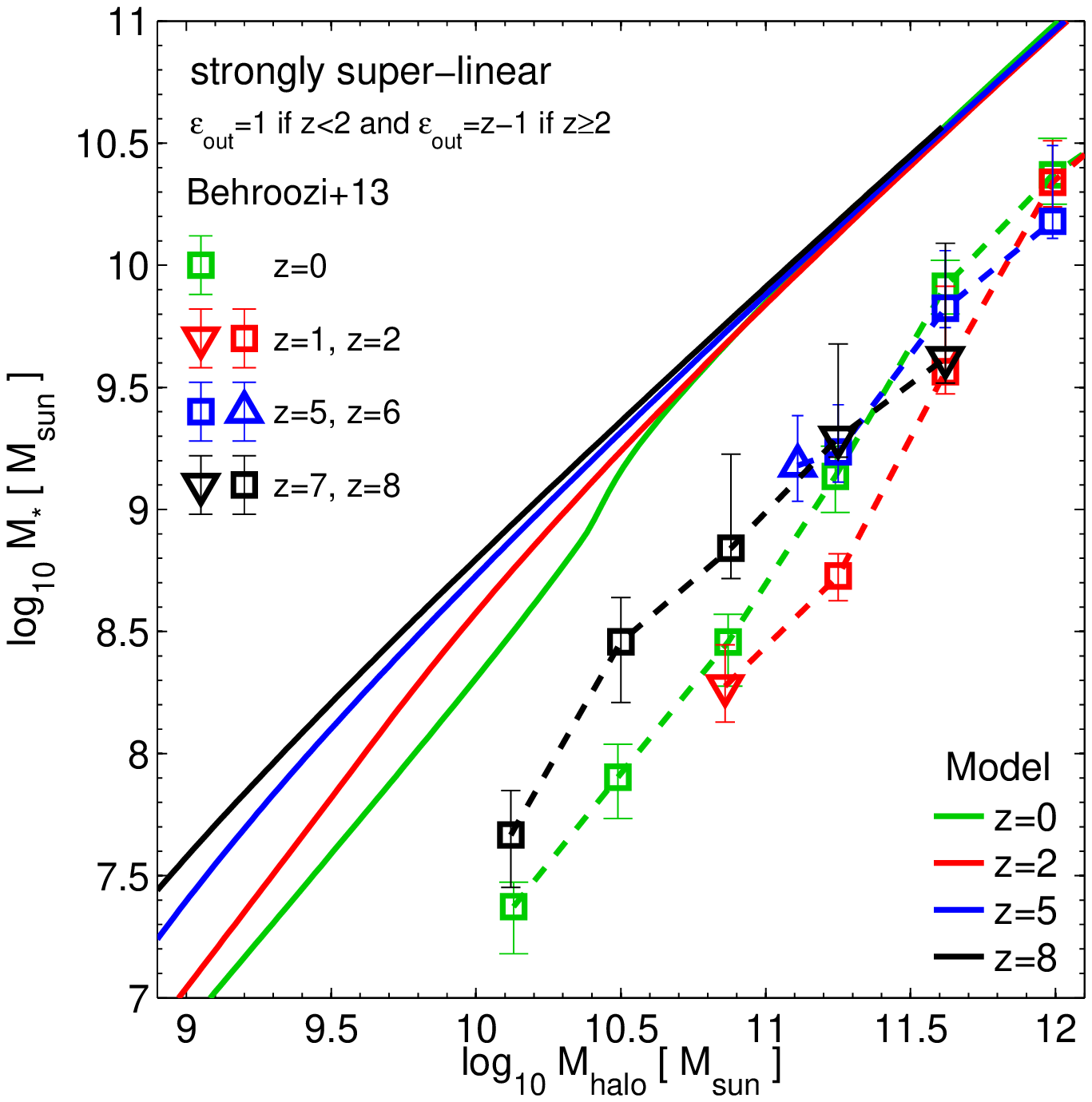}
\end{tabular}
\caption{Predictions of an alternative model in which the mass ejection parameter $\epsilon_{\rm out}$ increases at $z>2$. The star formation -- gas relation is assumed to be strongly super-linear (Equation \ref{eq:SGquad}). The panels show from left to right the cosmic star formation history, the gas-to-stellar mass ratio vs stellar mass, and the relation between stellar mass and halo mass. The lines and symbols are as in Fig.~\ref{fig:MadauPlot}, Fig.~\ref{fig:MH2}, and Fig.~\ref{fig:MhaloMstar}  for the left, middle, and right panel, respectively. The ad-hoc scaling of $\epsilon_{\rm out}$ with redshift cures the excess of SFRDs at high $z$ otherwise predicted for a strongly super-linear star formation -- gas relation, see Fig.~\ref{fig:MadauPlot2}. However, it does not help in bringing the  $M_{\rm halo}-M_*$ relation or the gas-to-stellar mass fractions in agreement with observations.}
\label{fig:AlternativeMassLoading}
\end{figure*}

Our findings come with a number of potential caveats. First, our approach is based on the assumption that a single, unchanging star formation -- gas relation, potentially parametrized by local galaxy properties such as the gas metallicity, or the free-fall time, etc., holds from early times until the present. Second, the model obviously idealizes many aspects of star formation in high redshift galaxies. For instance, it assumes that the gas is always arranged in an exponential disk (likely not true during major mergers) and that accreted halos donate all their cold gas to the central galaxy. Furthermore, since the model only makes predictions for the mean evolution of halos of a given mass, it ignores variations in the accretion histories and possible environmental dependences. Such, admittedly interesting, complexities are best studied in numerical simulations. However, demanding requirements in terms of simulation volume and resolution, see e.g., \cite{2010MNRAS.402.1536S}, limit the systematic exploration of the parameter space of feedback and star formation models using this approach. Nonetheless, we plan to compare the predictions of our model with dedicated high resolution cosmological simulations in future work. 

Additional caveats are:
\begin{itemize}
\item \emph{Model parameters:} We use the same set of parameters as in \cite{2012ApJ...753...16K}. Most of these parameters have a theoretical or empirical justification (we refer to reader to \citealt{2012ApJ...753...16K}). When we vary the model parameters within justifiable ranges, we find that the mass ejection rate into the IGM per SFR, $\epsilon_{\rm out}$, the metallicity of the IGM, $Z_{\rm IGM}$, and the spin of the halo, $\lambda$, have the largest effect on the predicted cosmic star formation history. The impact of these changes is largest if a linear star formation -- gas relation is used, and smaller if the relation is highly non-linear and, hence, star formation is accretion limited early on. However, even for a linear relation we can vary each of these parameters by a factor of $\sim{}3$ without any qualitative changes in the model predictions. For instance, increasing $\epsilon_{\rm out}$ by such a factor decreases the SFRD at $z\sim{}1-2$ by $\sim{}0.5$ dex, but leads to little change at higher redshifts (e.g., at $z=4$ the decrease in the SFRD is only 0.2 dex). Reducing $Z_{\rm IGM}$ by a factor 3 delays star formation in low mass halos and results in SFRDs that are reduced by a similar factor. Decreasing $\lambda$ from 0.07 to 0.05 increases the SFRD at $z\sim{}4-10$ by a factor $\sim{}2-3$. The predicted \emph{observable} cosmic star formation history is less affected ($\lesssim{}0.2$ dex typically). \\
We stress that none of the tested combination of model parameters would bring the predictions for the strongly super-linear star formation -- gas relation (equation \ref{eq:SGquad}) in agreement with the observations of the cosmic star formation history, gas-to-stellar fraction, and mass-metallicity relation. This is not surprising because the model predictions for a strongly super-linear star formation -- gas relation are primarily driven by the short gas depletion time. The latter implies that star formation in galaxies can keep up with the accretion of gas onto galaxies and results in relatively low gas fractions, high SFRs, and high metallicities at early times. In contrast, a linear star formation -- gas relation with a long gas depletion time ensures that the SFRs at high redshift fall short of the gas accretion rates and, hence, that galaxies have relatively large gas fractions, low metallicities, and low stellar masses.\\
The equilibrium approach presented in Section \ref{sect:EQ} allows a further assessment of the importance of the various model parameters. According to (\ref{eq:dotZeq1})--(\ref{eq:fseq}) galaxies have metallicities, gas fractions, and stellar fractions that depend on the model parameters and on the ratio $r$ between the star formation rate and gas accretion rate. For instance, the chosen value for $Z_{\rm IGM}$ will affect the metallicity of a galaxy if $r$ is small compared with $Z_{\rm IGM}$. However, since $r$ is in fact much larger than $Z_{\rm IGM}$ for galaxies at $z\sim{}0-10$, see Section \ref{sect:GASF}, the choice of $Z_{\rm IGM}$ does not strongly alter the predicted metallicities of galaxies in this redshift range. Similarly, we can understand why changing $\epsilon_{\rm out}$ within a factor of a few does not strongly affect galaxy properties at high redshifts. According to (\ref{eq:dotZeq1alt})--(\ref{eq:fseqalt}) only the combination $1+r'(1-R+\epsilon_{\rm out})$ enters the equilibrium predictions. Here, $r'$ is the ratio between the matter accretion time and the gas depletion time. At early times $r'\ll{}1$ and, hence, the precise value of $\epsilon_{\rm out}$ does not matter as long as $r'(1-R+\epsilon_{\rm out})\ll{}1$.

\item \emph{Missing physics:} Given the simplicity of the model it is possible that we are missing an important physical process that would affect the model predictions and potentially change our conclusions. For instance, a mechanism that increases the mass ejection rate of the winds with redshift would help to mitigate some of the problems of a highly non-linear star formation -- gas relation.  We demonstrate this in Fig.~\ref{fig:AlternativeMassLoading} where we show the model predictions for a hypothetical scenario in which $\epsilon_{\rm out}=1$ for $z<2$ and $\epsilon_{\rm out}=z-1$ for $z\geq{}2$. By choosing this particular redshift dependence of $\epsilon_{\rm out}$ even a strongly super-linear star formation -- gas relation can be brought into agreement with observations of the cosmic star formation history. However, this does not necessarily mean that other observational constraints are met as well. The figure shows, for instance, that this particular model does not reproduce the $M_{\rm halo}-M_*$ relation at any redshift and the gas-to-stellar mass ratios at $z\gtrsim{}1$. A simple, linear star formation -- gas relation with an $\H2$ depletion time of $2.5$ Gyr (as observed in the local universe) does not require the assumption of additional ad-hoc (and fine-tuned) physics to reproduce global galaxy properties from high redshifts until today, see Section \ref{sect:ZSG}.

\item \emph{Metal mixing}: We assume that metals in the disks of galaxies are efficiently mixed, i.e., that the gas has a spatially uniform metallicity. In reality, we expect some metallicity gradients, since accretion of metal-poor gas occurs preferentially at the outer edge of the disk. Numerical simulation by \cite{2012ApJ...749...36K} indicate that the metallicity difference is only modest for the majority of the gas in the disks of high redshift galaxies. In order to test the importance of metal mixing we modified the model to allow only local enrichment (i.e., metals produced in a given radius bin are added only to that radial bin). This somewhat extreme case of no-mixing lowers the predicted SFRDs by $\sim{}3$ if we use a linear star formation -- gas relation, but makes almost no difference if the relation is strongly non-linear.

\item \emph{Dust:} The model assumes that the dust-to-gas ratio (which determines the $\H2$ abundance) scales proportional to the metallicity. At low metallicities ($Z<0.1 Z_\odot$) it is unclear whether this assumption is justified. A potential concern is that at high $z$ the dust production may lag behind the metal production, limiting the formation of $\H2$, and thus star formation. Empirically, there is evidence for a substantial presence of dust in high redshift galaxies from absorption line studies of damped Ly-$\alpha$ systems at $z\sim{}1.5-3.5$ (e.g., \citealt{1994ApJ...426...79P, 2002ApJ...566...68P}) and from the thermal emission of dust in quasar hosts at $z\sim{}4-6$ (e.g., \citealt{2001A&A...374..371O, 2002A&A...384L..11B}). Both AGB stars (e.g., \citealt{2009MNRAS.397.1661V}) and supernovae (e.g., \citealt{2001MNRAS.325..726T}) have been proposed as formation channels for dust at such high redshifts. If supernovae dominate the dust production, we expect a dust production time scale of $\sim{}10^{7}$ yr, which is shorter than the gas accretion time ($\sim{}1-2\times{}10^{8}$ at $z=10$) and, hence, not a limiting factor.

\item \emph{Molecular hydrogen:} The $\H2$ abundances are estimated based on the equilibrium model by \cite{2009ApJ...693..216K}. For Milky-Way like, turbulent ISM conditions the $\H2$ formation times $t_{\H2,{\rm form}}$ are short ($\sim{}10^{6}$ yr, \citealt{2007ApJ...659.1317G}), but they could be higher in high redshift galaxies. With $t_{\H2,{\rm form}}\propto{}(Z\rho)^{-1}$ \citep{1971ApJ...163..165H} and gas densities in the high $z$ galaxies that are likely an order of magnitude higher than in the Milky Way, we expect formation times of $\lesssim{}10^{8}$ yr for gas with metallicities $Z\ge{}2\times{}10^{-5}$, which is sufficiently short to not limit star formation. In addition, numerical simulations of star formation in the ISM show that decreasing the metallicity from solar to a hundredth solar postpones the onset of star formation only by a few Myr \citep{2012MNRAS.426..377G}. Hence, we do not expect that star formation is limited by the $\H2$ formation time even in case of a highly non-linear star formation -- gas relation.

\end{itemize}

\begin{figure*}
\begin{tabular}{cc}
\includegraphics[width=80mm]{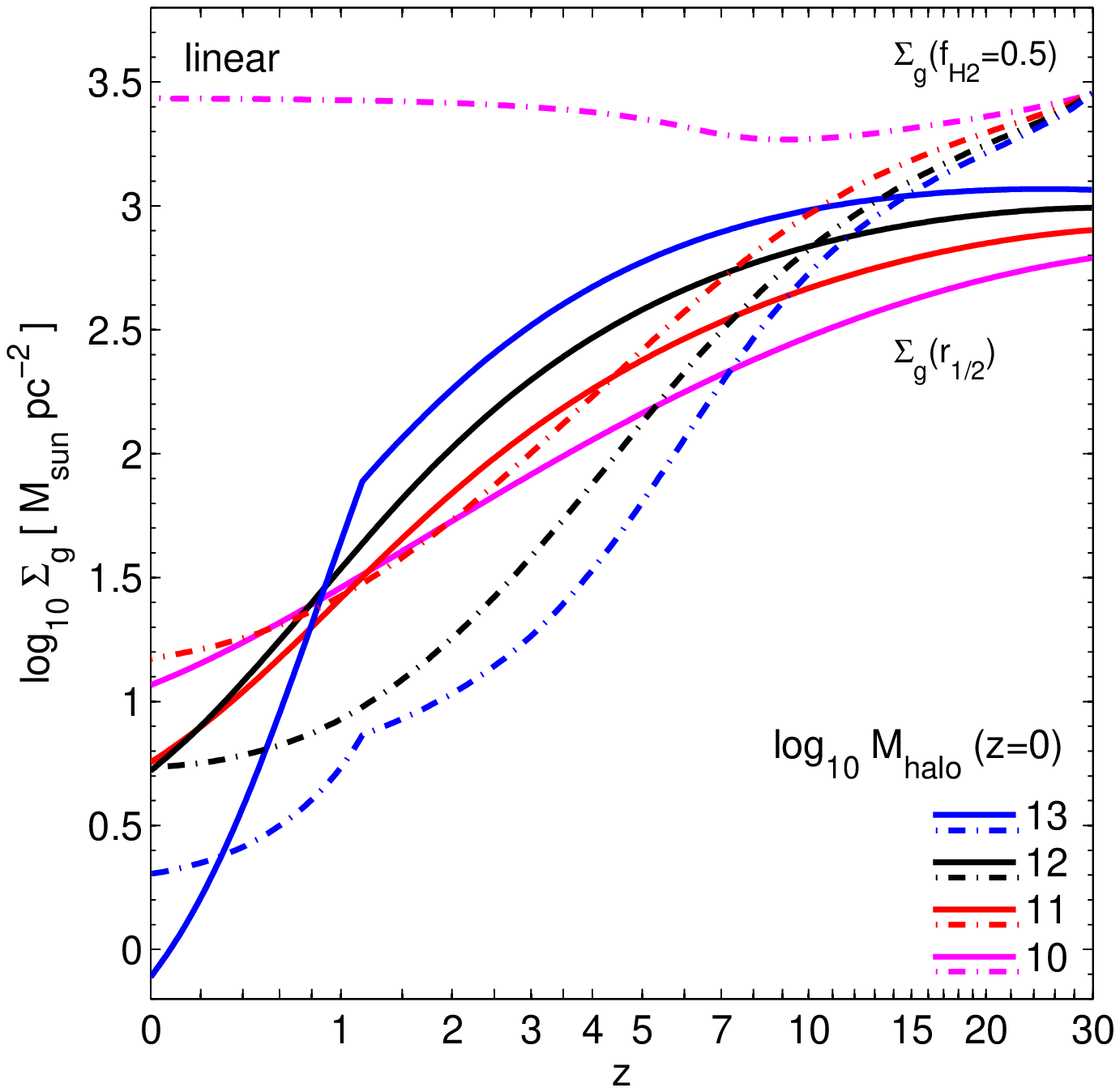} & 
\includegraphics[width=80mm]{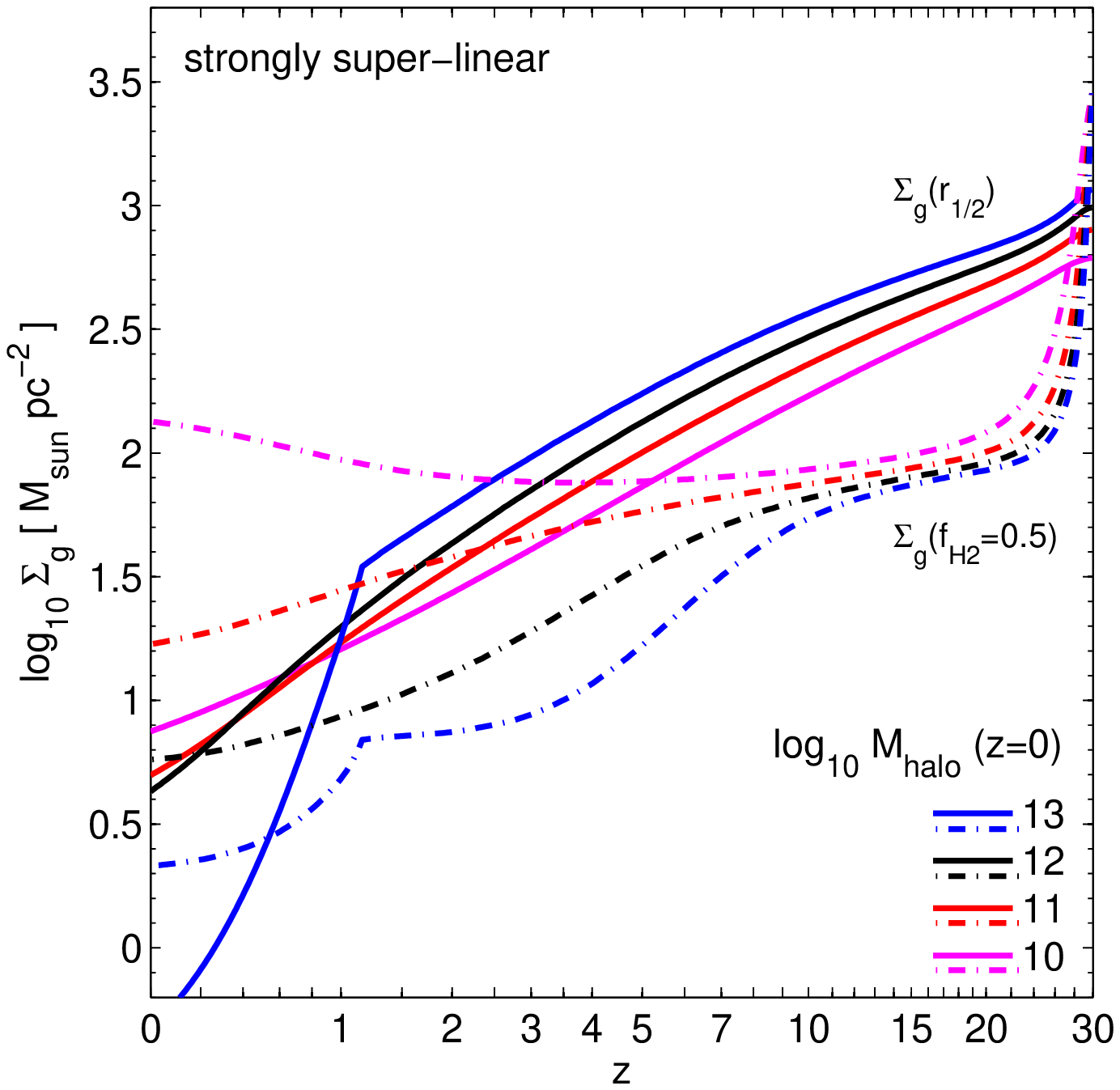} 
\end{tabular}
\caption{Evolution of the gas surface density at the disk half mass radius ($\Sigma_{\rm g}(r_{1/2})$) and evolution of the surface density at which the molecular fraction is 50\% ($\Sigma_{\rm g}(f_\H2=0.5)$) for galaxies in halos of different masses. (Left) for a linear star formation -- gas relation (Equation \ref{eq:SGlin}). (Right) for a strongly super-linear star formation -- gas relation (Equation \ref{eq:SGquad}). Solid and dot-dashed lines show how $\Sigma_{\rm g}(r_{1/2})$ and $\Sigma_{\rm g}(f_\H2=0.5)$, respectively, change across cosmic time for galaxies that, by $z=0$, reside in the center of a halo of the specified mass (see color-coding in the legend). The expansion of the universe, the consumption of gas, and the quenching of gas accretion result in a reduction of $\Sigma_{\rm g}(r_{1/2})$ with time. Star formation increases the metallicity and thus decreases $\Sigma_{\rm g}(f_\H2=0.5)$ (Equation \ref{eq:fH2}). A strongly super-linear star formation -- gas relation leads to early enrichment (to $\sim{}0.1$ $Z_\odot$ already at $z\gtrsim{}10$). Enrichment proceeds much more gradually in case of a linear $\Sigma_{\rm SFR}-\Sigma_\H2$ relation.}
\label{fig:gasSurfaceDensity}
\end{figure*}

\section{The drivers of galaxy evolution}
\label{sect:Driver}

So far we have demonstrated that the functional form of the star formation -- gas relation has a large impact on a variety of global galaxy properties, such as the overall metallicities, SFRs, and the gas-to-stellar fractions. In this section we analyze this result in more detail using the framework of the model introduced in Section \ref{sect:model} in order to learn more about the actual drivers of galaxy properties across cosmic history. To simplify the discussion, we will focus on only two of the introduced star formation -- gas relations: the linear star formation - $\H2$ relation (equation \ref{eq:SGlin}) and the strongly super-linear star formation - $\H2$ relation (equation \ref{eq:SGquad}). After discussing the role of metal enrichment (Section \ref{sect:metalenrich}), we show how the competition between star formation and gas accretion drives the evolution of basic galaxy properties (Section \ref{sect:GASF}). Finally, in Section \ref{sect:EQ} we present a novel equilibrium picture of galaxy evolution that captures the interplay between star formation, gas accretion, and outflows and in which average galaxy properties are determined by the functional form of the star formation -- gas relation. Galaxies remain in equilibrium since birth and may drop out of equilibrium only at very late times when the gas accretion onto their halos is quenched. In this picture the evolution of galaxy properties across cosmic time is a simple consequence of the modulation of the external accretion rate.

\subsection{Surface densities of galaxies and the role of metal enrichment}
\label{sect:metalenrich}

In this section we discuss why the functional form of the star formation -- gas relation has a large impact on the SFRD at high redshift. In Fig.~\ref{fig:gasSurfaceDensity} we show the evolution of the typical gas surface density (here $\Sigma_{\rm g}(r_{1/2})$, the surface density at the gas half mass radius) of galaxies residing in halos of different mass. At $z\sim{}20-30$ this surface density is typically in the range $300-1000$ $M_\odot$ pc$^{-2}$. As the universe expands, gas surface densities in galaxies decrease and reach values in agreement with observations of nearby galaxies ($\sim{}10$ $M_\odot$ pc$^{-2}$ for galaxies in halos with masses $\sim{}10^{11}-10^{12}$ $M_\odot$, e.g., \citealt{2002ApJ...569..157W}). Galaxies in halos above the virial shock quenching mass ($\gtrsim{}2\times{}10^{12}$ $M_\odot$) experience a fast decline in their gas surface density at low redshift caused by the shut-down of gas accretion. 

We also include in the figure the evolution of the critical surface density $\Sigma_{\rm g}(f_\H2=0.5)$ required to turn half the gas molecular. We note that there is a one-to-one correspondence between this critical surface density and the metallicity of the gas, see Equation (\ref{eq:fH2}). Initially the critical surface density is large $\sim{}3000$ $M_\odot$ pc$^{-2}$ (corresponding to $Z_{\rm IGM}=2\times{}10^{-5}$), but it decreases with time as metal enrichment by star formation boosts the molecular content of the gas. Clearly, the choice of the star formation -- gas relation has a large impact on the evolution of this critical surface density or, equivalently, the metallicity. If the star formation -- gas relation is linear, halos with $z=0$ masses below $\sim{}3\times{}10^{10}$ $M_\odot$ are unable to significantly increase the metallicity of the gas. In contrast, efficient enrichment can take place in such halos at early times ($z>10$) if the relation is strongly super-linear. 

\begin{figure*}
\begin{tabular}{ccc}
\includegraphics[width=53mm]{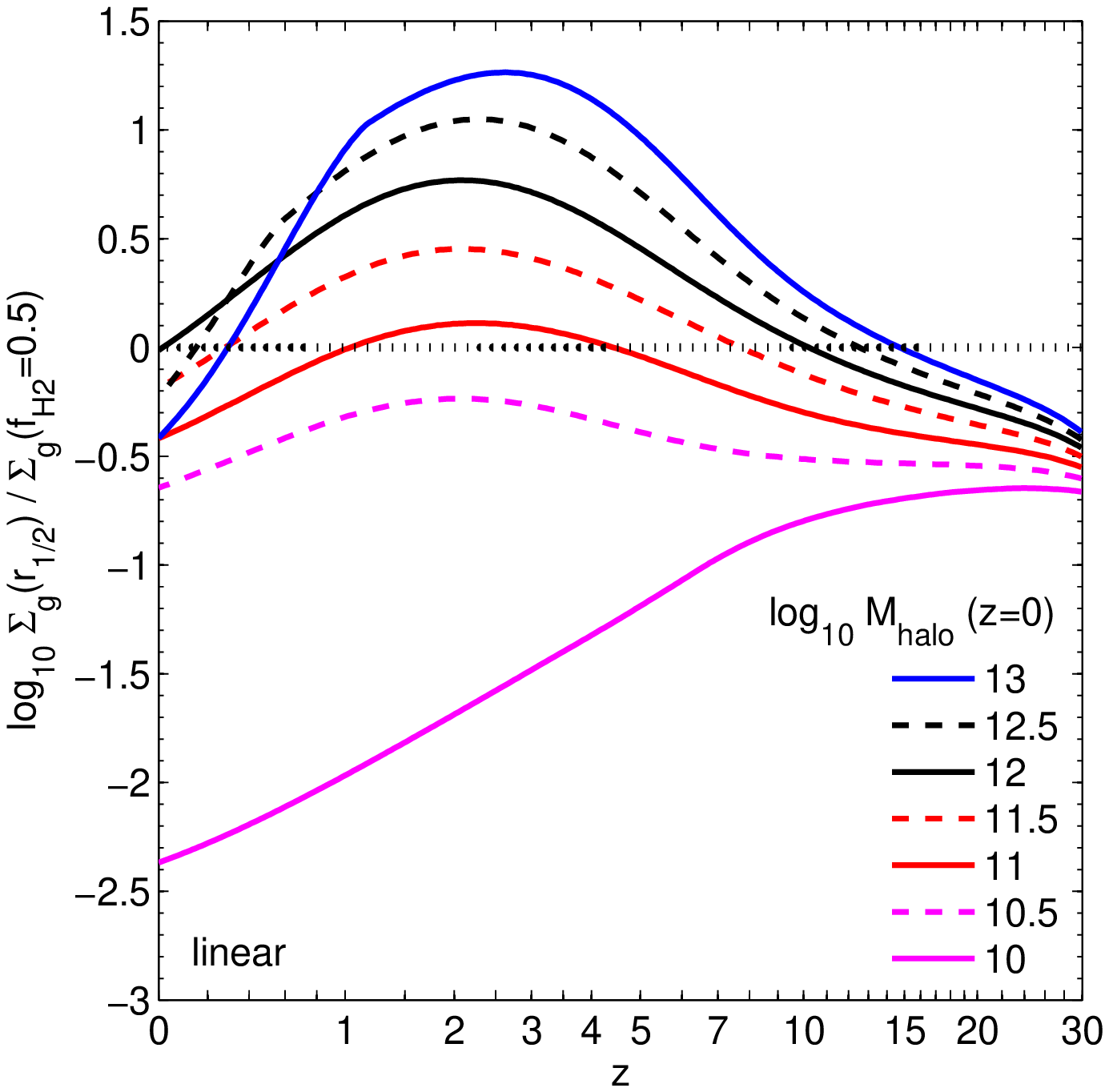} & 
\includegraphics[width=53mm]{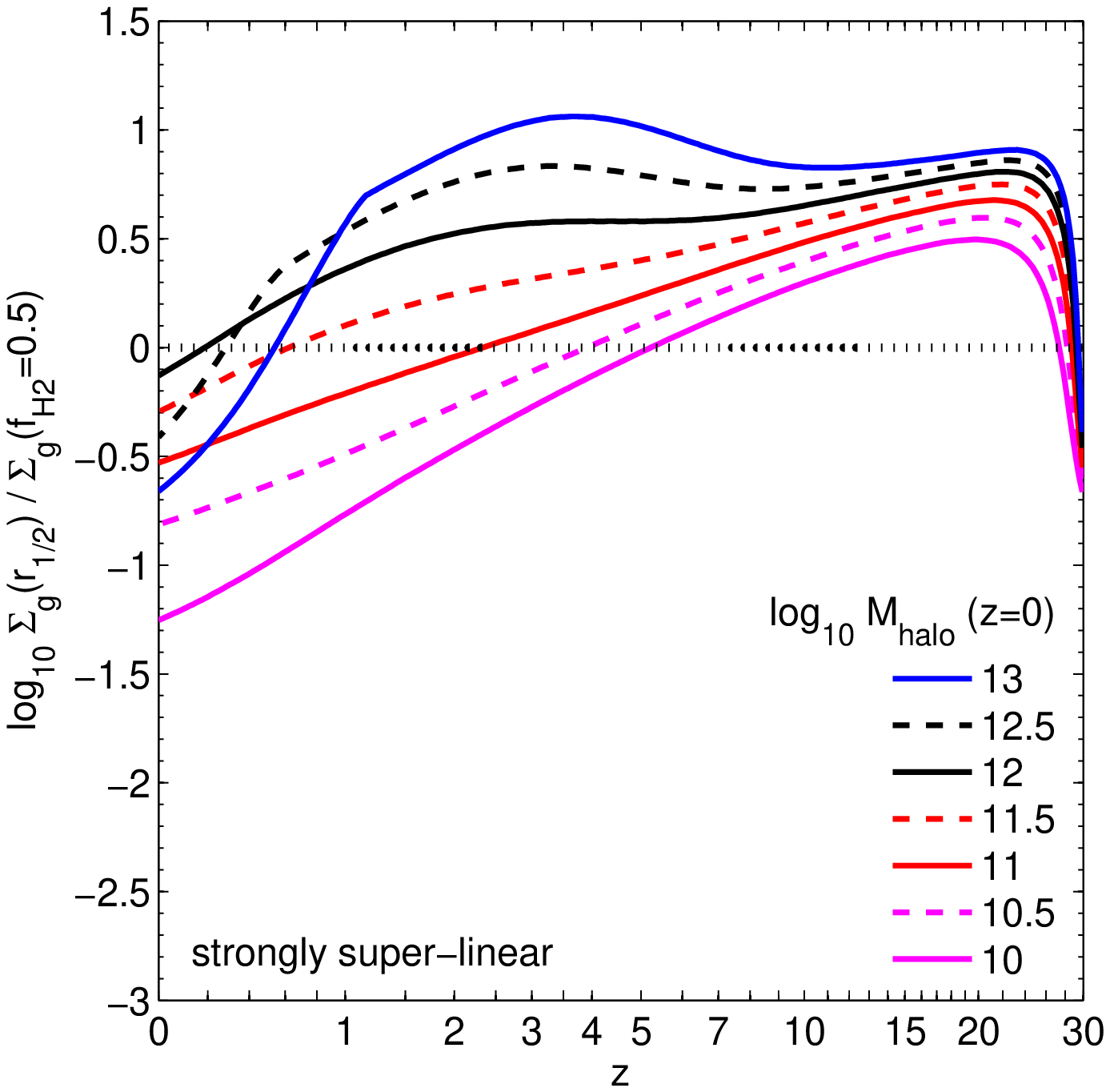} &
\includegraphics[width=53mm]{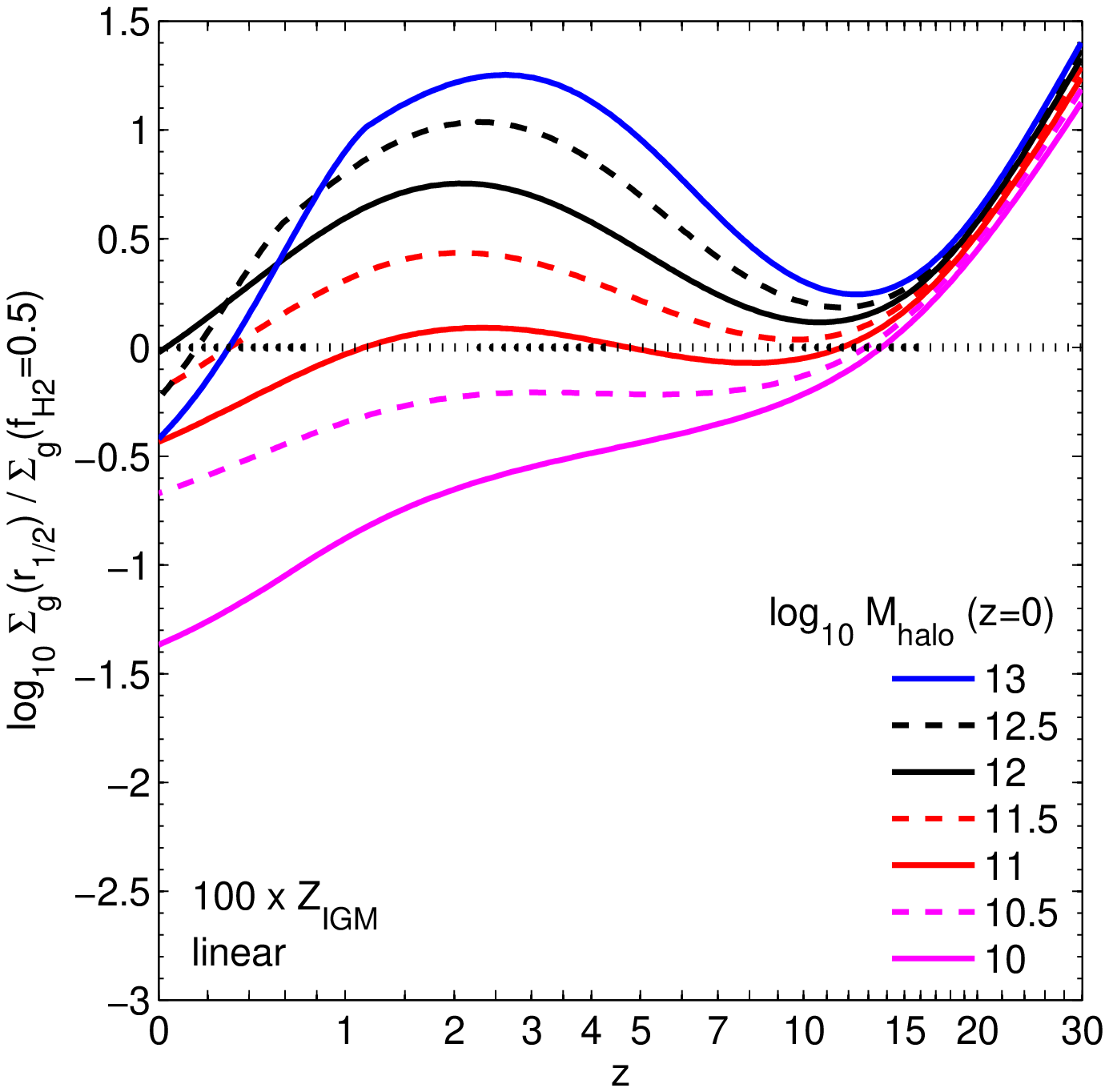}
\end{tabular}
\caption{Evolution of the ratio between the typical gas surface density in a disk, $\Sigma_{\rm g}(r_{1/2})$, and the surface density at which the molecular fraction is 50\%, $\Sigma_{\rm g}(f_\H2=0.5)$, see Fig.~\ref{fig:gasSurfaceDensity}. A ratio significantly below (above) unity means that most of the gas in the disk of a galaxy is atomic (molecular). (Left) for a linear star formation -- gas relation (Equation \ref{eq:SGlin}). Gas surface densities become sufficiently large to form molecules and stars only in massive halos at sufficiently late times. (Middle) for a strongly super-linear star formation -- gas relation (Equation \ref{eq:SGquad}). A strongly super-linear star formation -- gas relation allows the formation of fully molecular gas disks at earlier times and in halos of lower masses and, hence, boosts the star formation rate at the relevant, and observationally accessible, redshift range $z\sim{}4-10$. (Right) same as left panel, but with the metallicity of the gas in the galaxy disks initially set to $Z=0.1$ $Z_\odot$. The $z\lesssim{}10$ evolution is hardly affected by this dramatic change, demonstrating that the lack of early enrichment is not the reason that a linear star formation -- gas relation is able to reproduce the observed cosmic star formation history at $z\le{}10$.}
\label{fig:gasSurfaceDensity2}
\end{figure*}

In Fig.~\ref{fig:gasSurfaceDensity2} we compare the ratio between the typical gas surface density, $\Sigma_{\rm g}(r_{1/2})$, and the critical surface density, $\Sigma_{\rm g}(f_\H2=0.5)$. $\H2$-based star formation can only take place throughout the disk if this ratio is $\gtrsim{}1$. If the star formation -- gas relation is linear, gas in galaxies with $z=0$ halo masses below $\sim{}3\times{}10^{10}$ $M_\odot$ remains effectively un-enriched and thus atomic. Hence, star formation is almost completely quenched in halos of such mass. However, galaxies in more massive halos are able to turn a substantial fraction of their gas reservoir into $\H2$. In fact, the condition $\Sigma_{\rm g}(r_{1/2})\sim{}\Sigma_{\rm g}(f_\H2=0.5)$ is satisfied increasingly earlier for halos of higher mass, e.g., at $z\sim{}10$ ($z\sim{}4.5$) for halos with $z=0$ masses of  $10^{12}$ $M_\odot$ ($10^{11}$ $M_\odot$). If the star formation -- $\H2$ relation is strongly super-linear, galaxies generally reach metallicities of $Z\sim{}2\times{}10^{-3}$ (a tenth solar) quickly after a single Pop III star initially enriches the gas to $Z\sim{}2\times{}10^{-5}$. In this case, most galaxies, even those in low mass halos, have fully molecular gas disks and are able to form stars efficiently at early times. The metallicity-dependent quenching of star formation has thus little impact on the cosmic star formation history in this case. 

Returning to the discussion of the linear star formation -- gas relation, it is important to point out that the lack of early enrichment is \emph{not} the reason that $\lesssim{}3\times{}10^{10}$ $M_\odot$ halos are inefficient star formers at observationally accessible redshifts $z\lesssim{}10$. To show this we re-run the model for the linear relation with a 100 times higher initial gas metallicity of the halos, i.e., with $Z=2\times{}10^{-3}$ instead of $2\times{}10^{-5}$, see Fig.~\ref{fig:gasSurfaceDensity2}. Clearly, the ratio between typical and critical gas surface density is not affected in a significant way at $z\lesssim{}10$. In particular, galaxies with $z=0$ halo masses below $3\times{}10^{10}$ $M_\odot$ still do not contribute to star formation at these redshifts.
Hence, while star formation is regulated by the presence of $\H2$ and, thus, by the metal enrichment of the gas, the level of metal enrichment is determined primarily by the functional form of the star formation -- gas relation. A non-linear relation leads to fast enrichment and, hence, to galaxy disks that are fully molecular and have large star formation rates even in low mass halos at high redshifts. In contrast, a linear relation delays the enrichment of the gas and therefore limits star formation to sufficiently massive halos. 

This leaves us with one open question, namely how can the star formation -- gas relation have such a strong impact on the metal enrichment in halos? The metallicity evolution is determined by the SFR (increases metal mass and decreases gas mass; Equation \ref{eq:dotMZ}) and the gas accretion rate (increases gas mass and metal mass; Equation \ref{eq:dotMg}). Rearranging these equations it is easy to show (see appendix \ref{sect:Stability}) that the metallicity $Z$ is constant, i.e., $\dot{Z}=0$, if

\begin{equation}
r\equiv{}\frac{\dot{M}_{\rm *, form}}{\dot{M}_{\rm g, in}} = \frac{Z - Z_{\rm IGM}}{ y(1-R)(1-\zeta)} \label{eq:dotZeq0}.
\end{equation}
Given a SFR to gas accretion rate ratio $r$, the equilibrium metallicity is therefore
\begin{equation}
Z_{\rm eq}(r) = ry(1-R)(1-\zeta)+Z_{\rm IGM} \label{eq:dotZeq1}.
\end{equation}

We demonstrate in appendix \ref{sect:Stability} that this equilibrium solution is (linearly) stable under most circumstances. Intuitively, the stability of the equilibrium is a simple consequence of the interplay between gas accretion and metal enrichment via star formation. Gas inflows from the IGM dilute the metallicity of the ISM at a rate that is proportional to both the gas accretion rate \emph{and} to the difference between the current ISM metallicity and the metallicity of the accreted gas. This implies that the dilution rate of the ISM metallicity by gas inflows will be smaller or larger depending on the current metallicity of the ISM. In contrast, the metal production rate by stellar evolution processes is typically less sensitive to the ISM metallicity (see appendix \ref{sect:Stability} for a more elaborate discussion).  In equilibrium the dilution rate of metals via gas inflows matches the enrichment rate via star formation. Hence, whenever the current ISM metallicity is below its equilibrium value, the dilution rate is reduced relative to its value for $Z_{\rm eq}$, while the enrichment rate remains the same. Consequently, the ISM metallicity increases. Vice versa if the ISM metallicity is above its equilibrium value.

The answer to our question above is therefore that a change of the functional form of the star formation -- gas relation changes the ratio $r$ (assuming that the gas accretion rate and the gas mass of the galaxy remain unaffected) and, hence, the equilibrium metallicity of a galaxy. This argument will be made rigorous in Section \ref{sect:EQ}.

Equation (\ref{eq:dotZeq1}) has a few interesting properties. First, it demonstrates that galactic outflows of warm/cold ISM material (parametrized by $\epsilon_{\rm out}$) play no role for the ISM metallicity. The reason is that while outflows carry metals away they also remove a corresponding amount of gas leaving the equilibrium metallicity unchanged. But since the true metallicity of the ISM is likely close to $Z_{\rm eq}(r)$, it also remains unaffected. Second, $Z_{\rm eq}(r)$ depends on $\zeta$ (and thus on halo mass) and on $Z_{\rm IGM}$. Hence, an analysis of the mass or redshift dependence of the metallicities of high redshift galaxies (combined with measured SFRs and mass accretion rates, estimated, e.g., from their halo masses) could be used to learn more about supernova driven blow-outs (parametrized by $\zeta$) or the IGM enrichment ($Z_{\rm IGM}$).

In the next section, we study the connection between the star formation -- gas relation and the driver of the metallicity evolution in galaxies, the ratio of SFR to gas accretion rate.

\subsection{Gas Accretion and Star formation}
\label{sect:GASF}

\begin{figure*}
\begin{tabular}{cc}
\includegraphics[width=80mm]{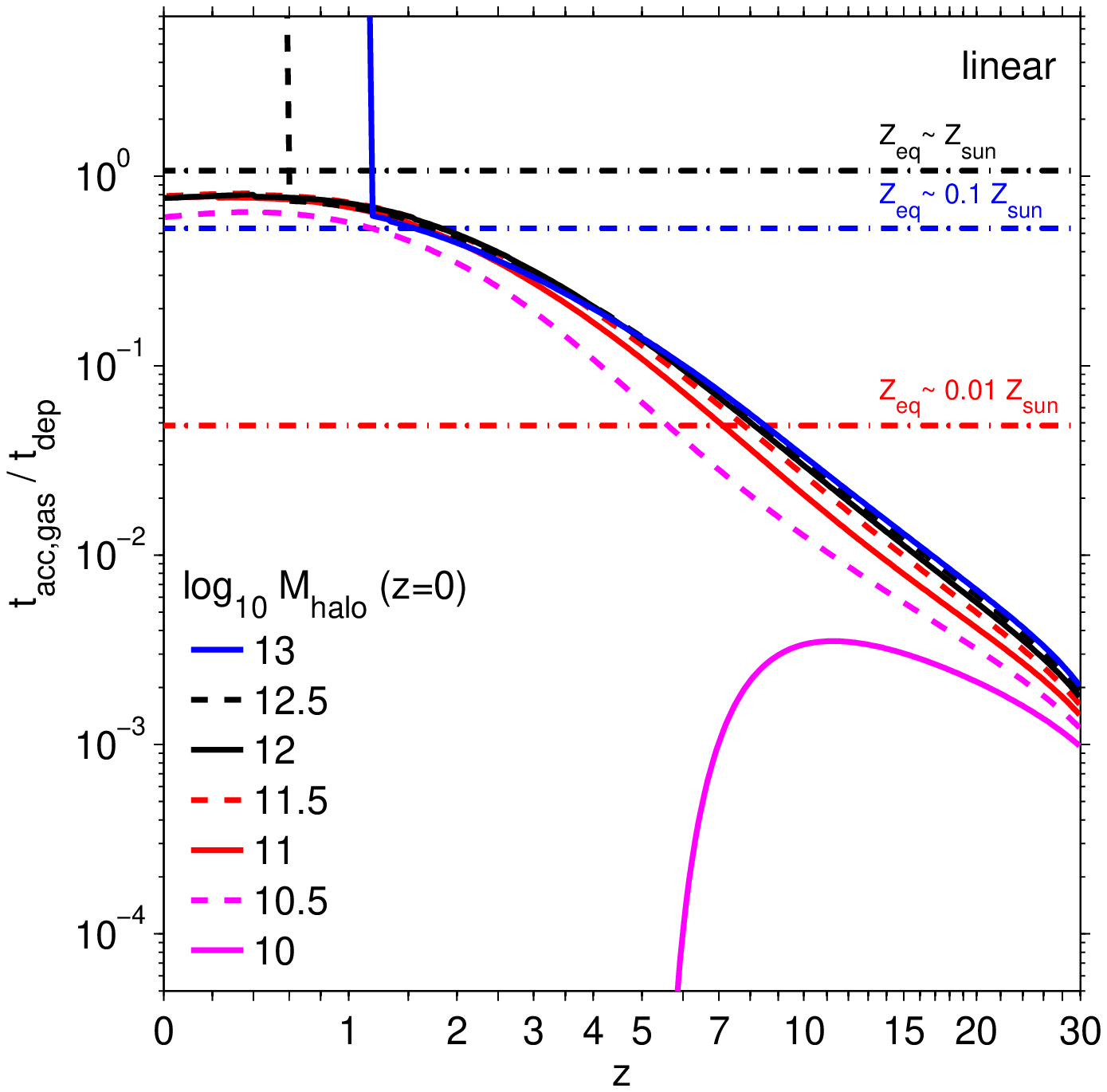} &
\includegraphics[width=80mm]{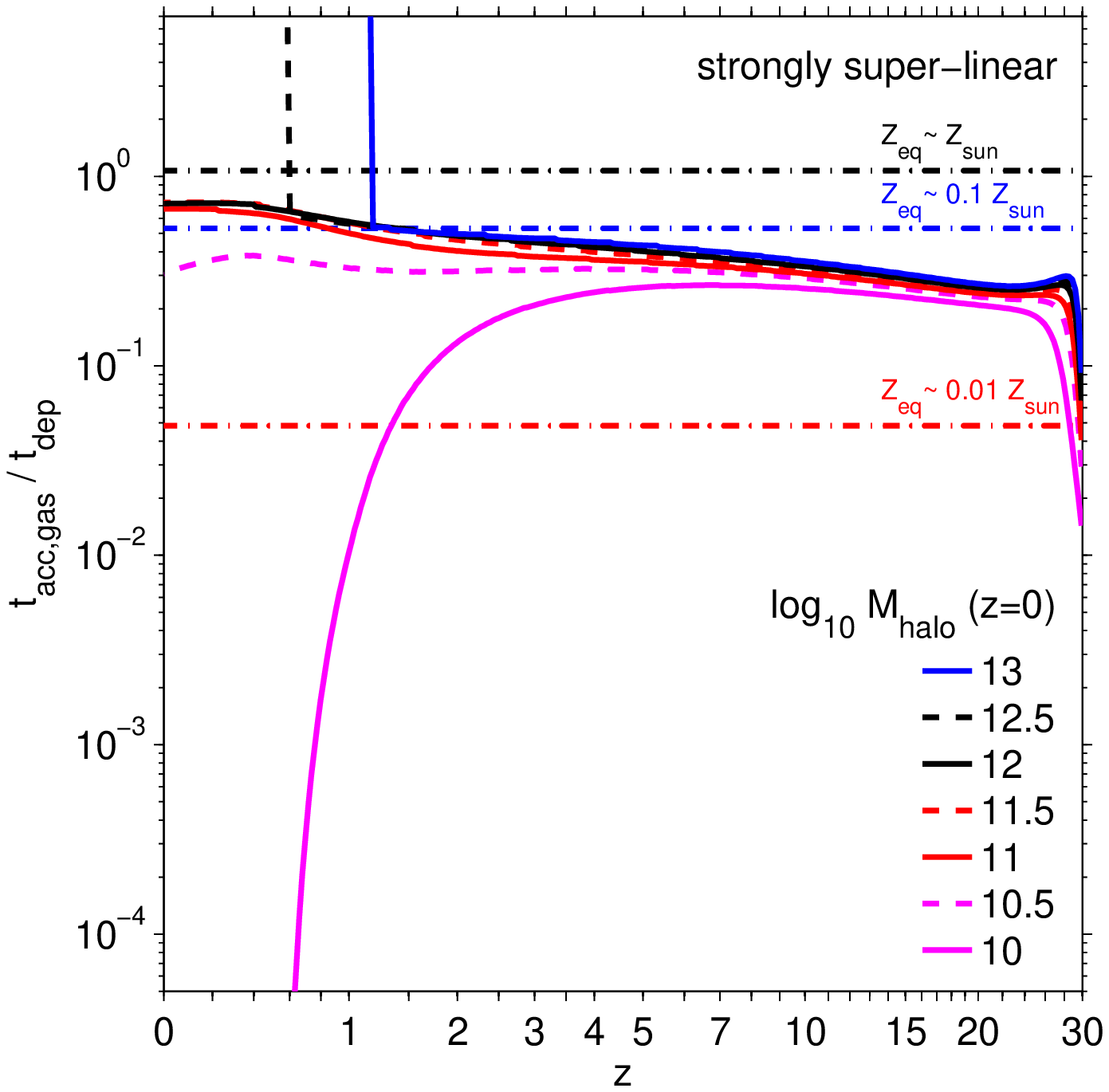} 
\end{tabular}
\caption{Ratio between gas accretion time (gas mass divided by gas accretion rate) and gas depletion time (gas mass divided by SFR), or equivalently the ratio between SFR and gas accretion rate, for galaxies in halos with present day masses in the range $10^{10}-10^{13}$ $M_\odot$ (see legend). $\Sigma_{\rm SFR}$ is assumed to scale either linearly (left panel) or quadratically (right panel) with $\Sigma_\H2$. Star formation is limited by gas accretion if the ratio is of the order of unity. The ratio gets very large in high mass halos at low redshift due to virial shock quenching of gas accretion, see Section \ref{sect:model}. The dot-dashed lines shows the ratio of accretion to depletion time that is required to keep the gas metallicity constant at the value indicated in the figure. These lines are computed using Equation (\ref{eq:dotZeq0}) with $\zeta=0.9$ for $Z=0.1-0.01$ $Z_\odot$ and $\zeta=0.5$ for $Z=Z_\odot$. A ratio of accretion to depletion time above a given dot-dashed line with label $Z$ indicates that metal enrichment via star formation outmatches the accretion of gas of low metallicity and, as a net effect, that the metallicity can raise above $Z$. A linear star formation -- gas relation (Equation \ref{eq:SGlin}) has a gas depletion time that is significantly larger than the gas accretion time at $z>3$. Hence, metal enrichment and star formation is delayed until late times. In contrast, a strongly super-linear star formation -- gas relation has a depletion time that is short enough for star formation to catch up with the accretion rate early on. This leads to a fast and sustained enrichment to $\sim{}0.1$ $Z_\odot$ at high redshift and, consequently, to high molecular fractions and substantial star formation over much of cosmic history.}
\label{fig:tacc_over_tSF}
\end{figure*}

Previously we have demonstrated that the functional form of the star formation -- gas relation has a profound impact on the SFR of galaxies at high redshifts and we have attributed this effect to the altered metal enrichment. In this section, we will show that the metal enrichment is intrinsically linked to the star formation -- gas relation via the balance between star formation and gas accretion in halos.

In Fig.~\ref{fig:tacc_over_tSF} we plot the evolution of the $r$, the ratio of the SFR and the gas accretion rate, or equivalently the ratio of the accretion time (gas mass divided by the gas accretion rate) to the gas depletion time (gas mass divided by SFR), for halos of a range of masses. At $z\gtrsim{}3$, a linear star formation -- gas relation (Equations \ref{eq:SGlin}) results in a gas depletion time that is significantly longer (it is 2.5 Gyr if all gas is molecular and even longer otherwise) than the gas accretion time ($\sim{}2-3\times{}10^7$ yr at $z=30$, $\sim{}1-2\times{}10^{8}$ yr at $z=10$, $2-7\times{}10^{8}$ yr at $z=5$). Hence, star formation is ``depletion limited'' at those redshifts. Changing the functional form of the star formation gas -- relation alters the gas depletion times, but has only a small effect on the gas accretion rates, which are determined primarily by the mass accretion rates onto dark matter halos. A strongly super-linear star formation gas -- relation, for instance, results in gas depletion times as low as $\sim{}2\times{}10^{8}$ yr at $z>10$. As a consequence, star formation is ``accretion limited'' (the SFR is of the same order as the gas accretion rate) over most of cosmic history in this case.

Equation (\ref{eq:dotZeq0}) allows us to understand the consequences of a depletion limited vs an accretion limited star formation. Consider a gas disk with a given metallicity $Z$. Such a gas disk will increase or decrease its metallicity until it matches the metallicity $Z_{\rm eq}$ specified by Equation (\ref{eq:dotZeq1}) for the given SFR to gas accretion rate ratio $r$. For a stellar mass loss fraction of $R=0.46$, a stellar metal yield of $y=0.069$, and $\zeta=0.9$ (a reasonable choice for all but the most massive halos at $z\sim{}4-10$), we obtain $Z_{\rm eq}=0.1$ $Z_\odot$ for $r=0.53$, but $Z=0.01$ $Z_\odot$ for $r=0.048$. Therefore, if star formation is accretion limited the gas will get enriched to $\sim{}0.1$ $Z_\odot$ or more on the gas accretion time scale. 
Accretion limited star formation, and the rapid enrichment to $Z\sim{}0.1$ $Z_\odot$, has indeed been seen in many hydrodynamical simulations with a super-linear star formation model, see, e.g., \cite{2006MNRAS.370..273D, 2011MNRAS.410.1703F}. The equilibrium metallicity is small if $r$ is small, i.e., if the depletion time is much longer than the gas accretion time. Hence, metal enrichment is delayed until $z\sim{}2-3$ in case of a  linear star formation -- gas relation with a sufficiently large ($>$Gyr) depletion time.

To summarize, the functional form of the star formation -- gas relation determines whether star formation in a given halo at a given time is depletion limited or accretion limited. This is crucial, because the evolution of the metallicity in a halo is driven by the competition between gas accretion and gas depletion (Equation \ref{eq:dotZeq1}). A linear star formation -- gas relation delays metal enrichment until late times and, hence, suppresses star formation at high redshifts. In contrast, a strongly super-linear star formation -- gas relation boosts metal enrichment and star formation at high redshifts.

A quantity similar to the ratio between SFR and gas accretion rate has been studied empirically by \cite{2012arXiv1209.3013B} using abundance matching techniques. In their work the baryon equivalent accretion rate (BEAR; $f_{\rm b}\dot{M}_{\rm halo}$) is used as a proxy for the unknown gas accretion rate. They find (see Fig.~1 in \citealt{2012arXiv1209.3013B}) that the SFR to BEAR ratio is roughly constant out to $z\sim{}3$ and speculate that it might be a time-independent for much of cosmic history.

\begin{figure}
\begin{tabular}{cc}
\centerincludegraphics[width=70mm]{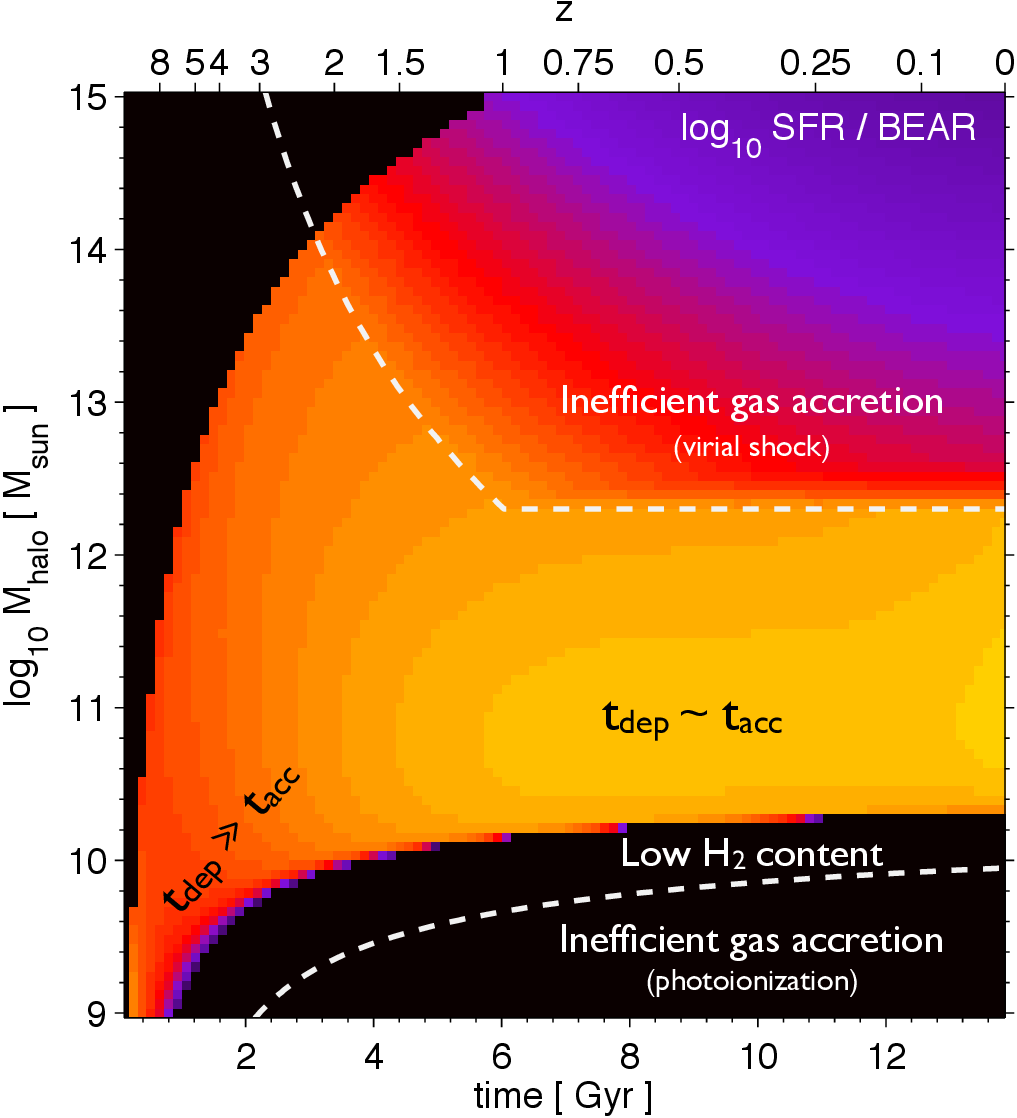} &
\centerincludegraphics[height=65.5mm]{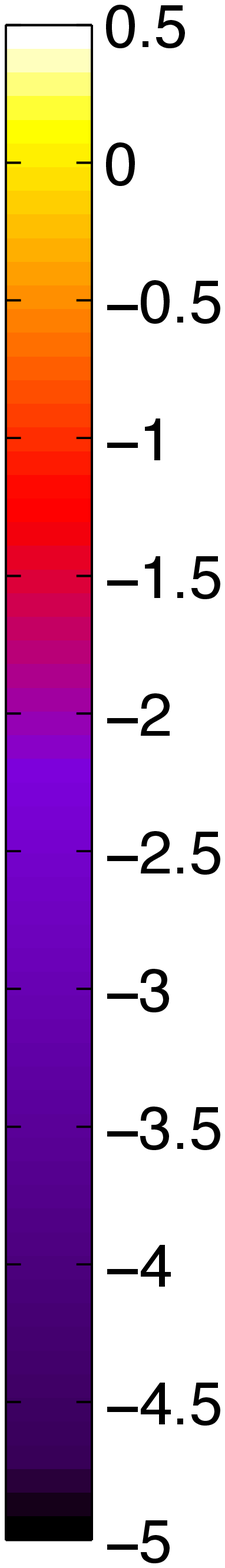}
\end{tabular}
\caption{Ratio between SFR and baryon equivalent accretion rate ($f_{\rm b}\dot{M}_{\rm halo}$, BEAR) of galaxies as function of cosmic time and halo mass for a linear $\Sigma_{\rm SFR}-\Sigma_\H2$ relation. Colors indicate $\log_{10} {\rm SFR} / {\rm BEAR}$, ranging from black (SFR much smaller than BEAR) to yellow (SFR $\sim{}$ BEAR). For $z\lesssim{}2$ the ratio peaks at a halo mass of approximately $10^{11}$ $M_\odot$ in reasonable agreement with observations (compare with Fig. 1 of \citealt{2012arXiv1209.3013B}). Galaxies in low mass halos ($M_{\rm halo}\lesssim{}2\times{}10^{10}$ $M_\odot$ at $z\sim{}0$) have their star formation quenched due the scarcity of molecular hydrogen in such halos.  In addition, halos stop accreting gas once they become massive enough (the upper dashed line) to support a virial shock that prevents cold gas accretion onto the galaxy \citep{2006MNRAS.368....2D}. At $z\lesssim{}1$ star formation is confined to a narrow range of halo masses of approximately a few times $10^{10}$ $M_\odot$ to $10^{12}$ $M_\odot$ \citep{2010ApJ...718.1001B, 2012ApJ...753...16K}.}
\label{fig:SFE}
\end{figure}

In Fig.~\ref{fig:SFE} we show our predictions for the SFR to BEAR ratio using a similar color scheme and axis ranges as in Fig.~1 of \cite{2012arXiv1209.3013B}. We compute baryon equivalent accretion rates based on the halo mass evolution formula provided in the appendix of \cite{2012arXiv1209.3013B}. We find that a linear, $\H2$-based star formation gas -- relation can reproduce their Fig.~1 reasonably well. Most importantly, the ratio between SFR and BEAR peaks around halo masses of $\sim{}10^{11}$ $M_\odot$ at $z\lesssim{}2$. Some features seen in Fig.~\ref{fig:SFE}, such as the complete lack of star formation in sufficiently low mass halos, may be related to the fact that the model follows the average evolution of galaxy properties and, at least in its current form, ignores the possibility of scatter caused by, e.g., the variability of the accretion rates or scatter in the disk sizes of galaxies.

\subsection{An equilibrium view on galaxy evolution}
\label{sect:EQ}

Not only does the ratio between SFR and gas accretion rate regulate the metallicity in galaxies (see Equation \ref{eq:dotZeq1}), it also determines the gas fraction $f_{\rm g}=M_{\rm g}/M_{\rm halo}$, stellar fraction $f_{\rm s}=M_{\rm *}/M_{\rm halo}$, and the metal fraction $f_{\rm Z}=M_{\rm Z}/M_{\rm halo}$ of galaxies.

Specifically, we can show using Equations \ref{eq:dotMgin}, \ref{eq:dotMstarin}, \ref{eq:dotMg}, and \ref{eq:dotMstar} that for a given ratio $r\equiv{}\frac{\dot{M}_{\rm *, form}}{\dot{M}_{\rm g, in}}$ \\
$\dot{f}_{\rm g}=0$ if $f_{\rm g}=f_{\rm g, eq}(r)$ with
\begin{equation}
f_{\rm g, eq}(r) = \left[1- r(1-R+\epsilon_{\rm out})\right]\epsilon_{\rm in}f_{\rm g, in}f_{\rm b}, \label{eq:fgeq}
\end{equation}
$\dot{f}_{\rm s}=0$ if $f_{\rm s}=f_{\rm s, eq}(r)$ with
\begin{equation}
f_{\rm s, eq}(r) = (1 - f_{\rm g, in})f_{\rm b} + \epsilon_{\rm in}f_{\rm g, in}f_{\rm b}(1-R)r, \label{eq:fseq}
\end{equation}
$\dot{f}_{\rm Z}=0$ if $f_{\rm Z}=f_{\rm Z, eq}(r)$ with
\begin{equation}
f_{\rm Z, eq}(r) = Z_{\rm eq}(r) f_{\rm g, eq}(r),  \label{eq:fZeq}.
\end{equation}
The equilibrium values of other properties, such as the gas-to-stellar mass ratio $f_{\rm gs}=f_{\rm g}/f_{\rm s}$ or the baron fraction $f_{\rm g}+f_{\rm s}$, can be obtained by combining these equations. For instance,
$\dot{f}_{\rm gs}=0$ if $f_{\rm gs}=f_{\rm gs, eq}(r)$ with
\begin{equation}
f_{\rm gs, eq}(r) = \frac{f_{\rm g, eq}(r)}{f_{\rm s, eq}(r)}  = \frac{1-(1-R+\epsilon_{\rm out})r}{\frac{1-f_{\rm g,in}}{\epsilon_{\rm in}f_{\rm g,in}}+(1-R)r}.  \label{eq:fgseq}
\end{equation}

We show in appendix \ref{sect:Stability} that the equilibrium state specified by Equations (\ref{eq:fgeq})-(\ref{eq:fZeq}) is (linearly) stable under most circumstances for SFR to gas accretion rate ratios less than $1/(1-R+\epsilon_{\rm out})$. If, however, the galaxy is in a condition in which the $\H2$ fraction is extremely sensitive to small changes of the metal abundance of the ISM \emph{and} star formation depends on the presence of molecular hydrogen, a run-away increase or decrease in $\H2$ abundance might take place. The latter ceases, and the galaxy reaches a new equilibrium, as soon as the $\H2$ fraction becomes sufficiently insensitive to changes in metal abundance. There can be no equilibrium state for $r>1/(1-R+\epsilon_{\rm out})$, because it would lead to negative equilibrium gas fractions, see Equation \ref{eq:fgeq}. However, $r>1/(1-R+\epsilon_{\rm out})$ is a physical possibility and applies, e.g., to massive galaxies at low redshifts that have their gas accretion quenched, but are still star forming. These galaxies are obviously not in an equilibrium state, but rely on their diminishing gas reservoir for star formation.

The stability of the equilibrium state implies, for instance, that if $f_{\rm g}<{}f_{\rm g, eq}(r)$ (and it is sufficiently close to the equilibrium value), the gas fraction increases and continues to do so until it reaches $f_{\rm g, eq}(r)$. Similarly, if $f_{\rm g}>f_{\rm g, eq}(r)$, the gas fraction decreases until it matches the equilibrium value. We can therefore draw important conclusions for the gas fractions and gas-to-stellar mass ratios from Fig.~\ref{fig:tacc_over_tSF}. A small SFR to gas accretion rate ratio of, e.g., $r=0.048$ corresponds to an equilibrium metallicity of $Z=0.01$ $Z_\odot$, to an equilibrium gas fraction $f_{\rm g,eq}/f_{\rm b}\sim{}0.83$ $(0.46)$ and to an equilibrium gas-to-stellar mass ratio $f_{\rm gs, eq}\sim{}6.8$ $(0.9)$. Here we assume\footnote{The parameter $f_{\rm g, in}$ can affect the predicted equilibrium gas fractions and gas-to-stellar mass ratios noticeably. The values chosen here are for illustration purposes, but we stress that we compute $f_{\rm g, in}$ self-consistently in the model (see Section \ref{sect:model}).} $f_{\rm g, in}=0.9$ ($f_{\rm g, in}=0.5$) and the default parameters of the model: $\epsilon_{\rm in}=1$, $R=0.46$, $\epsilon_{\rm out}=1$. Hence, depletion limited star formation implies low metallicities, large gas fractions, and large gas-to-stellar mass ratios. In contrast, equilibrium metallicities are higher and equilibrium gas fractions and gas-to-stellar mass ratios are much lower if star formation is accretion limited. For instances, a SFR to gas accretion rate ratio $r=0.53$ corresponds to $Z_{\rm eq}=0.1$ $Z_\odot$, $f_{\rm g,eq}/f_{\rm b}\sim{}0.17$ $(0.09)$ and $f_{\rm gs, eq}\sim{}0.46$ $(0.14)$ for $f_{\rm g, in}=0.9$ ($f_{\rm g, in}=0.5$).

In the context of our model the baryonic properties of galaxies (of given halo mass and at a given redshift) are specified by $f_{Z}$, $f_{g}$, and $f_{s}$. In more abstract terms, the set $(f_{Z},f_{g},f_{s})$ represents the baryonic state of a galaxy. Equations (\ref{eq:fgeq}) - (\ref{eq:fZeq}) denote a curve\footnote{This is only approximately true, because some of the terms in the Equations (\ref{eq:fgeq}) - (\ref{eq:fZeq}) dependent on more than just $r$, e.g., on the halo mass. This also implies that different galaxies will in general follow somewhat different curves of fixed points during their evolution. see Fig.~\ref{fig:eqModel}.} of fixed points parametrized by $r$ in this baryonic state space. We demonstrate in the appendix that if $(f_{Z},f_{g},f_{s})$ is reasonably close to one of the fixed points it will approach it. As the cosmic accretion rate onto a halo changes, $r$ changes as well and the galaxy transits smoothly from one equilibrium state to the next. Strong perturbations, such as major mergers, will also allow a galaxy to leave its current fixed point, but we would expect that such events are relatively rare. Hence, galaxy evolution is largely an evolution along this curve of fixed points driven by the external modulation of the gas accretion rate.

\begin{figure*}
\begin{tabular}{cc}
\includegraphics[width=80mm]{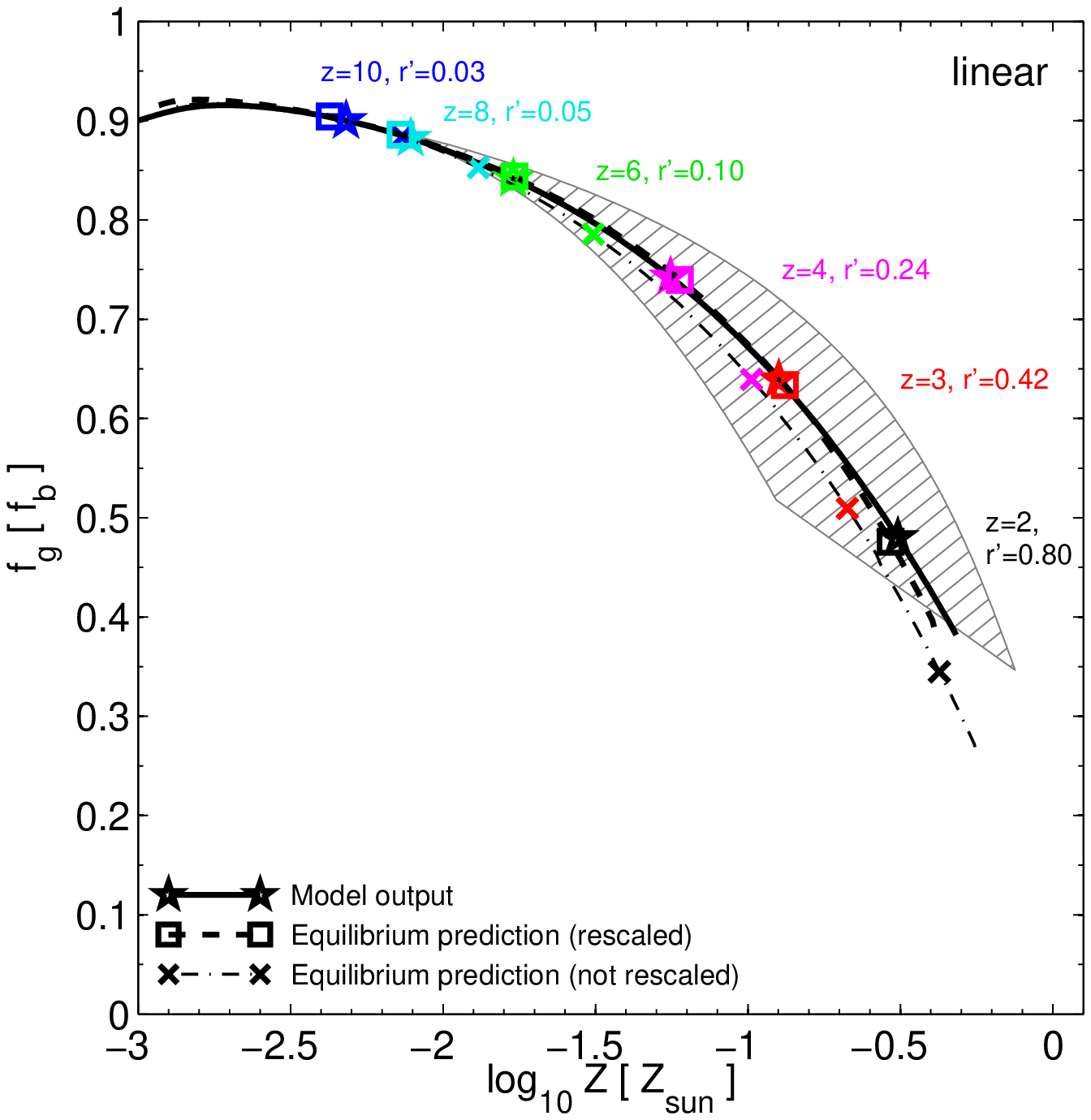} &
\includegraphics[width=80mm]{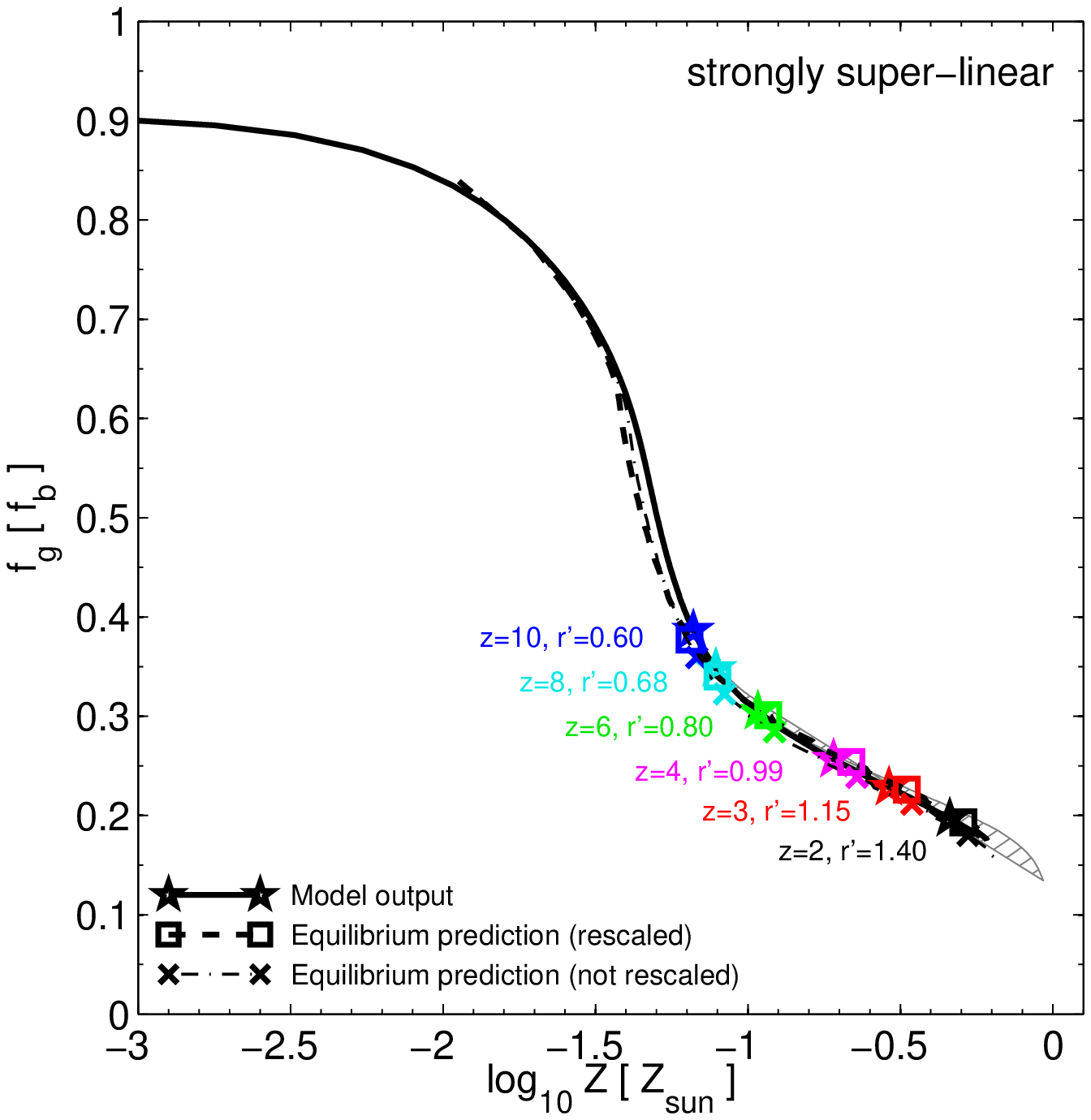}
\end{tabular}
\caption{Comparison of the gas fractions and metallicities at $z\geq{}1.5$ as predicted by the dynamical model (see section \ref{sect:model}) and by the equilibrium scenario (see section \ref{sect:EQ}) for a linear $\Sigma_{\rm SFR}-\Sigma_\H2$ relation (equation \ref{eq:SGlin}; left panel) and a strongly super-linear relation (equation \ref{eq:SGquad}; right panel). Solid lines show the predictions of the dynamical model for a galaxy residing in the progenitor of a $10^{12}$ $M_\odot$ (at $z=0$) halo. Shaded areas show the corresponding results for halos with $z=0$ masses of $10^{11}$ $M_\odot$ (lower boundary) and $10^{13}$ $M_\odot$ (upper boundary), respectively. Symbols highlight predictions at the specific redshifts 10, 8, 6, 4, 3, and 2 as indicated in the figure. Dot-dashed lines show the equilibrium predictions based on a straightforward application of equations (\ref{eq:dotZeq1alt}) and (\ref{eq:fgeqalt}) with $r'=t_{\rm acc}/t_{\rm dep}$. The value of $r'$ is indicated in the figure. Rescaling $r'$ can substantially improve the accuracy of the equilibrium formalism, see text. The dashed line shows the result of rescaling $r'$ by a factor 0.45 (left panel) and 0.9 (right panel), respectively. There is excellent agreement between the predictions of the dynamical model (solid line) and those of the rescaled equilibrium scenario (dashed line).}
\label{fig:eqModel}
\end{figure*}

Instead of using $r$ it is also possible to phrase the equilibrium metallicities, gas fractions, and stellar fractions in terms of the ratio $r'$ between the \emph{matter} accretion time ($M_{\rm halo}/\dot{M}_{\rm halo}$) and the gas depletion time ($M_{\rm g}/\dot{M}_{\rm *,form}$). Using $r'$ allows for a conceptually clean separation between gravitational processes (determining matter accretion times) and baryonic physics (determining gas depletion times). However, as we show now, both formulations are in fact equivalent and the conversion from one formulation to the other is straightforward. Using the definitions of $r$ and $r'$ and equation (\ref{eq:dotMgin}) it is easy to show that $f_{\rm g} = \frac{r}{r'} \epsilon_{\rm in}f_{\rm g,in}f_{\rm b}$. Hence, in equilibrium, see (\ref{eq:fgeq}),
\[
r' = \frac{r}{1-r(1-R+\epsilon_{\rm out})},\,\text{and}\,r = \frac{r'}{1+r'(1-R+\epsilon_{\rm out})}.
\]
Plugging the latter equation into (\ref{eq:dotZeq1}), (\ref{eq:fgeq}), and (\ref{eq:fseq})  results in the following alternative formulation for the equilibrium baryonic state of a galaxy
\begin{flalign}
Z_{\rm eq}(r') &= \frac{r'y(1-R)(1-\zeta)}{1+r'(1-R+\epsilon_{\rm out})}+Z_{\rm IGM}, \label{eq:dotZeq1alt} \\
f_{\rm g, eq}(r') &= \frac{ \epsilon_{\rm in}f_{\rm g, in}f_{\rm b} }{1+ r'(1-R+\epsilon_{\rm out})}, \label{eq:fgeqalt}\\
f_{\rm s, eq}(r') &= (1 - f_{\rm g, in})f_{\rm b} + \frac{ r' \epsilon_{\rm in}f_{\rm g, in}f_{\rm b}(1-R) }{1+r'(1-R+\epsilon_{\rm out})}. \label{eq:fseqalt}
\end{flalign}

The idealized equilibrium picture that we outline here assumes that galaxies can adapt on short timescales to changes in the SFR to gas accretion rate ratio $r$. The matter accretion time $M_{\rm halo}/\dot{M}_{\rm halo}$ is a natural timescale for gas fraction, stellar fractions, and metallicities to reach their equilibrium values, but the precise value depends on the parameters of the model, e.g., on the exponent of the star formation -- gas relation, see appendix \ref{sect:Stability}. Since the accretion time is a significant fraction of the Hubble time at any given redshift, we expect that the stellar fractions, gas fractions and metallicities lag somewhat behind their equilibrium predictions. Hence, we expect that gas fractions are somewhat underestimated by the equilibrium formalism, while stellar fractions and metallicities are overestimated, as seen in Fig.~\ref{fig:eqModel}. In addition, the derivation of the equilibrium equations assumes that model parameters and halo masses remain fixed, which is not always a good assumption.

In order to test whether the equilibrium picture can be used to make quantitative predictions we show the $f_{\rm g}$-$Z$ projection of the baryonic state space in Fig.~\ref{fig:eqModel}. We plot both the metallicities and gas fraction as computed by the dynamical model, and also the corresponding equilibrium quantities based on Equations  (\ref{eq:dotZeq1alt}) and (\ref{eq:fgeqalt}).  

The equilibrium formalism describes the overall trend between $f_{\rm g}-Z$ well, but for a given $r'$ it can underpredict the gas fractions and overpredict the metallicities. The disagreement can be reduced if $r'$ (or $r$) is scaled by a factor $\sim{}0.45[2-(2-n)^2]$ before applying equations  (\ref{eq:dotZeq1alt})-(\ref{eq:fgeqalt}) (or equations (\ref{eq:dotZeq1})-(\ref{eq:fgeq})). Here, $n\in[1,2]$ is the power-law exponent $n$ of the star formation -- gas relation.

Our equilibrium picture of galaxy formation differs from the ``equilibrium condition'' of a steady state model \citep{2008MNRAS.385.2181F, 2010ApJ...718.1001B, 2010MNRAS.405.1690D, 2012MNRAS.421...98D} in which the SFR in a given halo is balanced by the net gas accretion rate. Most importantly, we demonstrate that galaxies can be in a stable equilibrium even if the ``equilibrium condition'' is not even remotely satisfied. Specifically, galaxies can have little star formation and can massively build up their gas reservoirs and, yet, be in a proper equilibrium state. 

The ``equilibrium condition'' of the steady state model is equivalent to the assumption that the gas \emph{mass} in a galaxy remains constant. In contrast, our equilibrium condition is that the gas \emph{fraction} remains constant. The special state with $r=1/(1-R+\epsilon_{\rm out})$ (or equivalently $r'\rightarrow\infty{}$) satisfies the ``equilibrium condition'' of the steady state model. Indeed a similar relation for $r$ is often assumed to hold generally in the steady state approach, see e.g., Equation 2 of  \cite{2012MNRAS.421...98D}. However, as Equation (\ref{eq:fgeq}) shows the equilibrium gas fraction, and, hence, gas mass, in this equilibrium state is zero, and, hence, gas-rich, star forming galaxies at high redshift will not be in this particular state.

The equilibrium view presented in this section shows why metallicities, gas fractions, and stellar fractions of galaxies are strongly correlated. They are all coupled via a single parameter, which can be chosen to be $r$, the ratio between the SFR and the gas accretion rate, see equations (\ref{eq:dotZeq1}) - (\ref{eq:fseq}), or $r'$, the ratio between the matter accretion time and the gas depletion time, see equations (\ref{eq:dotZeq1alt}) - (\ref{eq:fseqalt}). The latter formulation makes the following point particularly transparent. The functional form of the star formation -- gas relation determines the gas depletion time while gravitational processes alone determine the matter accretion time. \emph{Hence, for a given matter accretion rate the functional form of the star formation -- gas relation determines the equilibrium properties (incl. the metallicity, gas fraction, stellar fraction) of a given galaxy. Galaxy properties evolve with time primarily due to the modulation of the cosmic accretion rate.}

\section{Summary and Discussion}
\label{sect:Summary}

We have studied the implications of the functional form of the star formation -- gas relation for the cosmic star formation history, for gas accretion in galaxies, and for the evolution of global galaxy properties using a model that follows the average evolution of galaxies in growing dark matter halos across cosmic time \citep{2012ApJ...753...16K}. We specifically contrasted the predictions for a linear  star formation -- gas relation \citep{2011ApJ...730L..13B} with those for a strongly super-linear relation \citep{2011ApJ...731...41O}, but also tested a few other functional forms suggested in the literature \citep{2009ApJ...699..850K, 2012ApJ...745...69K, 2012arXiv1210.1218S}. In the following we express our statements in terms of a linear vs a super-linear star formation -- gas relation. However, as we argued in section \ref{sect:EQ}, the physically relevant characteristic of the star formation -- gas relation is not the actual value of the slope, but the rather the duration of the gas depletion time w.r.t. the matter accretion time. The adopted linear star formation -- gas relation has a constant (and relatively long) depletion time, while the strongly super-linear relation has a galaxy-averaged depletion time that is typically comparable to or even shorter than the matter accretion time. Our main findings are as follows:

\begin{itemize}
\item We show that a linear  $\Sigma_{\rm SFR}-\Sigma_\H2$ relation with a $\sim{}2.5$ Gyr depletion time results in a mass -- metallicity relation, in gas-to-stellar mass ratios, and in a stellar-to-halo mass relation that is in reasonable agreement with observations. In contrast, the short gas depletion times that result from adopting a strongly super-linear star formation -- gas relation lead to a stark mismatch between model predictions and observations.

\item The functional form of the star formation -- gas relation strongly affects the cosmic star formation history at $z\gtrsim{}4$. A linear star formation -- $\H2$ relation with a sufficiently long gas depletion time suppresses star formation in halos with $z=0$ masses below a few times $\sim{}10^{10}$ $M_\odot$. If, however, the star formation -- gas relation is highly non-linear, star formation can occur in halos with $z=0$ masses even below $\sim{}10^{10}$ $M_\odot$. As a result, SFRDs can change by factors of $\sim{}30$ at $z=10$ ($\sim{}5$ at $z=6$), when switching between a linear and a strongly super-linear star formation -- $\H2$ relation, see Fig.~\ref{fig:MadauPlot} and \ref{fig:MadauPlot2}.

\item  The functional form of the star formation -- gas relation also strongly affects the \emph{observable} cosmic star formation history, i.e., the SFRD of galaxies with magnitudes accessible to observations (here assumed to be $M_{\rm UV,AB}<-17.7$, \citealt{2012ApJ...754...83B}). A linear star -- $\H2$ formation relation results in an observable cosmic star formation history compatible with observational data, while a strongly super-linear relation severely overpredicts the observable SFRDs at high redshift.

\item Whether star formation is based on $\H2$ or on total gas leaves little imprint on the \emph{observable} cosmic star formation history. However, the \emph{total} (including galaxies down to the faintest magnitudes) cosmic star formation history can be affected depending on the functional form of the star formation -- gas relation. If the relation is linear, $\H2$-based star formation will be suppressed in low mass (and $\H2$-poor) halos and the total SFRD will be reduced significantly, e.g, by more than an order of magnitude at $z=8$. However, if the star formation -- gas relation is strongly super-linear, then low mass halos will be enriched early on. Hence, low mass halos will contain a significant reservoir of $\H2$ and star formation in such halos will contribute significantly to the SFRD independently of whether star formation is $\H2$-based or not.

\item For the adopted linear $\Sigma_{\rm SFR}-\Sigma_\H2$ relation (Equation \ref{eq:SGlin}), the gas depletion time exceeds the gas accretion time ($\sim{}$ free-fall time of the halo) at high $z$. During this time the SFRs are orders of magnitude smaller than the gas accretion rates. Star formation becomes accretion limited, i.e., SFRs $\sim{}$ gas accretion rates, only at relatively recent epochs ($z\lesssim{}2-3$). We show that these predictions are in reasonable agreement with abundance matching predictions by \cite{2012arXiv1209.3013B}. In contrast,  star formation is close to accretion limited already at very high redshifts if the star formation -- gas relation is strongly super-linear.

\item The evolution of metallicities, gas fractions, and stellar fractions are driven by the ratio between the SFR and gas accretion rate or, equivalently, by the ratio between gas depletion time (set by the star formation -- gas relation) and the matter accretion time (set by gravity). For a given star formation -- gas relation galaxies evolve towards a (linearly stable) equilibrium state with a gas fraction, stellar fraction and metallicity determined only by the accretion rate. The equilibrium concept presented in this work differs from the steady state model suggested in the literature (e.g., \citealt{2012MNRAS.421...98D}) and shows that galaxies can be in equilibrium even if their gas masses are growing quickly. We argue that galaxy evolution can be interpreted as an evolution along a set of equilibria driven primarily by the change in the cosmic accretion rate.

\end{itemize}

As pointed out above we find that the observed cosmic star formation history is well reproduced if the star formation -- gas relation is linear and has an $\H2$ depletion time of $\sim{}2.5$ Gyr. Given the simplicity of the model and the uncertainties in the observational data we do not rule out the possibility that the relation is actually slightly super- or sub-linear (e.g., has an exponent in the range $\sim{}0.85-1.3$) or has a somewhat different depletion time. However, our findings are difficult to reconcile with the idea of a highly super-linear, especially quadratic, star formation -- gas relation.

Many of our findings for  a \emph{linear} star formation -- gas relation are almost identical to those found by \cite{2012ApJ...753...16K} for a \emph{non-linear} star formation -- gas relation, similar to Equation (\ref{eq:SGK09}), with an exponent of 1.33 at high gas surface densities. However, their star formation -- gas relation uses a very large transition surface density ($\Sigma_{\rm g}=850$ $M_\odot$ pc$^{-2}$ compared with $\Sigma_{\rm g}=85$ $M_\odot$ pc$^{-2}$ used in Equation \ref{eq:SGK09}) at which the scaling changes from sub-linear to super-linear. This matters because gas surface densities of high redshift galaxies are of the order of a few 100 $M_\odot$ pc$^{-2}$, see Fig.~\ref{fig:gasSurfaceDensity}. Hence, a transition surface density as large as $\Sigma_{\rm g}=850$ $M_\odot$ pc$^{-2}$ implies that a large 
fraction of the gas in galaxy disks is subject to a sub-linear star formation -- gas relation and that $\Sigma_{\rm SFR}$ scales super-linearly with $\Sigma_\H2$ only in the central regions (where surface densities are larger by a factor $e^{1.7}\sim{}5.5$ than the surface densities at the half-mass radius). Averaging over sub- and super-linear regions results in a close-to-linear relation between $\H2$ masses and SFRs in galaxies, hence, to similar predictions as for the strictly linear star formation -- gas relation given by Equation (\ref{eq:SGlin}). 

\cite{2012ApJ...753...16K} argue that star formation based on $\H2$ is an essential ingredient to bring the predicted SFRDs at high redshifts in agreement with observations. We find that metal-dependent quenching of star formation based on $\H2$ has only a small impact (0.3 dex at $z\sim{}9$ and less at lower redshift) on the \emph{observable} SFRDs at high redshift, because it affects primarily faint galaxies below the current detection limits. However, metal-dependent quenching plays an important role for the total (i.e., including galaxies below the detection limits) SFRD \emph{if} the star formation -- gas relation is not highly super-linear. This is in agreement with \cite{2012ApJ...753...16K} who effectively use an approximately linear  star formation -- gas relation (see above) to study the total SFRD. In case of a strongly super-linear relation, however, star formation in galaxies is almost always close to being accretion limited even at $z\sim{}10$. Therefore, galaxies get enriched early on and SFRs proceed at a rate that is effectively identical (to within a factor of 2 or so) to the gas accretion rates, independently on whether a $\Sigma_{\rm SFR}-\Sigma_\H2$ or a $\Sigma_{\rm SFR}-\Sigma_{\rm g}$ relation is used.

\cite{2010MNRAS.402.1536S} analyze the star formation history of the universe using large scale cosmological simulations. The star formation in their simulations is modeled via a moderately non-linear ``Schmidt law'', i.e., a star formation -- gas relation based on volumetric densities. Star formation is allowed to take place only in regions with densities above a threshold that is either fixed (the default model) or scales with metallicity. Both recipes result in essentially identical cosmic star formation histories, indicating that metal-dependent quenching of star formation might not very important. However, compared with \cite{2012ApJ...753...16K} their threshold is much lower than the density required for $\H2$ formation \citep{2011ApJ...728...88G} and the metallicity scaling is much weaker ($\propto{}\log{}Z^{-1}$, \citealt{2004ApJ...609..667S}) than the one expected to regulate the $\H2$ abundance (approximately $\propto{}Z^{-1}$, see Equations \ref{eq:fH2}--\ref{eq:tauc}). It is therefore difficult to properly judge the implications of their findings in the context of $\H2$ based star formation.

\cite{2012ApJ...749...36K} study the evolution of the cosmic star formation history in simulations that follow the abundance of $\H2$ directly (using Equations \ref{eq:fH2}--\ref{eq:tauc}) and compare these to simulations that use a fixed density threshold for star formation. The chosen threshold (50 cm$^{-3}$)  corresponds approximately to the transition from the atomic to the molecular phase for gas of solar metallicity. The Schmidt law used in their simulations is moderately non-linear. Hence, we would expect that the two sets of simulations result in very similar predictions for the SFRs of galaxies bright enough to be observable.  Indeed, \cite{2012ApJ...749...36K} find that the SFRDs of galaxies with stellar masses larger than $10^{7.75}$ $M_\odot$ change by (only) a factor of $\sim{}2$. The differences are even smaller when galaxies are selected based on their UV luminosity ($M_{\rm UV, AB}<-17.7$; M. Kuhlen private communication). 

Another interesting question is the importance of stellar feedback for the regulation of star formation at high $z$. \cite{2010MNRAS.402.1536S} found that increasing the exponent of the star formation -- gas relation or decreasing the depletion time has little impact on the cosmic star formation history at $z\lesssim{}6$ and the authors use this fact to infer that galaxies are able to regulate their star formation via feedback processes. However, their default star formation -- gas relation is non-linear and, hence, increasing its exponent or decreasing its depletion time further has little impact if SFRs are already close to the accretion limit. 

\cite{2011MNRAS.416.1566L} predict, based on semi-analytical modeling, that the cosmic star formation history is relatively insensitive to the choice of the  star formation -- gas relation. However, they do not test for a linear star formation -- gas relation in which star formation depends on a metallicity dependent $\H2$ fraction. In fact, the tested relations are either independent of the gas metallicity or are moderately super-linear or both. For such relations we expect that star formation is limited by the gas accretion rate out to relatively high redshift and thus not very sensitive to the precise functional form of the adopted star formation -- gas relation. In addition \cite{2011MNRAS.416.1566L} consider the total cosmic star formation history (cf. red lines in Fig.~\ref{fig:MadauPlot2}) which is less sensitive to the choice of the star formation -- gas relation than the star formation history of galaxies accessible to current observations (cf. black lines in Fig.~\ref{fig:MadauPlot2}). A similar study by \cite{2012MNRAS.424.2701F} that includes a linear $\H2$-based star formation -- gas relation finds that the cosmic star formation history at high redshift differs depending on the choice of the star formation -- gas relation (see their Fig. 2), in agreement with our results.

We argue against the interpretation that galaxy properties and the cosmic star formation history are regulated by feedback driven large scale galactic outflows. The reason is, as we pointed out in Section \ref{sect:EQ}, that \emph{for a given star formation -- gas relation} the matter accretion rate onto the halo is the main regulatory mechanism that determines the gas masses, SFRs, metallicities, and stellar masses of galaxies. Of course, the strength of outflows can affect the predicted equilibrium properties, see equations (\ref{eq:dotZeq1alt}) -- (\ref{eq:fseqalt}), but only if the mass loading of the outflows is sufficiently large ($\epsilon_{\rm out}\gtrsim{}1/r'$). Of course, star formation could well be feedback-regulated within the ISM. In particular, the existence of a star formation -- gas relation that holds on $\sim{}$ kpc scales may be a consequence of the interplay between turbulence, gravity, and feedback in the first place (e.g., \citealt{2011MNRAS.417..950H}).

We have demonstrated that the functional form of the star formation -- gas relation has a particularly strong impact on galaxy properties at high redshifts. Especially the gas fractions of galaxies and the mass-metallicity relation differ widely depending on the adopted star formation -- gas relation. Ongoing and upcoming observational efforts carried out with, e.g., the Atacama Large Millimeter/sub-millimeter Array, will be crucial to constrain the functional form of the star formation -- gas relation and should be able to test many of the specific predictions of our model.

\acknowledgements
I thank the referee for a variety of suggestions that helped to improve this paper. I thank Peter Behroozi, Claude-Andr\'{e} Faucher-Gigu\`{e}re, Michael Kuhlen, and Matt McQuinn for insightful discussions regarding $\H2$-based star formation in dwarf galaxies, the suppression of accretion due to a photo-ionizing background, and abundance matching techniques. I am also grateful to Nick Gnedin and Michael Kuhlen for helpful comments on an early draft of this paper. I acknowledge support for this work by NASA through Hubble Fellowship grant HF-51304.01-A awarded by the Space Telescope Science Institute, which is operated by the Association of Universities for Research in Astronomy, Inc., for NASA, under contract NAS 5-26555. This work made extensive use of the NASA Astrophysics Data System and {\tt arXiv.org} preprint server.

\vspace{0.5cm}
\appendix

\section{Equilibria and Stability Analysis}
\label{sect:Stability}
 
 The time derivatives of the metallicity $Z$, metal fraction $f_{Z}$, gas fraction $f_{g}$, and stellar fraction $f_{s}$ are 
 \begin{flalign}
 \dot{Z}&=\frac{d}{dt}\frac{M_{Z}}{M_g}= \frac{1}{M_g}\left(\dot{M}_Z - Z\dot{M}_g\right), \label{eqA:dotZ} \\
 \dot{f_X}&=\frac{d}{dt}\frac{M_{X}}{M_h}= \frac{1}{M_h}\left(\dot{M}_X - f_X\dot{M}_h\right) \equiv{} g_{X}(f_{Z},f_{g},f_{s}),
 \end{flalign} 
 here $X\in{}\{Z,g,s\}$ and we introduced the functions $g_{Z}$, $g_{g}$, and $g_{s}$.
 
Combining Equation (\ref{eqA:dotZ}) with Equations (\ref{eq:dotMg}) and (\ref{eq:dotMZ}) we obtain 
 \begin{flalign}
  \dot{Z} &= \frac{1}{M_g}\left[ y(1-R)(1-\zeta)\dot{M}_{\rm *,form}  - (Z - Z_{\rm IGM})\dot{M}_{\rm g,in} \right] \nonumber \\
  &= \frac{\dot{M}_{\rm g,in}}{M_g}\left[ y(1-R)(1-\zeta) r - (Z - Z_{\rm IGM}) \right]. \label{eqA:dotZ2}
 \end{flalign}
 The second equality assumes $\dot{M}_{\rm g,in}\neq{}0$ and we introduced $r=\dot{M}_{\rm *,form}/\dot{M}_{\rm g,in}$.
 Clearly, aside from trivial cases such as $\dot{M}_{\rm *,form}=0$ \emph{and} $Z=Z_{\rm IGM}$, the metallicity can only be constant 
 if the equilibrium condition (\ref{eq:dotZeq0},\ref{eq:dotZeq1}) is satisfied, i.e, for
 $Z_{\rm eq}=  y(1-R)(1-\zeta) r + Z_{\rm IGM}$.
  
Let us assume that the metallicity is perturbed slightly relative to its equilibrium value, i.e., $Z=Z_{\rm eq}+\delta{}$ with a small perturbation $\delta$.
Let us also introduce the shorthands
 $A=1-R+\epsilon_{\rm out}$, $B=y(1-R)(1-\zeta)$, $C=\epsilon_{\rm in}f_{\rm g,in}f_{\rm b}$, $D=(1-f_{\rm g,in})f_{\rm b}$ to simplify the notation.
The evolution of $\delta{}$ can be obtained from the linearized version of Equation (\ref{eqA:dotZ2}).
\begin{equation}
\dot{\delta} = \delta{}\,\frac{\dot{M}_{\rm g,in}}{M_g}\left[ B \frac{\partial{}r}{\partial{}Z}(Z_{\rm eq}) - 1 \right].
\end{equation}
Hence, we find that $\delta{}\rightarrow{}0$ if $B \frac{\partial{}r}{\partial{}Z} < 1$. In particular, if, for a given gas configuration, star formation proceeds independently of its metallicity, the equilibrium metallicity is always stable against perturbations. However, if the SFR changes strongly with metallicity the equilibrium can become unstable. In this case the perturbation initially grows, i.e., gas is enriched or depleted, but eventually $\frac{\partial{}r}{\partial{}Z}$ becomes small (e.g., when all the gas is molecular) and the system approaches a new equilibrium. We note that this derivation ignores changes to the gas masses (or stellar masses), and hence, it does not prove that $Z_{\rm eq}$ is the metallicity of an equilibrium state that is also stable against these perturbations, which is what we show next.

Consider a galaxy that resides in a halo with mass $M_{\rm halo}$ and that accretes gas and stars according to the Equations (\ref{eq:dotMg}) and (\ref{eq:dotMstar}). The accretion rates are governed by gravitation and not affected by changes in the gas fraction or the metallicity of the galaxy. Also, the scale radius of the gas disk (proportional to the virial radius of the halo) is not affected by such changes (at least in the context of the model of Section \ref{sect:model}). We assume that the star formation in the galaxy is described by a power-law relationship
\begin{equation}
\dot{M}_{\rm *, form} = N \left[ f_g f_\H2(f_{Z}) \right]^n. \label{eqA:SFR}
\end{equation}
Here, $N$ is a normalization constant (which includes everything that does not change when the gas fraction or metal fraction changes) and $n$ is the power-law exponent of the star formation -- gas relation. The assumption that the $\H2$ fraction of the gas depends on $f_{Z}$ and not on $Z$ deserves a comment. Equation (\ref{eq:fH2}) shows that the $\H2$ fraction depends primarily on the dust optical depth (Equation \ref{eq:tauc}). The latter is proportional to  $Z\Sigma_{\rm g}\propto{}f_{Z}$. 

Let $f_{X}=f_{X,{\rm eq}}+\delta_{X}$ be a small perturbation around the equilibrium state $(f_{Z,{\rm eq}},f_{g,{\rm eq}},f_{s,{\rm eq}})$ (see Equations \ref{eq:fgeq} and \ref{eq:fgseq}). The evolution of the perturbation is governed by the following system of equations
\begin{flalign*}
\dot{f}_{X} &= \dot{\delta_X} = g_{X}(f_{Z},f_{g},f_{s}) \\
&\approx{} \sum_{Y\in\{Z,g,s\}}\delta_{Y}\frac{\partial{}g_{X}}{\partial{}f_{Y}}(f_{Z,{\rm eq}},f_{g,{\rm eq}},f_{s,{\rm eq}}).
\end{flalign*}
If all eigenvalues of the Jacobian $J_{X,Y} = \frac{\partial{}g_{X}}{\partial{}f_{Y}}$ have a real value less than zero, then a system in the state $(f_{Z,{\rm eq}},f_{g,{\rm eq}},f_{s,{\rm eq}})$ is stable against small perturbations.

In the following we use the additional shorthands $E=\frac{\partial{}\ln{f_\H2}}{\partial{}\ln{}f_{Z}}(f_{Z,{\rm eq}})$ and $T=M_{\rm halo}/\dot{M}_{\rm halo}$. We find that 
\begin{flalign*}
J_{Z,Z}&=\frac{1}{T}\left[ \left(f_{Z,{\rm eq}} - C Z_{\rm IGM}\right) \left( \frac{nE}{ f_{Z,{\rm eq}}} - \frac{A}{Bf_{g,{\rm eq}}-Af_{Z,{\rm eq}}} \right) -1 \right] \\
J_{Z,g}&=\frac{1}{T}(f_{Z,{\rm eq}} - C Z_{\rm IGM}) \left[ \frac{n}{f_{g,{\rm eq}}} + \frac{f_{Z,{\rm eq}}}{f_{g,{\rm eq}}}\frac{A}{Bf_{g,{\rm eq}}-Af_{Z,{\rm eq}}}   \right] \\
J_{Z,s}&=0 \\
J_{g,Z}&=- \frac{1}{T}\left[ \frac{nE}{f_{Z,{\rm eq}}}(C - f_{g,{\rm eq}}) \right] \\
J_{g,g}&=- \frac{1}{T}\left[ \frac{n}{f_{g,{\rm eq}}}(C - f_{g,{\rm eq}}) + 1 \right] \\
J_{g,s}&=0 \\
J_{s,Z}&=\frac{1}{T}\left[ \frac{nE}{f_{Z,{\rm eq}}}( f_{s,{\rm eq}} - D ) \right] \\
J_{s,g}&=\frac{1}{T}\frac{n}{f_{g,{\rm eq}}}( f_{s,{\rm eq}} - D ) \\
J_{s,s}&=-\frac{1}{T}.
\end{flalign*}
Replacing $(f_{Z,{\rm eq}}$, $f_{g,{\rm eq}}$, $f_{s,{\rm eq}})$ with the expressions (\ref{eq:fgeq}) -- (\ref{eq:fZeq}) allows us to write $J_{X,Y}$ as a function of the following parameters: $n$, $A$, $B$, $E$, $Z_{\rm IGM}$, and $r$, the ratio between SFRs and gas accretion rate, \emph{for the given equilibrium state}. In particular, $C$ and $D$ drop out, which shows that the stability of the equilibrium does not depend on the values of $f_{\rm g, in}$ or $\epsilon_{\rm in}$. 
\begin{flalign*}
J_{Z,Z}&=\frac{1}{T}\left[  \frac{ n E B r^2 A + (n E Z_{\rm IGM} A + B - n E B) r + Z_{\rm IGM} }{ (B r + Z_{\rm IGM}) (A r - 1) }  \right]\\
J_{Z,g}&=\frac{1}{T}\left[  r \frac{  (n  B  A - B A) r - n B + n Z_{\rm IGM} A - Z_{\rm IGM}  A } { Ar - 1}  \right]\\
J_{Z,s}&=0 \\
J_{g,Z}&= \frac{1}{T} \left[ \frac{n E A r } { (B r + Z_{\rm IGM}) (A r - 1) } \right] \\
J_{g,g}&=\frac{1}{T} \left[  \frac{1 + (n - 1) A r} {A r - 1 } \right]\\
J_{g,s}&=0 \\
J_{s,Z}&=-\frac{1}{T}\left[ \frac{27}{50} \frac{r n E}{ (B r + Z_{\rm IGM}) (A r -1)}   \right] \\
J_{s,g}&=-\frac{1}{T}\left[ \frac{27}{50} \frac{nr}{A r -1} \right] \\
J_{s,s}&=-\frac{1}{T}.
\end{flalign*}
Clearly, $J$ has a singularity at $Ar - 1$ as expected from Equation ($\ref{eq:fgeq}$) and it has an eigenvalue -1. The convergence to the equilibrium solution occurs on the timescale $T$ if the other two eigenvalues have real components that are smaller or equal to this eigenvalue. The general expression for the eigenvalues of $J$ is rather lengthy and we do not reproduce it here. Instead, we compute the real part of the eigenvalues numerically using the default parameters for $A$,  $B$ and $Z_{\rm IGM}$. The parameter $B$ depends on $\zeta$ and thus on halo mass. We adopted $\zeta=0$ which results in the most stringent limit on stability.

\begin{figure}
\begin{tabular}{c}
\includegraphics[width=80mm]{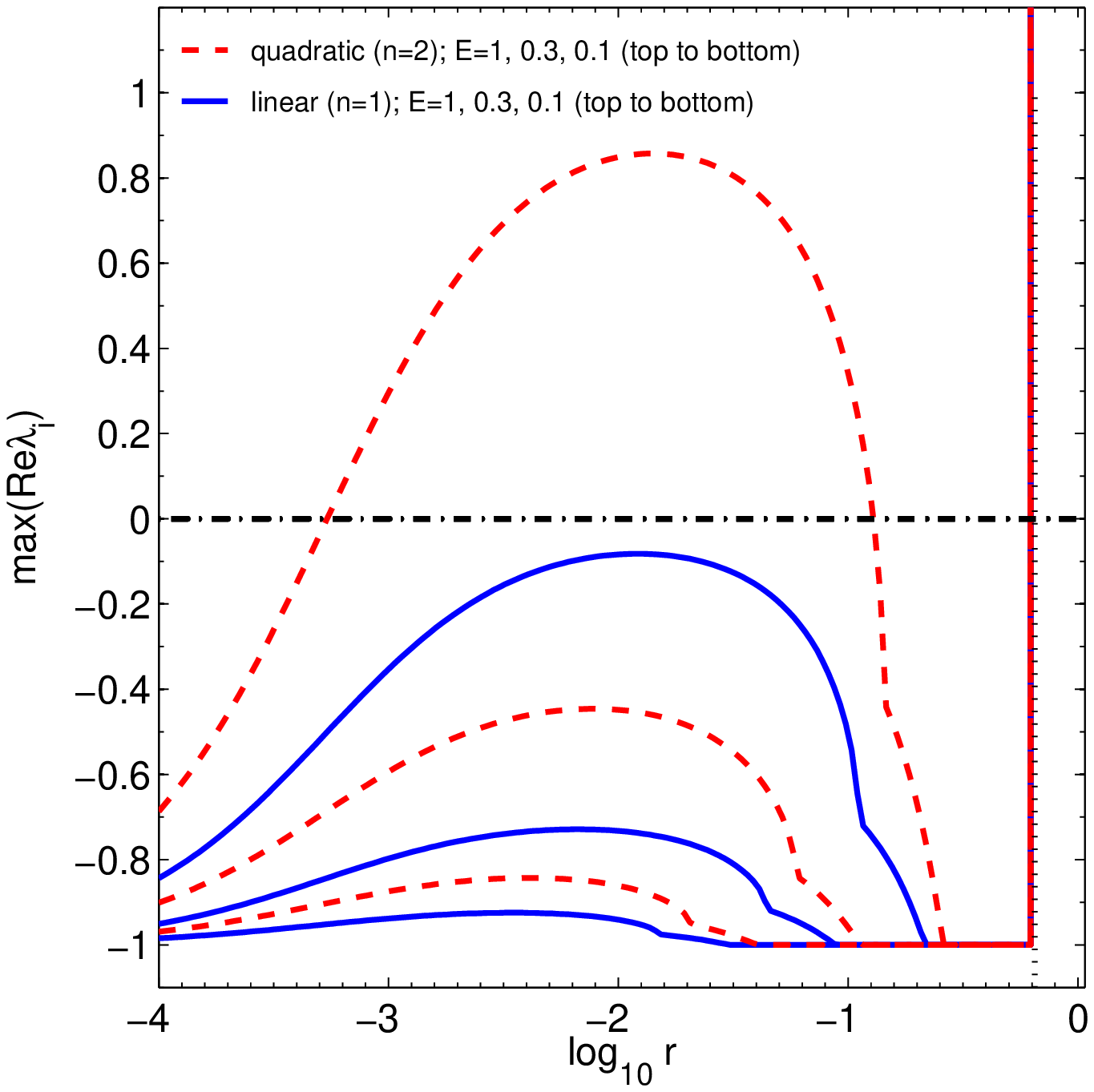}
\end{tabular}
\caption{Maximum of the real part of the eigenvalues of $J\,T$ as function of the ratio between SFR and gas accretion rate. The equilibrium is stable if the maximum of the real part of the eigenvalues is less than zero (horizontal dot-dashed line). Results for a linear  ($n=1$) star formation -- $\H2$ relation are shown by the three solid lines. The individual lines correspond to $E=1$, $E=0.3$, and $E=0.1$ from top to bottom. Dashed lines show the corresponding results for a quadratic star formation -- $\H2$ relation ($n=2$). The equilibrium is stable for $r\lesssim{}1/A$ (vertical dotted line) if $E\lesssim{}0.5-1$ (depending on $n$), i.e., as long as the $\H2$ fraction is not extremely sensitive to changes in the metal fraction.}
\label{figA:Stability}
\end{figure}

We show in Fig.~\ref{figA:Stability} the result of our numerical analysis for $n=1$ and $n=2$. The maximum of the real part of the eigenvalues is less than zero (and hence the equilibrium is stable) for any $r<1/A$ as long as $E$ is sufficiently small ($E\lesssim{}0.5-1$). Hence, as long as the $\H2$ fraction is not extremely sensitive to changes in the metal fraction, the equilibrium is stable.

\end{document}